\documentclass{aa}  

\usepackage{graphicx}
\usepackage{txfonts}
\usepackage{natbib}
\bibpunct{(}{)}{;}{a}{}{,}
\begin{document} 

\newcounter{subtable}

   \title{The chemical composition of $\alpha$ Cen AB revisited
\thanks{Based on observations collected at the La Silla Observatory, ESO (Chile) with the HARPS and FEROS spectrographs.}
\thanks{Table \ref{tab_appendix_EWs} is only available in electronic form at the CDS via anonymous ftp to {\tt cdsarc.u-strasbg.fr (130.79.128.5)} or via {\tt http://cdsweb.u-strasbg.fr/cgi-bin/qcat?J/A+A/???/???}}
}

   \author{Thierry Morel}

   \institute{Space sciences, Technologies and Astrophysics Research (STAR) Institute, Universit\'e de Li\`ege, Quartier Agora, All\'ee du 6 Ao\^ut 19c, B\^at. B5C, B4000-Li\`ege, Belgium\\
              \email{morel@astro.ulg.ac.be}             }

   \date{Received 29 March 2018; accepted 22 April 2018}

 \abstract{The two solar-like stars $\alpha$ Cen A and B have long served as cornerstones for stellar physics in virtue of their immediate proximity, association in a visual binary, and masses that bracket that of the Sun. The recent detection of a terrestrial planet in the cool, suspected tertiary Proxima Cen now makes the system also of prime interest in the context of planetary studies. It is therefore of fundamental importance to tightly constrain the properties of the individual stellar components. We present a fully self-consistent, line-by-line differential abundance analysis of $\alpha$ Cen AB based on high-quality HARPS data. Various line lists are used and analysis strategies implemented to improve the reliability of the results. Abundances of 21 species with a typical precision of 0.02-0.03 dex are reported. We find that the chemical composition of the two stars is not scaled solar (e.g. Na and Ni excess, depletion of neutron-capture elements), but that their patterns are strikingly similar, with a mean abundance difference (A -- B) with respect to hydrogen of --0.01$\pm$0.04 dex. Much of the scatter may be ascribed to physical effects that are not fully removed through a differential analysis because of the mismatch in parameters between the two components. We derive an age for the system from abundance indicators (e.g. [Y/Mg] and [Y/Al]) that is slightly larger than solar and in agreement with most asteroseismic results. Assuming coeval formation for the three components belonging to the system, this implies an age of about $\sim$6 Gyrs for the M dwarf hosting the terrestrial planet Proxima Cen b. After correction for Galactic chemical evolution effects, we find a trend between the abundance ratios and condensation temperature in $\alpha$ Cen A akin to that of the Sun. However, taking this finding as evidence for the sequestration of rocky material locked up in planets may be premature given that a clear link between the two phenomena remains to be established. The similarity between the abundance pattern of the binary components argues against the swallowing of a massive planet by one of the stars after the convective zones have shrunk to their present-day sizes.}


\maketitle
%
%-------------------------------------------------------------------

\section{Introduction}\label{sect_introduction}

Binary components with similar characteristics (e.g. where the magnitude of atomic diffusion effects is nearly identical) are expected to exhibit the same photospheric chemical composition. However, in some rare cases, dedicated surveys have hinted at differences in bulk metallicity for stars in binaries that, if real, might be related to the ingestion of metal-rich, rocky material \citep[][]{desidera04,desidera06}. Much more convincing evidence is being provided thanks to the development of new techniques and dramatic improvements in data quality. Small differences (at the $\sim$0.01-dex level) in the abundance patterns of binary components have been revealed in the last few years, the most documented examples being 16 Cyg \citep[e.g.][]{ramirez11,tucci_maia14,nissen17} and XO-2 \citep[e.g.][]{biazzo15,ramirez15,teske15}. Interestingly, the deviations are different for volatile and refractory elements, which is interpreted as a consequence of the sequestration of rocky material in their direct environment and/or the swallowing of planetary material. It has also been argued that such abundance anomalies might constitute the signature of planetary formation and be used to identify terrestrial planet-host candidates \citep[e.g.][]{melendez12}. Very precise stellar abundance analyses of binaries therefore hold the promise of providing valuable insights into the formation and evolution of planetary systems. 

Despite being our nearest neighbour at only $\sim$1.3 pc, little is known about the existence or lack thereof of planets in $\alpha$ Centauri. This triple system is therefore a target of choice for a detailed abundance study. The main pair is made up of a solar analogue (\object{$\alpha$ Cen A}, HR 5459, HD 128620, HIP 71683; G2 V) and a cooler secondary (\object{$\alpha$ Cen B}, HD 128621, HIP 71681; K1 V). It is a long-period, eccentric binary seen almost edge-on \citep[$\cal P_\mathrm{orb}$ $\sim$ 79.9 yr, $e$ $\sim$ 0.52, and $i$ $\sim$ 79$\degr$;][]{pourbaix16}. The inner pair is believed to be weakly gravitationally bound \citep[e.g.][]{kervella17b} to a distant, faint tertiary sharing the same proper motion (\object{Proxima Cen}, GJ 551; M5.5 V). 

A candidate terrestrial planet (\object{Proxima Cen b}) potentially orbiting within the habitable zone of the third component was recently discovered through radial-velocity (RV) monitoring \citep{anglada_escude16}. As we discuss in the following, it is likely that $\alpha$ Cen AB do not host giant planets. There is no conclusive evidence for lower-mass planets either despite theoretical arguments suggesting that terrestrial systems might have formed on stable orbits in the inner pair \citep[e.g.][]{quintana02,guedes08,quarles18}. The detection  of a transit signal in photometric data requires a very favourable geometrical configuration, while revealing the reflex motion of a planet in the sub-Neptune mass regime through RV variations remains extremely challenging in magnetically active, solar-like stars. Any indication, even indirect, of the presence of putative planets in $\alpha$ Cen AB based on stellar abundances is therefore valuable. It is also timely in view of the major observational efforts currently being undertaken to find potentially habitable worlds in the system.\footnote{See, e.g. ESO press release at: \\ {\tt https://www.eso.org/public/unitedkingdom/news/eso1702/}} Furthermore, improving the basic parameters of $\alpha$ Cen AB turns out to be relevant for a better characterisation of the properties of Proxima Cen b \citep[see, e.g.][]{barnes16} given that achieving stringent constraints on some fundamental quantities (e.g. chemical composition, age) is fraught with difficulties in late-M dwarfs. 

From a different perspective,  $\alpha$ Cen AB also hold great potential for stellar physics thanks to the wide array of accurate and weakly model-dependent observations available. The two stars have long been used as testbeds for stellar interior and atmosphere models \citep[e.g.][]{guenther00,kervella17a} or, more recently, as benchmarks for the {\it Gaia} mission \citep[][]{heiter15} and massive spectroscopic surveys \citep[e.g.][]{pancino17}. Tightly constraining their parameters is therefore worthwhile for a wide range of issues in stellar physics. For instance, accurate non-seismic constraints allow more robust inferences to be made about the internal structure of solar-like stars from a modelling of their $p$-mode oscillations \citep[e.g.][]{eggenberger04}.

\section{Goals of this study}\label{sect_goals}

Not surprisingly given their brightness, $\alpha$ Cen AB have been the subject of numerous abundance studies in the past. However, although the metal-rich nature of the system has been known for decades \citep[e.g.][]{french71}, we show below that there are still significant study-to-study discrepancies in the detailed chemical composition of the two components \citep[as also noted by][]{porto_de_mello08}. Furthermore, to the best of our knowledge only a few spectroscopic investigations appear to have derived the abundance pattern of {\it both} components in a fully self-consistent way. Even fewer have performed a strictly differential analysis with respect to the Sun that allows one to reach a much higher level of accuracy for solar analogues, as extensively discussed in the recent literature \citep[e.g.][]{nissen15,melendez09}.

 Our objective is to carry out an in-depth abundance study reaching a precision that would allow us to firmly assess the level of similarity in the stellar abundance patterns and to detect the signature of planetary formation, if any. In addition, following the recent recognition that some abundance ratios are very sensitive to age \citep[e.g.][]{nissen15}, we aim at putting constraints on the evolutionary state of the system solely based on abundance indicators. To reach these goals, we have enforced a strict line-by-line differential analysis using various line lists (totalling about 450 spectral features) to reach a typical precision of 0.02-0.03 dex for an unprecedented number of chemical elements. 

\section{Observational material}\label{sect_observations}

\subsection{HARPS data}\label{sect_observations_harps}

Except for oxygen (see Sect.~\ref{sect_observations_feros}), we made use of high-resolution HARPS spectra retrieved from the online library of {\it Gaia} FGK benchmarks \citep{blanco_cuaresma14b}.\footnote{Available at {\tt https://www.blancocuaresma.com/s/benchmarkstars}. Two spectra are available for $\alpha$ Cen A. We decided to use the exposure with a slightly lower S/N because of a better blaze correction.} A solar reflected spectrum (co-addition of Ceres, Ganymede, and Vesta spectra) to be used as reference for the differential analysis was also retrieved. As discussed by \citet{melendez12} and \citet{bedell14}, the use of a solar reflected spectrum obtained with the same instrument maximises the precision of abundance analyses. As they also showed, co-adding spectra from various reflecting sources is not an issue, as long as they are obtained with an identical instrumental set-up. 

The spectra cover the spectral domain 480-680 nm, with a gap between 530.4 and 533.8 nm. Difficulties related to the placement of the continuum level are encountered at shorter wavelengths because of strong line crowding, while the presence of strong telluric bands severely restricts the number of useful lines in the red. The signal-to-noise ratio (S/N) of the spectra ranges from 381 to 514 across the whole spectral range \citep{blanco_cuaresma14b}. The spectra are provided in the laboratory rest frame and with the initial reduction steps (e.g. merging of the orders, wavelength calibration) already performed \citep[see][for further details]{blanco_cuaresma14b}. They were normalised to the continuum by fitting low-order cubic spline or Legendre polynomials to the line-free regions using standard tasks implemented in the IRAF\footnote{{\tt IRAF} is distributed by the National Optical Astronomy Observatories, operated by the Association of Universities for Research in Astronomy, Inc., under cooperative agreement with the National Science Foundation.} software. 

The spectra have a nominal resolving power, $R$, of about 115\,000. The incentive and starting point of our study consisted in determining the parameters of $\alpha$ Cen A based on a spectrum degraded to $R$ = 65\,000 as part of a hare-and-hound campaign for the preparation of the PLATO\,2.0 mission \citep{rauer14}. For this exercise, spectra were provided to different research groups without any prior knowledge of the stars to be analysed in order to assess to what extent the reference parameters (discussed below) are recovered as a function of, for example, resolving power. We decided to proceed with the full abundance analysis of $\alpha$ Cen AB using data with this spectral resolution. As shown in Sect.~\ref{sect_uncertainties}, this choice has no discernable impact on our results. The convolution was directly performed online with the iSpec software \citep{blanco_cuaresma14a}. 

The observations of $\alpha$ Cen AB were secured on 8 April 2005 \citep{blanco_cuaresma14b} when the components were widely separated ($\sim$10\arcsec). Furthermore, the fibre entrance aperture projected on the sky of HARPS in the high-resolution (HAM) mode is only 1.0\arcsec. There is therefore no contamination of either spectrum by that of the other component.

\subsection{FEROS data}\label{sect_observations_feros}

No attempts were made to derive the oxygen abundance from \ion{[O}{i]} $\lambda$630.0 because this feature is very weak and its strength too uncertain. We use the \ion{O}{i} triplet at $\sim$777.4 nm
instead. Because it is not covered by the HARPS spectra, we base our analysis on the weighted (by the S/N) average of numerous exposures available in the FEROS archives ($R$ $\sim$ 47\,000). For the Sun, asteroid spectra were considered. All the spectra were normalised as described in Sect.~\ref{sect_observations_harps}. The spectra of $\alpha$ Cen AB were obtained during the period 2004--2007, when the binary separation was once again much larger than the diameter of the fibre aperture (2.0\arcsec).

\section{Methods of analysis} \label{sect_methods}

The stellar parameters and abundances of 21 metal species were self-consistently determined from the spectra using a curve-of-growth analysis, MARCS model atmospheres \citep{gustafsson08}, and the 2017 version of the line-analysis software MOOG originally developed by \citet{sneden73}. 

\subsection{Line selection} \label{sect_line_selection}

Our results are sensitive to a number of assumptions. Chief among them is the choice of the line list and of the family of model atmospheres. As discussed in Sect.~\ref{sect_uncertainties}, although the use of Kurucz models leads to relatively modest differences, the selection of the diagnostic lines appears more critical. To investigate this aspect further, we carried out the analysis using ten line lists commonly used in the literature for solar-type stars \citep{bensby14,biazzo12,chen00,feltzing01,jofre14,jofre15,melendez14,morel14,reddy03,sousa08}. The line lists of \citet{jofre14,jofre15} contain lines of iron and other metals, respectively. For the former, we adopted the so-called FGDa line list. For the latter, we made use of the ``golden'' set of lines for FG dwarfs. They have been shown by \citet{jofre14,jofre15} to be appropriate for the analysis of both $\alpha$ Cen A and B. The ten line lists widely differ in their basic properties (e.g. number of features, source of oscillator strengths and damping parameters). It is beyond the scope of this paper to discuss these differences in detail, but a number of points are worth mentioning. First, the study of \citet{feltzing01} concentrated on metal-rich stars and, as a result, the features are generally weaker than those in the other lists. On the contrary, the lists of \citet{chen00} and, to a lesser extent, \citet{jofre14} are mainly made up of strong lines, which makes the determination of microturbulence, $\xi$, quite hazardous. Second, the line list of \citet{morel14} was primarily built for the analysis of cooler red giants. However, there is a very significant overlap ($\sim$95\%) with the other line lists, indicating that it is also appropriate for solar-like dwarfs. Indeed, the few features not included in other line lists were found to be outliers (likely because of blends occurring at higher temperatures) and eventually rejected. Finally, the line list of \citet{sousa08} only contains iron features.

Hyperfine structure (HFS) and isotopic splitting were taken into account for Sc, V, Mn, Co, and Cu using atomic data from the Kurucz database\footnote{Available at {\tt http://kurucz.harvard.edu/linelists.html}.} and assuming the Cu isotopic ratio from \citet{asplund09}. The corrections are not significant for the other odd-$Z$ elements or Ba. The {\tt blends} driver in MOOG was used for the analysis. Although the differential HFS corrections can dramatically vary from one line to another, they are on average not very large for $\alpha$ Cen A (at most --0.11 dex for \ion{Mn}{i}). However, they are much more significant in $\alpha$ Cen B for a given ion, with mean corrections amounting to --0.06, --0.02, --0.15, --0.15, and --0.01 dex for \ion{Sc}{i}, \ion{Sc}{ii}, \ion{V}{i}, \ion{Co}{i}, and \ion{Cu}{i}, respectively.

The equivalent widths (EWs) were measured manually assuming Gaussian profiles (multiple fits were used for well-resolved blends). Despite being extremely tedious and time consuming, manual measurements have to be preferred over those obtained from (semi)automatic procedures. Because the HFS corrections are generally uncertain, we only retained Sc, V, Mn, Co, and Cu lines that have a profile that is nearly Gaussian. This ensures that they suffer as little as possible from HFS broadening, but it should be kept in mind that the necessarily imperfect treatment of this effect constitutes an additional source of uncertainty. To minimise difficulties related to strong spectral features (i.e. uncertain EW measurements and damping parameters, non-linear part of curve of growth), lines with RW = $\log$ (EW/$\lambda$) $>$ --4.80 were removed following, for example, \citet{jofre14}. The only exceptions were the strong \ion{Mg}{i} $\lambda$571.1 and \ion{Zn}{i} $\lambda$481.0 lines when they were the only magnesium and zinc features measurable. In addition, lines significantly affected by telluric features based on the atlas of \citet{hinkle00} were discarded. The selected spectral lines and corresponding EW measurements are presented in Table \ref{tab_appendix_EWs}. 

\subsection{Determination of stellar parameters and abundances} \label{sect_parameter_determination}

We carried out a strictly differential, line-by-line analysis relative to the Sun \citep[see, e.g.][]{melendez09}. For the solar analysis, $T_\mathrm{eff}$ and $\log g$ were held fixed to 5777 K and 4.44 dex, respectively, whereas the microturbulence was left as a free parameter. Such a differential analysis with respect to a reference star having similar parameters minimises systematic errors arising either from the data treatment (e.g. continuum placement), physical effects (e.g. inaccuracies of model atmospheres, departures from local thermodynamic equilibrium [LTE]), or uncertain atomic data. It also ensures the highest level of consistency because exactly the same set of lines is used for the star under study and that used as reference. However, we caution that our targets, especially $\alpha$ Cen B, have parameters significantly deviating from solar. Therefore, this is expected to limit the precision of the analysis. 

It is customary in abundance studies of binaries to perform a differential analysis of the two components relative to each other. This is because the stars have parameters that are often closer to each other than they are to solar \citep[e.g.][]{teske16b}. However, here we do not attempt to follow this approach, as we find it unlikely that it would improve the precision of our results. First, $\alpha$ Cen B is as dissimilar in terms of parameters to $\alpha$ Cen A as it is to the Sun: although the metallicity is identical, the $T_\mathrm{eff}$ and $\log g$ discrepancy is even larger. Second, we would loose the advantage of having a reference star with perfectly known parameters. 

The model parameters ($T_\mathrm{eff}$, $\log g$, $\xi$, and [Fe/H]) were iteratively modified until the following conditions were simultaneously fulfilled: (1) the \ion{Fe}{i} abundances exhibit no trend with lower excitation potential (LEP) or RW; (2) the mean abundances derived from the \ion{Fe}{i} and \ion{Fe}{ii} lines are identical; and (3) the iron abundance is consistent with the model values. As the abundances of the $\alpha$ elements are found to be solar within the errors, no models with enhancements of these elements were considered. This approach, which enforces both excitation and ionisation balance of iron and does not make use of any priors, is referred to as the ``unconstrained analysis'' in the following. 

As described in Sect.~\ref{sect_line_selection}, and similarly to \citet{morel13}, we used several line lists to improve the precision of our analysis. The results obtained for a given quantity (either a stellar parameter or an abundance ratio) based on the $i$th line list, $x_i$, were weighted by their total uncertainties, $\sigma_i$, to obtain the final values, $\overline{x}$, with the well-known formulae:

\begin{equation}
\overline{x} = {\Sigma (x_i/\sigma_i^2) \over \Sigma (1/\sigma_i^2)} \hspace*{0.5cm} \mathrm{and}
\label{equat_mean}
\end{equation}

\begin{equation}
\sigma(\overline{x}) = \left[ {1 \over \Sigma (1/\sigma_i^2)} \right]^{1/2} \mathrm{.}
\label{equat_error_mean}
\end{equation}

As summarised by \citet{heiter15}, the parameters of both stars are accurately known thanks to nearly model-independent techniques: $T_\mathrm{eff}$ from interferometry and $\log g$ from asteroseismology. The former offers an opportunity to assess the accuracy of our $T_\mathrm{eff}$ estimates (Sect.~\ref{sect_results_parameters}). We adopt as reference the values recently determined by \citet{kervella17a} from combining their VLTI/PIONIER measurements of the limb-darkened linear radii with the bolometric fluxes of \citet{boyajian13}: 5795$\pm$19 and 5231$\pm$21 K for $\alpha$ Cen A and B, respectively. These values are fully consistent with those reported by \citet{heiter15} based on different interferometric measurements \citep{kervella03,bigot06}. On the other hand, it is becoming increasingly popular to use the asteroseismic gravity as prior for the spectroscopic analysis \citep[e.g.][]{huber13}. We therefore also conducted a constrained analysis whereby $\log g$ is frozen to the asteroseimic value. We assume the recommended values quoted by \citet{heiter15}: $\log g$ = 4.32$\pm$0.02 and 4.53$\pm$0.02 dex for $\alpha$ Cen A and B, respectively. They are computed from scaling relations (see their Eq. 3) making use of the frequency of maximum oscillation power, $\nu_\mathrm{max}$, determined from RV time series by \citet{kjeldsen08}. The $\log g$ values are compatible within the errors with those derived by grid-based seismic studies \citep[e.g.][]{creevey13} or through a combination of interferometric, astrometric, and spectroscopic measurements yielding the stellar radii and masses \citep[e.g.][]{kervella17a}.\footnote{We do not adopt the potentially more precise estimates provided by the latter technique, as there are significant discrepancies in dynamical stellar masses between various studies (compare, e.g. \citealt{pourbaix16} and \citealt{kervella16}).}

\begin{figure}[h!]
\centering
\includegraphics[trim=30 165 30 80,clip,width=\hsize]{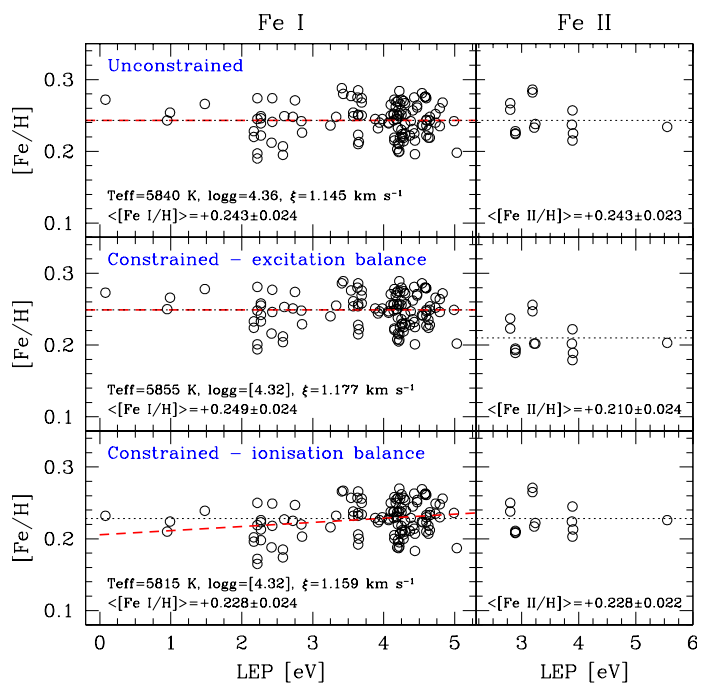}
\caption{Illustration of the three methods employed in the case of the differential analysis of $\alpha$ Cen A with the line list of \citet{bensby14}. {\it Upper panels:} unconstrained analysis requiring both excitation and ionisation equilibrium of iron; {\it middle panels}: constrained analysis assuming excitation balance of \ion{Fe}{i}; {\it bottom panels}: constrained analysis assuming ionisation balance of iron. The dotted horizontal lines indicate the mean iron abundances, while the red dashed lines show the fit to the \ion{Fe}{i} abundances as a function of LEP.}
\label{fig_methods}
\end{figure}

One consequence of enforcing a constrained analysis in the present case is that excitation and ionisation balance of iron can no longer be simultaneously fulfilled. We explore the two possible ways to proceed in the following; namely, adjusting $T_\mathrm{eff}$ to either satisfy excitation equilibrium of the \ion{Fe}{i} lines or iron ionisation equilibrium. We note that in the former case, \ion{Fe}{i} and \ion{Fe}{ii} yield discrepant mean abundances. The three situations that can be encountered are illustrated in Fig.~\ref{fig_methods}.

\subsection{Computation of uncertainties}\label{sect_uncertainties}

Various sources of abundance uncertainties were considered. First, there are those related to the determination of the atmospheric parameters ($T_\mathrm{eff}$, $\log g$, and $\xi$). The effect on the abundances was examined by altering one of the parameters by its uncertainty, while keeping the other two fixed. To estimate the uncertainty in $T_\mathrm{eff}$, for instance, we considered the range over which the slope of the relation between the \ion{Fe}{i} abundances and LEP is consistent with zero within the uncertainties. As the parameters of the model are interdependent, however, changes in one of them are necessarily accompanied by variations in the other two, and covariance terms also need to be taken into account. Accordingly, in turn two of the parameters were also adjusted while the third one was varied by the relevant uncertainty. Once again, the analysis was repeated using this new set of parameters to estimate the impact on the abundances. Second, we explored to what extent the choice of another family of model atmospheres affects the results. To this end, we redetermined the abundances using the line list of \citet{melendez14}, but with Kurucz models computed with the ATLAS9 code ported under Linux \citep{sbordone05}. The largest differences are found for $\alpha$ Cen B, but even in that case they appear to be small (Kurucz -- MARCS): $\Delta T_\mathrm{eff}$ = +5 K, $\Delta \log g$ = +0.02 dex, and abundance ratios deviating by less than 0.02 dex. Third, the line-to-line scatter was taken into account. A rather generous value of 0.05 dex (it is typically 0.03 dex) was assumed when only one line was used for a given ion. The final uncertainty was taken as the quadratic sum of all these various errors.

We made use of HARPS spectra with a resolving power degraded from $R$ $\sim$ 115\,000 to 65\,000 (see  Sect.~\ref{sect_observations_harps}). We repeated the unconstrained analysis of $\alpha$ Cen A using the original spectra and the line list of \citet{melendez14}, but found negligible differences compared to the default results (e.g. abundances deviating by less than 0.01 dex).

\section{Results}\label{sect_results}

\subsection{Stellar parameters}\label{sect_results_parameters}

\begin{figure*}[h!]
\begin{minipage}[t]{0.5\textwidth}
\centering
\includegraphics[trim=40 185 205 150,clip,width=0.95\textwidth]{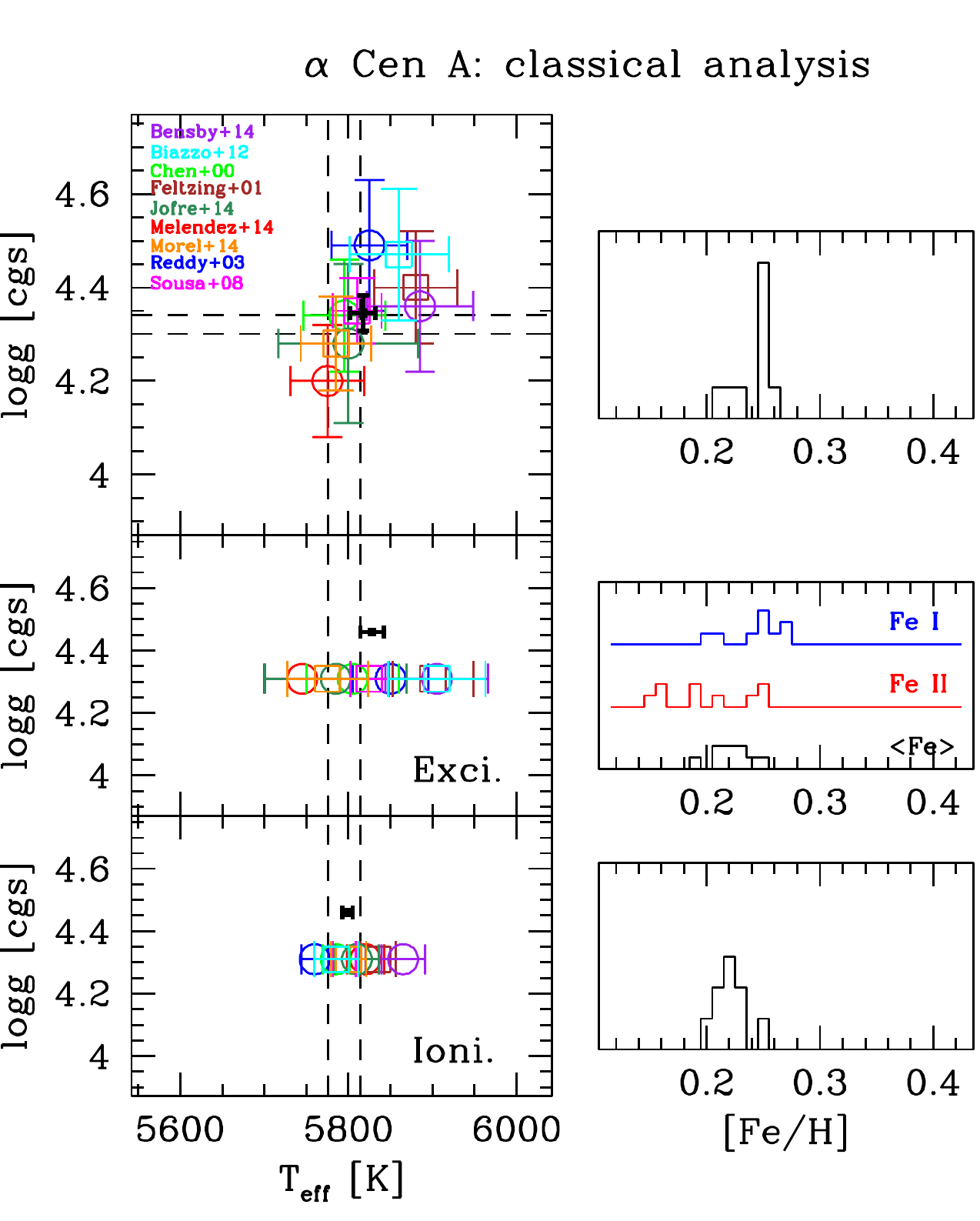}
\end{minipage}
\begin{minipage}[t]{0.5\textwidth}
\centering
\includegraphics[trim=40 185 205 150,clip,width=0.95\textwidth]{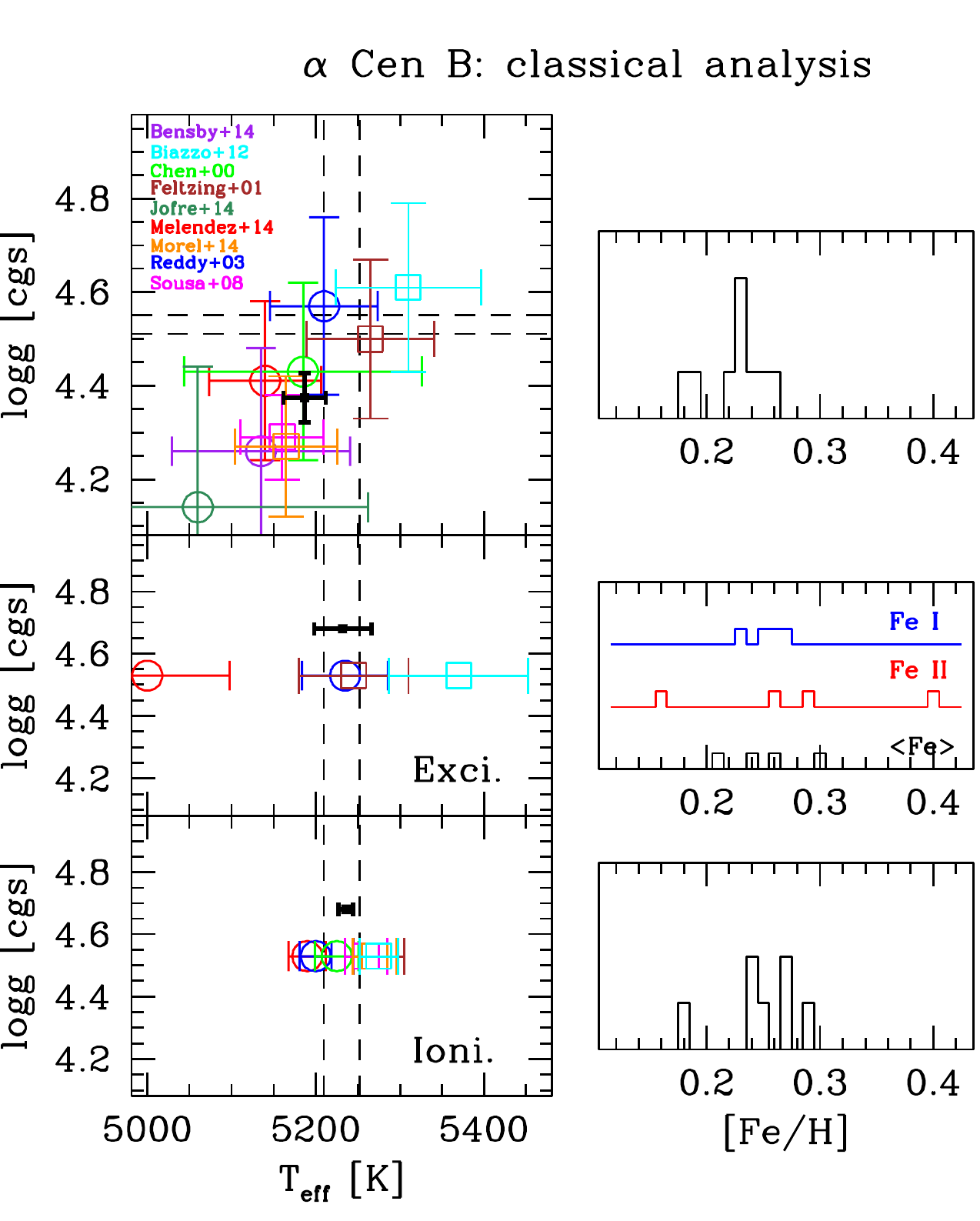}
\end{minipage}
\begin{minipage}[t]{0.5\textwidth}
\centering
\includegraphics[trim=40 180 205 140,clip,width=0.95\textwidth]{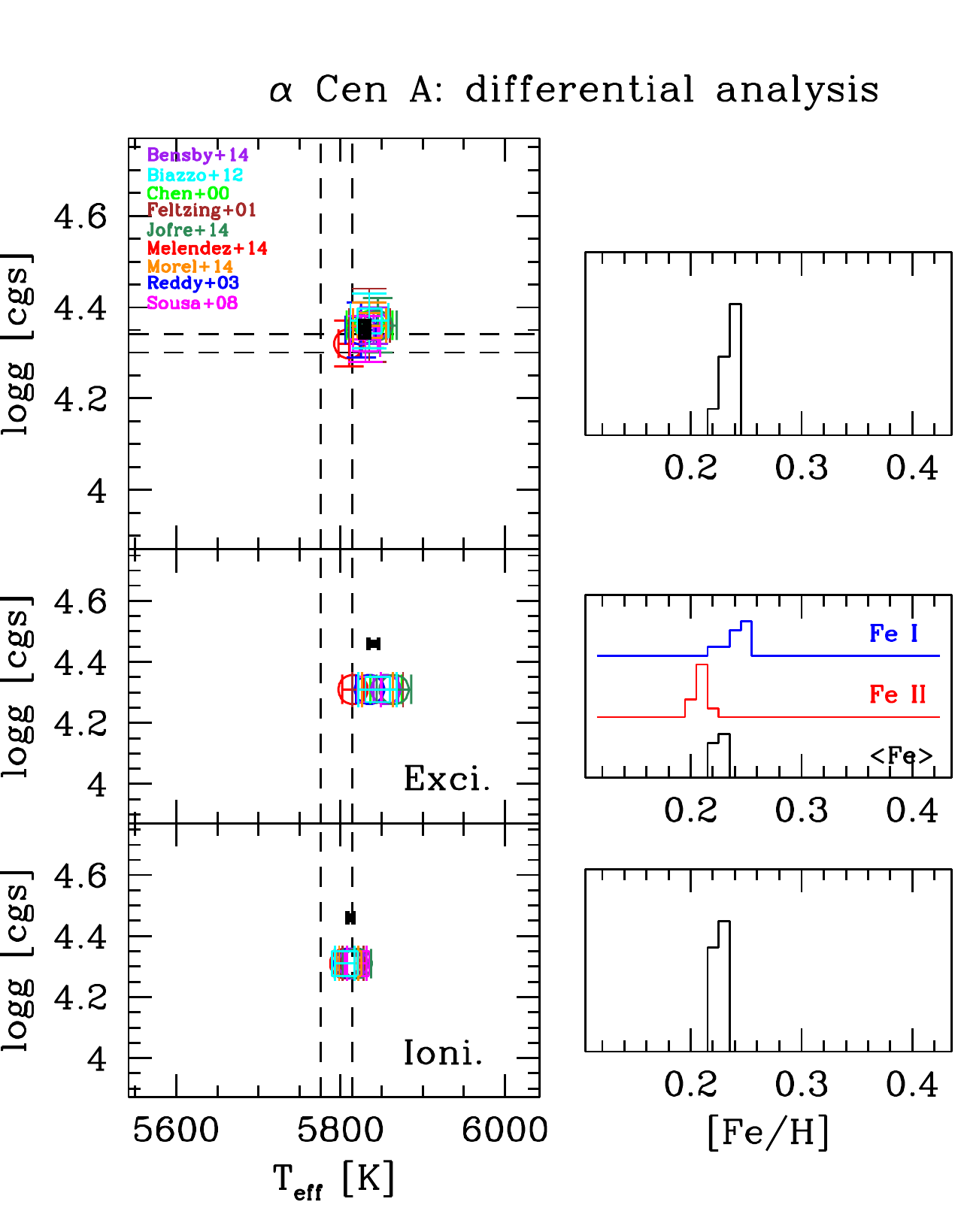}
\end{minipage}
\begin{minipage}[t]{0.5\textwidth}
\centering
\includegraphics[trim=40 180 205 140,clip,width=0.95\textwidth]{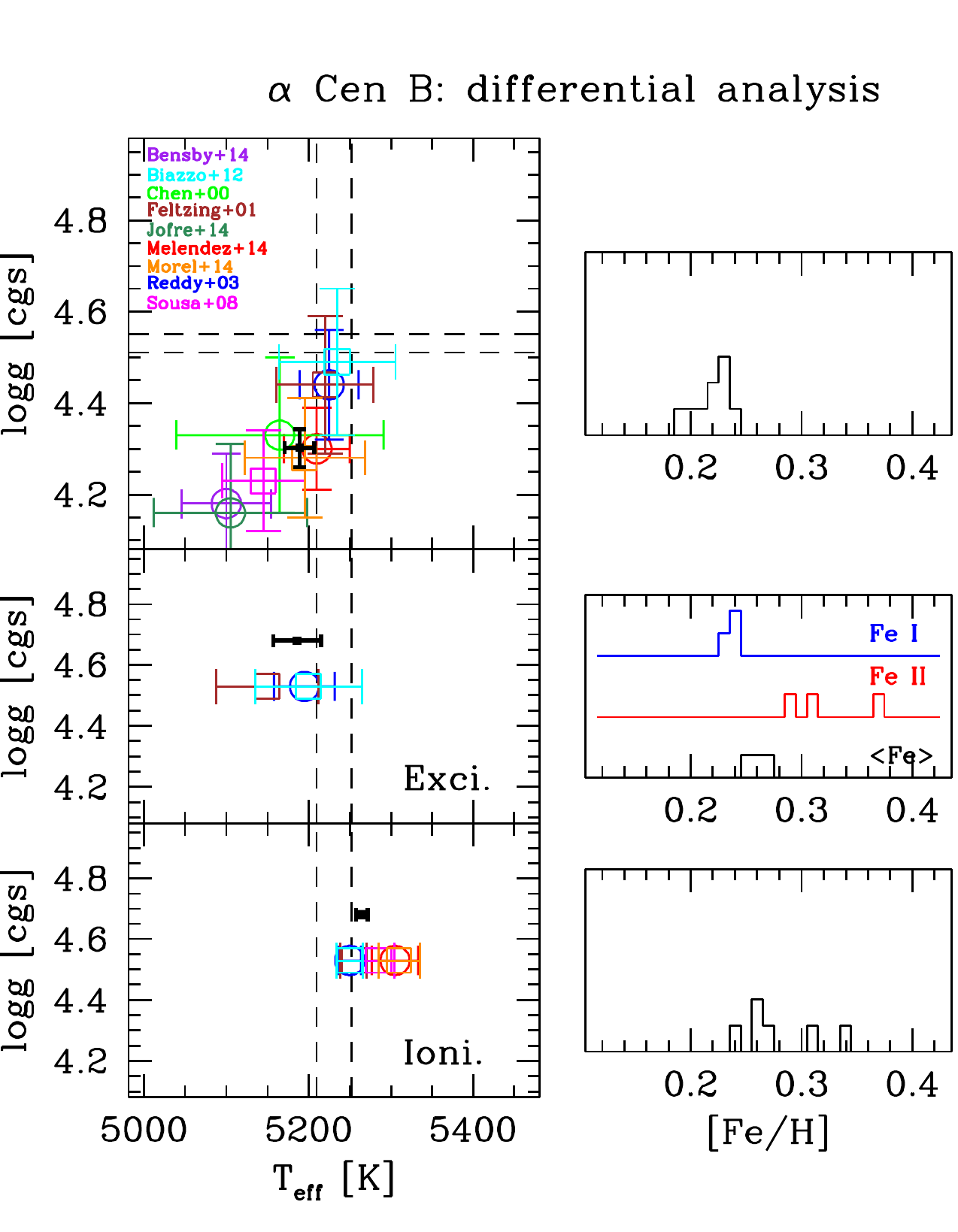}
\end{minipage}
\caption{Results of the classical ({\it top}) and differential ({\it bottom}) analyses of $\alpha$ Cen A ({\it left}) and $\alpha$ Cen B ({\it right}). Various analyses are considered (see Sect.~\ref{sect_parameter_determination}): unconstrained ({\it top panels}), constrained assuming excitation balance ({\it middle panels}), and constrained assuming ionisation balance ({\it bottom panels}). The $T_\mathrm{eff}$ and $\log g$ values obtained for each line list are shown with a different colour. The vertical and horizontal dashed lines indicate the 1-$\sigma$ range encompassed by the reference $T_\mathrm{eff}$ and $\log g$ values (Sect.~\ref{sect_parameter_determination}). The black error bars indicate the weighted mean of all the results (the position along the $y$-axis in the middle and bottom panels is shifted for clarity). The full results can be found in Tables \ref{tab_appendix_classical_parameters} and \ref{tab_appendix_differential_parameters}.}
\label{fig_parameters}
\end{figure*}

As a preamble, it was mentioned previously that the parameters of the targets differ significantly from solar. It is therefore unclear as to whether a differential analysis relative to the Sun actually improves the precision/accuracy of the results. To investigate this point, a standard analysis (i.e. without a joint analysis of the solar spectrum) was also undertaken. The results are presented in Table \ref{tab_appendix_classical_parameters}. We compare in Fig.~\ref{fig_parameters} ({\it upper panels}) our $T_\mathrm{eff}$ and $\log g$ estimates to those obtained from less model-dependent techniques, namely interferometry and asteroseismology, respectively (see Sect.~\ref{sect_parameter_determination}). Let us first discuss $\alpha$ Cen A. For the three methods considered, there is an overall good agreement between our results and the reference ones.\footnote{We do not expect a perfect coincidence between our excitation/ionisation temperatures, which relate to the physical conditions prevailing in the iron line-formation zone, and the interferometric value, which is more directly tied to the definition of effective temperature.} Even though the choice of the line list can lead to quite different parameters in the unconstrained case (covering a full range of 110 K and 0.29 dex for $T_\mathrm{eff}$ and $\log g$, respectively), the metallicity exhibits a relatively small scatter with all values being identical to within 0.06 dex. The constrained analysis assuming ionisation balance yields more precise $T_\mathrm{eff}$ values (by a factor $\sim$2.5), but this is not the case when enforcing excitation balance. The situation is quite different for $\alpha$ Cen B. First, the choice of the line list has a more profound effect on the results of the unconstrained analysis. Overall, there is also evidence for a systematic underestimation of $\log g$ at the $\sim$0.2-dex level. Even in that case, however, [Fe/H] is nearly identical and as precisely determined as in $\alpha$ Cen A. There is an outstanding dependence between $T_\mathrm{eff}$ and $\log g$ that illustrates the well-known degeneracy between the determination of these two quantities through spectroscopy alone. One would hope that fixing $\log g$ would break the degeneracy and increase the precision of the $T_\mathrm{eff}$ determination. This is indeed what is observed when ionisation equilibrium is enforced. In sharp contrast, fulfilling excitation balance provides very unsatisfactory results. In most cases, it is even impossible to reach convergence because the \ion{Fe}{i} abundances exhibit a noticeable trend as a function of the line strength whatever the microturbulence adopted. Similar problems have been encountered in the literature for relatively cool dwarfs \citep[e.g.][]{jofre14}. 

As can be seen in Fig.~\ref{fig_parameters} ({\it lower panels}) in the case of $\alpha$ Cen A, the choice of the line list becomes basically irrelevant when a differential analysis between two stars with similar parameters is performed. Nearly identical parameters are obtained and the metallicity distributions are strongly peaked. This is especially true when ionisation balance is assumed, that is, the [Fe/H] values differ by less than a mere 0.015 dex. Once again, a different picture is obtained for $\alpha$ Cen B. Most importantly, there does not seem to be much benefit in performing a differential analysis (compare lower and upper panels). This distinct behaviour compared to $\alpha$ Cen A may reflect the fact that, $\alpha$ Cen B being significantly cooler than the Sun, systematic effects (e.g. inadequacies in the atmosphere structure) are not efficiently erased through a differential analysis. The iron line lists selected are made up of features whose formation may be quite different in terms of, for example, depth in the photosphere or sensitivity to departures from LTE. This could explain why the results depend on the choice of the diagnostic lines even in the case of a differential analysis. The mismatch in terms of parameters with respect to the Sun could also be one of the reasons contributing to a line-to-line scatter almost twice as large in $\alpha$ Cen B compared to $\alpha$ Cen A (typically 0.036 vs 0.019 dex). In general, as for the standard analysis, a constrained analysis requiring ionisation balance performs much better. 

In summary, we can draw the following conclusions: 
\begin{itemize}
\item
The spectroscopic parameters are in general in satisfactory agreement with the presumably more accurate values derived from interferometry and asteroseismology, but there is evidence that $\log g$ is underestimated on average by $\sim$0.2 dex in $\alpha$ Cen B.
\item
There is a dramatic improvement in the precision of the results when a differential analysis is enforced for $\alpha$ Cen A. In contrast, there is apparently little benefit, if any, for $\alpha$ Cen B. This is interpreted as being due to parameters that depart significantly from solar in the latter case.
\item
The choice of the line list has very little effect on the results when two stars with similar parameters are analysed in a differential way.
\item
Fulfilling iron ionisation balance increases the precision of the results for the constrained analysis of Sun-like dwarfs. This approach should clearly be preferred over that requiring excitation equilibrium, especially for cool stars where convergence issues might be encountered.
\item
All analyses yield consistent metallicities and lead to a robust value, [Fe/H] $\sim$ +0.23 dex, which comfortably lies within the commonly accepted range for this system (0.20-0.25 dex; Sect.~\ref{sect_abundance_patterns}).
\end{itemize}

Now, the question arises as to what parameters should be adopted for the abundance analysis. First, based on the arguments presented above the results of the standard analysis can be regarded as less precise (at least for $\alpha$ Cen A) and are therefore discarded. Second, we ignore the constrained results assuming excitation equilibrium. A sound assumption could then be to adopt the results of the differential, constrained analysis assuming ionisation balance. However, as can be seen in Fig.~\ref{fig_parameters}, the [Fe/H] values for $\alpha$ Cen B show a relatively large spread and reach suspiciously high values (up to +0.34 dex). Although we have argued that the constrained results are generally more {\it precise}, it should also be kept in mind that they are not necessarily more {\it accurate}. The common belief that freezing $\log g$ improves the performance of spectroscopic analyses has been questioned \citep[e.g.][]{smalley14}. This is an important issue in the era of large stellar samples with asteroseismic data, which has certainly not yet received the attention it deserves. However, this cannot be meaningfully addressed with our limited dataset. We finally decided for the determination of the metal abundances to proceed with the parameters derived from the differential, unconstrained analysis (Table \ref{tab_appendix_differential_parameters}). We note that our philosophy differs from that sometimes adopted in the literature \citep[e.g. in the framework of the {\it Gaia}-ESO survey;][]{smiljanic14} in that our abundances are not derived assuming a single set of recommended parameters. Instead, the abundances for a given line list are derived adopting the corresponding parameters, and are eventually combined to yield the final values.

\subsection{Chemical abundances}\label{sect_results_chemical_abundances}

The abundances obtained for each line list are given for $\alpha$ Cen A and B in Tables \ref{tab_appendix_abundances_alfCenA} and \ref{tab_appendix_abundances_alfCenB}, respectively. We only consider in the following the final abundances obtained from averaging these values (Table \ref{tab_final_results}), as described in Sect.~\ref{sect_parameter_determination}. The Fe, Si, Ca, Sc, Ti, and Cr abundances are derived from lines pertaining to two ionisation stages. For iron, the final value we adopt is the average weighted by the inverse variance of the \ion{Fe}{i}- and \ion{Fe}{ii}-based abundances. For the other elements, we only consider abundances yielded by \ion{Si}{i}, \ion{Ca}{i}, \ion{Sc}{ii}, \ion{Ti}{i}, and \ion{Cr}{i}, as they are based on more features and exhibit a reduced line-to-line scatter. Ionisation balance is fulfilled for all of these species. The only exceptions are calcium and chromium in $\alpha$ Cen B. The origin of the \ion{Ca}{ii} and \ion{Cr}{ii} overabundances is unclear, but might arise from blends affecting the very few weak diagnostic lines. The non-LTE corrections for the features investigated are expected to be small \citep{mashonkina17,bergemann_cescutti10}.

\begin{table*}[h!]
%\scriptsize
\caption{Final results before and after corrections for Galactic chemical evolution (GCE; see Sect.~\ref{sect_discussion_Tc_trends}).}\label{tab_final_results} 
\centering
\begin{tabular}{l|crr|crr|r}
\hline\hline
\multicolumn{1}{c}{}         & \multicolumn{3}{c}{$\alpha$ Cen A}  & \multicolumn{3}{c}{$\alpha$ Cen B} & \multicolumn{1}{c}{A -- B}\\
                             & \multicolumn{1}{c}{$N$} & \multicolumn{1}{c}{Before GCE} & \multicolumn{1}{c}{After GCE} & \multicolumn{1}{|c}{$N$} & \multicolumn{1}{c}{Before GCE} & \multicolumn{1}{c|}{After GCE} & \multicolumn{1}{c}{}\\
\hline
$T_\mathrm{eff}$ [K]          &  9 &   \multicolumn{2}{c|}{5829$\pm$6}     &  9 & \multicolumn{2}{c|}{5189$\pm$18}     & \multicolumn{1}{c}{...}\\$\log g$ [cgs]               &  9 &   \multicolumn{2}{c|}{4.35$\pm$0.02}  &  9 & \multicolumn{2}{c|}{4.30$\pm$0.05}   & \multicolumn{1}{c}{...}\\
$\xi$ [km s$^{-1}$]          &  9 &   \multicolumn{2}{c|}{1.265$\pm$0.012} &  9 & \multicolumn{2}{c|}{0.950$\pm$0.039} &\multicolumn{1}{c}{...}\\
\hline
$[$Fe/H$]$\tablefootmark{a}  &  9 &   0.237$\pm$0.007 & \multicolumn{1}{c|}{...} & 9 &   0.221$\pm$0.016       & \multicolumn{1}{c|}{...} &   0.012$\pm$0.018      \\
$[$C/Fe$]$                   &  2 &   0.025$\pm$0.018 & --0.018$\pm$0.036       & 2 & --0.001$\pm$0.045        & --0.044$\pm$0.054       &   0.037$\pm$0.049        \\
$[$O/Fe$]$                   &  6 & --0.046$\pm$0.014 & --0.062$\pm$0.019       & 6 & --0.074$\pm$0.040        & --0.090$\pm$0.042       &   0.030$\pm$0.043        \\
$[$Na/Fe$]$                  &  7 &   0.094$\pm$0.010 &   0.058$\pm$0.027       & 7 &   0.128$\pm$0.036        &   0.092$\pm$0.044       & --0.029$\pm$0.038        \\
$[$Mg/Fe$]$                  &  6 &   0.013$\pm$0.023 & --0.004$\pm$0.026       & 6 &   0.044$\pm$0.041        &   0.027$\pm$0.043       & --0.028$\pm$0.048        \\
$[$Al/Fe$]$                  &  7 &   0.044$\pm$0.013 &   0.013$\pm$0.025       & 7 &   0.077$\pm$0.031        &   0.046$\pm$0.038       & --0.030$\pm$0.034        \\
$[$Si/Fe$]$\tablefootmark{b} &  8 &   0.024$\pm$0.009 &   0.009$\pm$0.014       & 8 &   0.034$\pm$0.018        &   0.019$\pm$0.021       & --0.007$\pm$0.021        \\
$[$Ca/Fe$]$\tablefootmark{b} &  8 & --0.020$\pm$0.010 & --0.017$\pm$0.011       & 8 &   0.001$\pm$0.030        &   0.004$\pm$0.030       & --0.024$\pm$0.032        \\
$[$Sc/Fe$]$\tablefootmark{b} &  4 &   0.029$\pm$0.014 & --0.002$\pm$0.026       & 5 &   0.036$\pm$0.018        &   0.005$\pm$0.028       & --0.011$\pm$0.024        \\
$[$Ti/Fe$]$\tablefootmark{b} &  8 &   0.016$\pm$0.011 &   0.005$\pm$0.014       & 8 &   0.063$\pm$0.033        &   0.052$\pm$0.034       & --0.044$\pm$0.035        \\
$[$V/Fe$]$                   &  5 &   0.019$\pm$0.017 &   0.017$\pm$0.017       & 5 &   0.073$\pm$0.050        &   0.071$\pm$0.050       & --0.049$\pm$0.053        \\
$[$Cr/Fe$]$\tablefootmark{b} &  8 &   0.011$\pm$0.011 &   0.017$\pm$0.012       & 7 &   0.043$\pm$0.033        &   0.049$\pm$0.033       & --0.038$\pm$0.036        \\
$[$Mn/Fe$]$                  &  2 &   0.034$\pm$0.019 &   0.024$\pm$0.021       & 0 & \multicolumn{1}{c}{...} & \multicolumn{1}{c|}{...} & \multicolumn{1}{c}{...}\\      
$[$Co/Fe$]$                  &  4 &   0.051$\pm$0.019 &   0.040$\pm$0.020       & 4 & --0.019$\pm$0.044        & --0.030$\pm$0.045       &   0.065$\pm$0.048        \\
$[$Ni/Fe$]$                  &  8 &   0.049$\pm$0.009 &   0.028$\pm$0.018       & 8 &   0.057$\pm$0.019        &   0.036$\pm$0.024       & --0.009$\pm$0.022        \\
$[$Cu/Fe$]$                  &  2 &   0.058$\pm$0.020 &   0.016$\pm$0.035       & 2 &   0.070$\pm$0.034        &   0.028$\pm$0.045       & --0.012$\pm$0.040        \\
$[$Zn/Fe$]$                  &  4 &   0.029$\pm$0.027 &   0.001$\pm$0.033       & 4 &   0.075$\pm$0.031        &   0.047$\pm$0.037       & --0.044$\pm$0.042        \\
$[$Y/Fe$]$                   &  3 & --0.034$\pm$0.013 &   0.008$\pm$0.032       & 3 &   0.047$\pm$0.029        &   0.089$\pm$0.041       & --0.080$\pm$0.032        \\
$[$Zr/Fe$]$                  &  1 &   0.033$\pm$0.053 &   0.064$\pm$0.057       & 1 &   0.122$\pm$0.060        &   0.153$\pm$0.064       & --0.090$\pm$0.081      \\
$[$Ba/Fe$]$                  &  5 & --0.052$\pm$0.025 &   0.017$\pm$0.055       & 5 & --0.042$\pm$0.025        &   0.027$\pm$0.055       & --0.017$\pm$0.036      \\
$[$Ce/Fe$]$                  &  1 & --0.065$\pm$0.055 & --0.034$\pm$0.059       & 0 & \multicolumn{1}{c}{...} & \multicolumn{1}{c|}{...} & \multicolumn{1}{c}{...}\\
\hline
\end{tabular}
\tablefoot{The values are the average ones derived from the differential, unconstrained analysis (Tables \ref{tab_appendix_differential_parameters} to \ref{tab_appendix_abundances_alfCenB}). $N$ gives the number of line lists the value is based on. We note that the values in the last column are slightly different from the straight subtraction of those corresponding to $\alpha$ Cen A and B because not exactly the same features were used after trimming the line lists (Sect.~\ref{sect_results_chemical_abundances}). \\
\tablefoottext{a}{Weighted average of the \ion{Fe}{i}- and \ion{Fe}{ii}-based abundances.}
\tablefoottext{b}{Only based on one ion (Sect.~\ref{sect_results_chemical_abundances}).}
}
\end{table*}

As explained in Sect.~\ref{sect_parameter_determination}, we did not carry out a differential analysis of one binary component relative to the other. To assess any differences in their chemical properties, we therefore simply subtracted the abundances of the two stars with respect to the Sun. However, we did trim the line lists to a common set of features before this operation, even though it leads to negligible differences (at most 0.011 dex). The results are given in Table \ref{tab_final_results}.

The surface gravity of $\alpha$ Cen B appears to be slightly underestimated (Sect.~\ref{sect_results_parameters}). To estimate the impact on the abundances, we considered two groups of results: those obtained with two line lists that lead to $\log g$ close to the seismic value \citep{biazzo12,reddy03} and two that lead to $\log g$ underestimated by $\sim$0.2 dex \citep{morel14,melendez14}. In all cases, $T_\mathrm{eff}$ is close to the reference value. The abundance ratios of the two groups differ on average by less than 0.01 dex. It is therefore unlikely that the bias in $\log g$ strongly affects the conclusions presented in the following.

The impact of departures from LTE can be estimated on a star-to-star basis for about half of the elements studied. We made use of {\tt Spectrum Tools}\footnote{Available online at: {\tt http://nlte.mpia.de}.} to interpolate the \ion{Fe}{i}, \ion{Mg}{i}, \ion{Si}{i}, \ion{Ti}{i}, \ion{Mn}{i}, and \ion{Co}{i} corrections for the relevant parameters (taken from Table \ref{tab_final_results}, except for the $\log g$ of $\alpha$ Cen B for which we adopt the asteroseismic value because the spectroscopic one is slightly underestimated). We restrict ourselves to lines used in both components. The non-LTE calculations are discussed in \citet{bergemann12}, \citet{bergemann15}, \citet{bergemann13}, \citet{bergemann11}, \citet{bergemann08}, and \citet{bergemann10}, respectively. The differential corrections ($\alpha$ Cen A or B relative to the Sun) for \ion{Fe}{i} are nearly negligible (less than 0.01 dex) and are known to be even smaller for \ion{Fe}{ii} \citep[e.g.][]{bergemann12}. This supports the assumption of LTE for the determination of the stellar parameters. The departures affecting \ion{Na}{i} were evaluated in the same way, but with the interactive tool {\tt INSPECT}.\footnote{Available online at: {\tt http://www.inspect-stars.com}.} The calculations are described in \citet{lind11}. For the \ion{O}{i} triplet, we adopt the values computed for $\alpha$ Cen AB by \citet{ramirez13}. Other literature sources used are: \citet[][\ion{C}{i}]{takeda05a}, \citet[][\ion{Al}{i}]{nordlander17}, \citet[][\ion{Ca}{i}]{mashonkina17}, \citet[][\ion{Zn}{i}]{takeda05b}, and \citet[][\ion{Ba}{ii}]{korotin15}. In these last cases, the coarseness of the grids only allows a rough estimate to be derived (e.g. the star-to-star \ion{Al}{i} corrections only reflect the dependence with $T_\mathrm{eff}$; \citealt{nordlander17}). Furthermore, we caution that our mean differential corrections are only representative given that data are only available for a subset of all the lines used. These corrections appear to be small and are discussed further in Sect.~\ref{sect_discussion_Tc_trends}.

It was shown by \citet{thompson17} that the strengths of the temperature-sensitive photospheric features in $\alpha$ Cen B slightly vary depending on the activity level. Our spectrum was obtained when the star was in a moderately active state corresponding to an activity level a few times that of the Sun, as diagnosed by the ratio of the X-ray to bolometric luminosities \citep{ayres15}. It is unlikely that activity-related phenomena significantly bias our results, but the analysis of various spectra taken along the $\sim$8-yr activity cycle would certainly be illuminating. This issue is irrelevant for $\alpha$ Cen A, as it was caught during a deep activity minimum \citep{ayres15}.

\section{Discussion}\label{sect_discussion}

\subsection{Chemical properties of $\alpha$ Cen AB}\label{sect_abundance_patterns}

We find that the abundance pattern of both stars is {\it not} scaled solar. Several elements are significantly enhanced (e.g. Na, Ni; see below), whereas others (especially the heaviest ones) are underabundant. The depletion of the neutron-capture elements (e.g. Y, Ba) will be used in Sect.~\ref{sect_ages} to put constraints on the age of the system. The slight ($\sim$0.06 dex) oxygen depletion is also noteworthy, but it is in line with the behaviour of the \ion{O}{i} triplet-based LTE abundances as a function of [Fe/H] seen in FGK stars \citep{ramirez13}. Broadly speaking, all elements have abundances as expected for nearby, solar-like dwarfs of that metallicity, including an upturn in [Na/Fe] and [Ni/Fe] for supersolar [Fe/H] \citep[e.g.][]{adibekyan12,bensby14,brewer16}.

\begin{figure}[h!]
\centering
\includegraphics[trim=20 260 35 185,clip,width=\hsize]{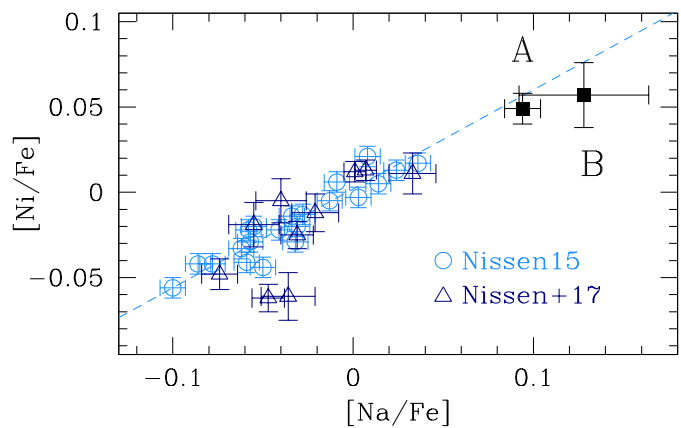}
\caption{Variation of [Ni/Fe] as a function of [Na/Fe] for solar analogues \citep{nissen15} and stars in the {\it Kepler} asteroseismic LEGACY sample \citep{lund17,silva_aguirre17} analysed by \citet{nissen17}. All stars have --0.15 $\lesssim$ [Fe/H] $\lesssim$ +0.15. The dashed line shows the linear fit derived by \citet{nissen15}. The position of $\alpha$ Cen A and B is indicated as filled squares.}
\label{fig_Na_Ni}
\end{figure}

Our study confirms a sodium and nickel excess in both stars (e.g. \citealt{porto_de_mello08} and \citealt{neuforge97}). There exists an intriguingly tight relation between the [Ni/Fe] and [Na/Fe] abundance ratios in solar analogues at near-solar metallicity, which may be related to the fact that the Na and Ni yields of Type II supernovae are both functions of the neutron excess \citep{nissen15}. A correlation is also seen in the data of \citet{ramirez14} for late F dwarfs and metal-rich solar analogues. As shown in Fig.~\ref{fig_Na_Ni}, the two components of $\alpha$ Cen have larger [Ni/Fe] and [Na/Fe] values than solar analogues at near-solar metallicity, but seem to follow the same linear trend.

The elemental number ratios C/O and Mg/Si recently received particular attention because they have been claimed to control the mineralogy of extrasolar terrestrial planets \citep[e.g.][and references therein]{suarez_andres18}. A knowledge of the C/O and Mg/Si ratios in $\alpha$ Cen AB would constrain the value in Proxima Cen, with potential consequences for our understanding of the structure and composition of its planet \citep{brugger16}. Relative to solar, we determine C/O = 1.18$\pm$0.07 and 1.18$\pm$0.17 in $\alpha$ Cen A and B, respectively. For Mg/Si, we obtain 0.98$\pm$0.06 and 1.02$\pm$0.11, respectively. Correcting for departures from LTE (Sect.~\ref{sect_results_chemical_abundances}) leads to negligible differences, except for the C/O ratio in $\alpha$ Cen B, which is lowered by $\sim$15\%. Our results suggest that any putative terrestrial planets in $\alpha$ Cen have a composition not vastly different from those in our solar system. 

\begin{figure*}[h!]
\begin{minipage}[t]{0.33\textwidth}
\centering
\includegraphics[trim=52 150 65 70,clip,width=0.95\textwidth]{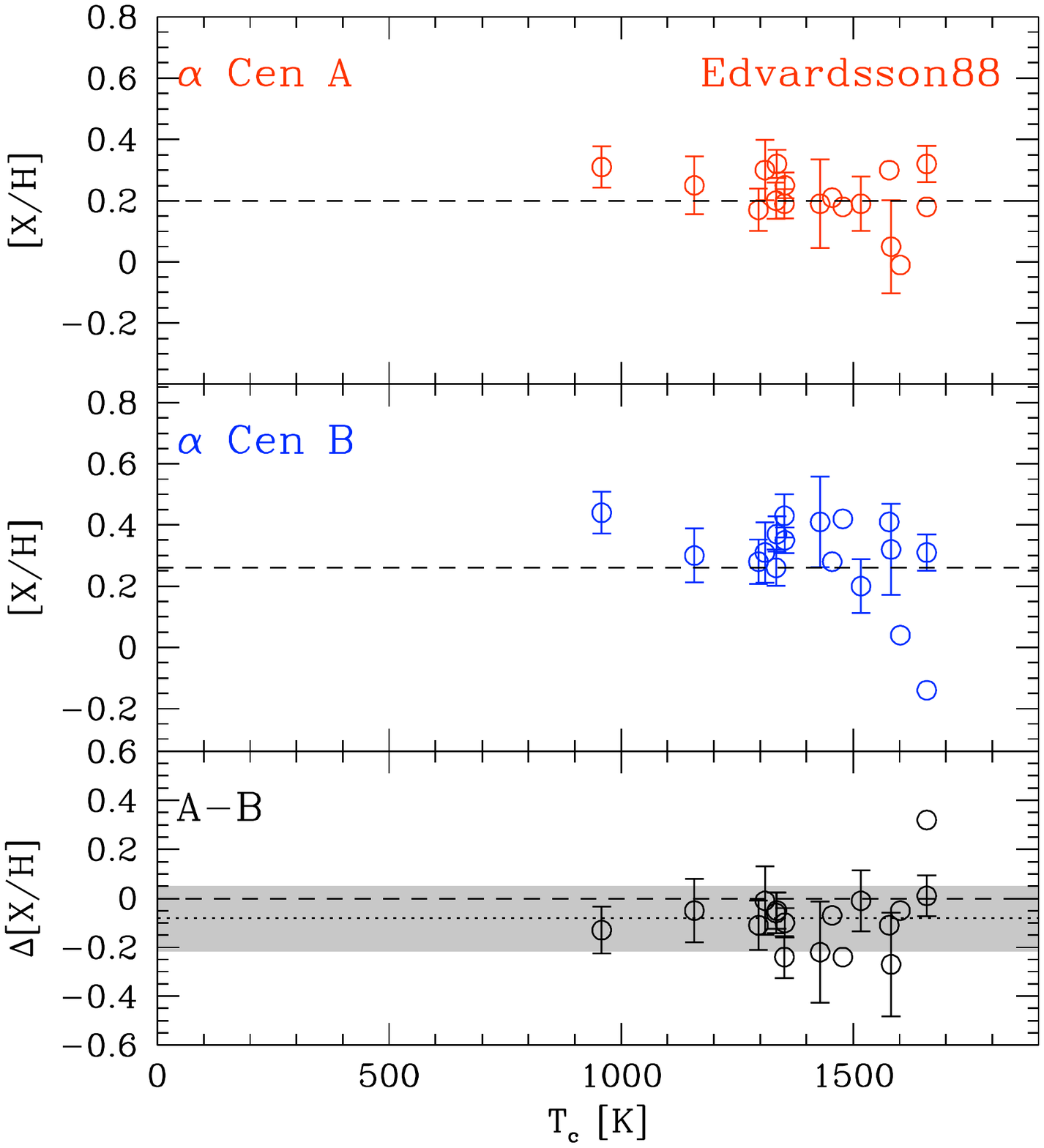}
\end{minipage}
\begin{minipage}[t]{0.33\textwidth}
\centering
\includegraphics[trim=52 150 65 70,clip,width=0.95\textwidth]{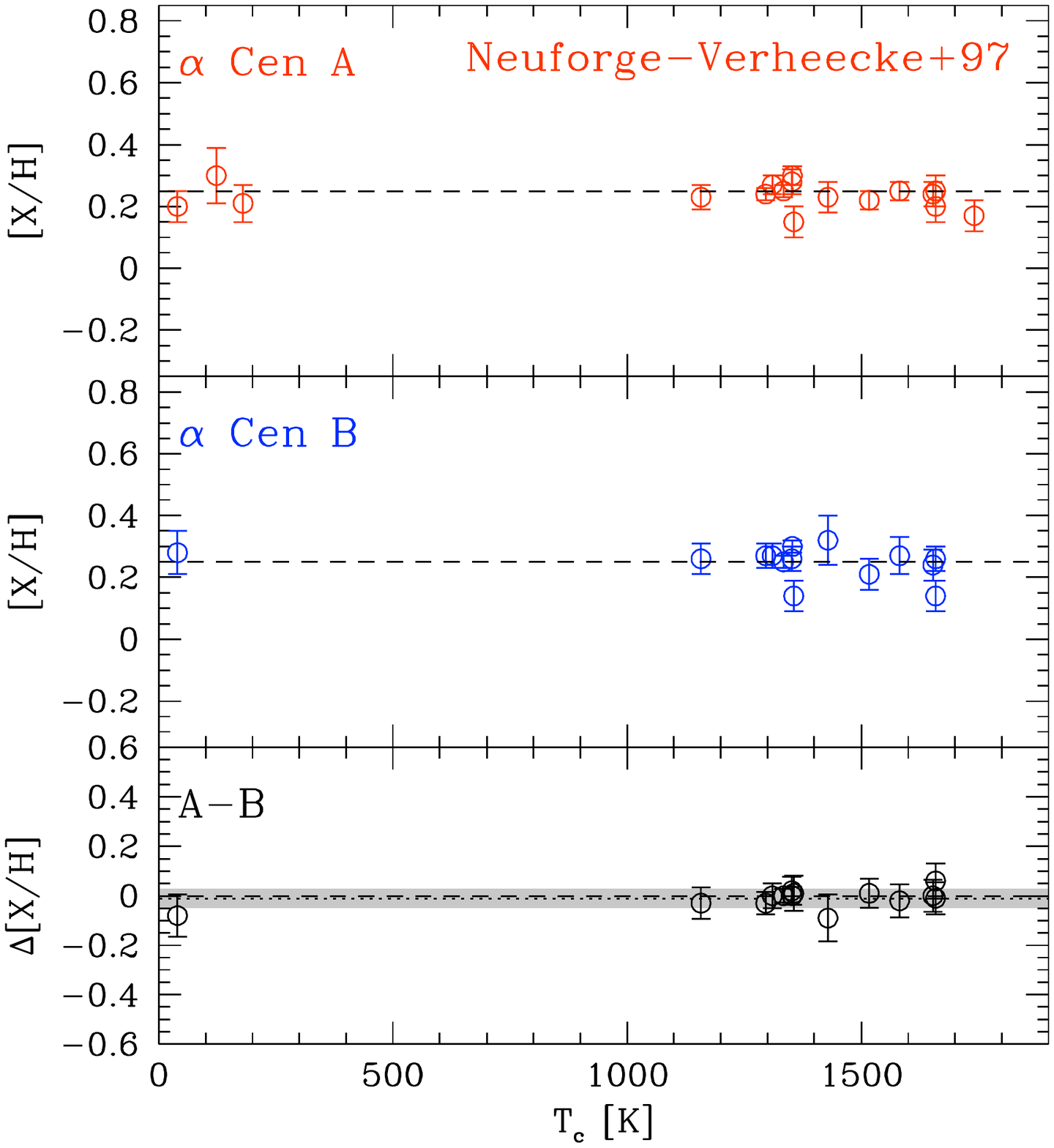}
\end{minipage}
\begin{minipage}[t]{0.33\textwidth}
\centering
\includegraphics[trim=52 150 65 70,clip,width=0.95\textwidth]{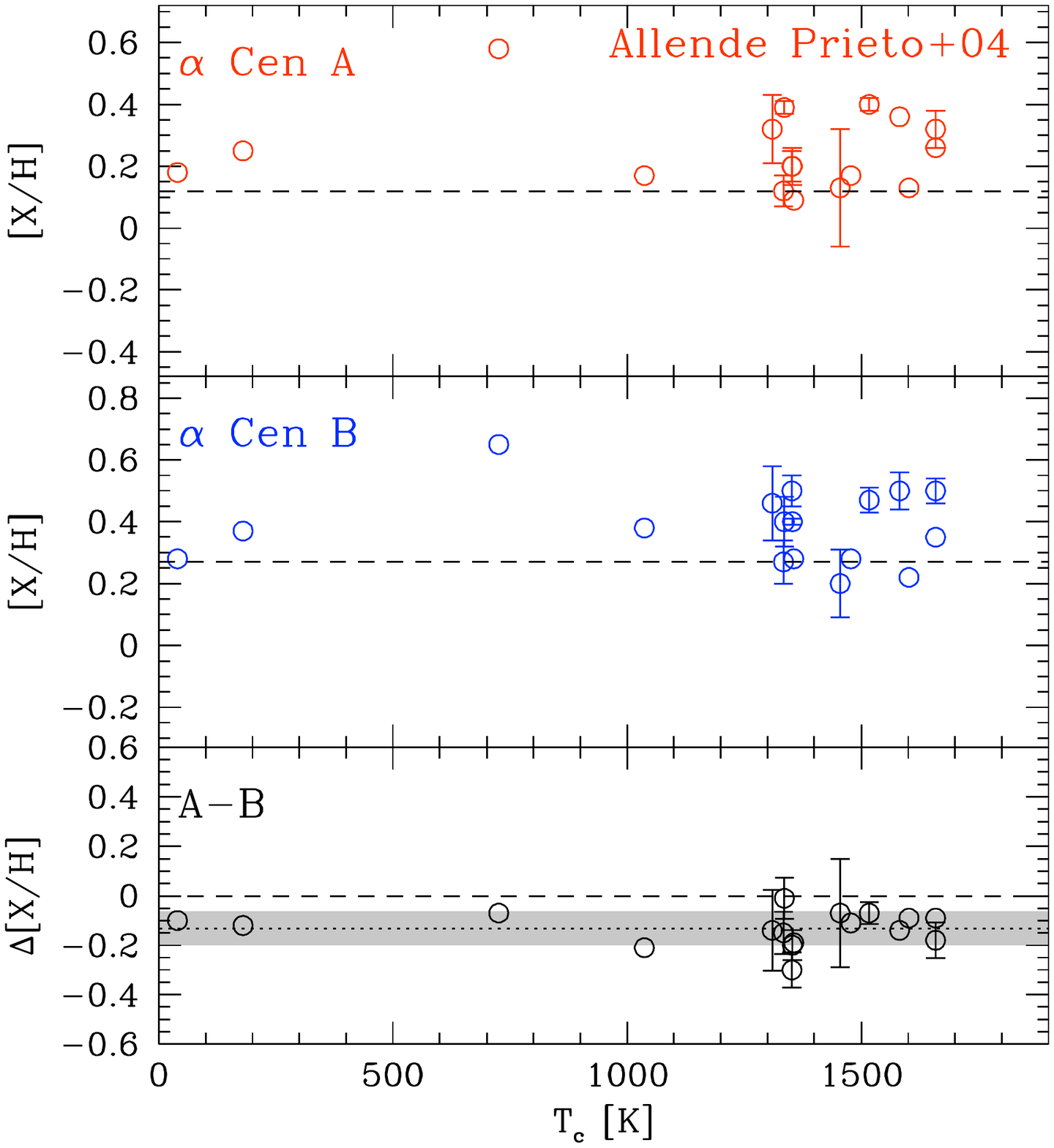}
\end{minipage}
\begin{minipage}[t]{0.33\textwidth}
\centering
\includegraphics[trim=52 150 65 70,clip,width=0.95\textwidth]{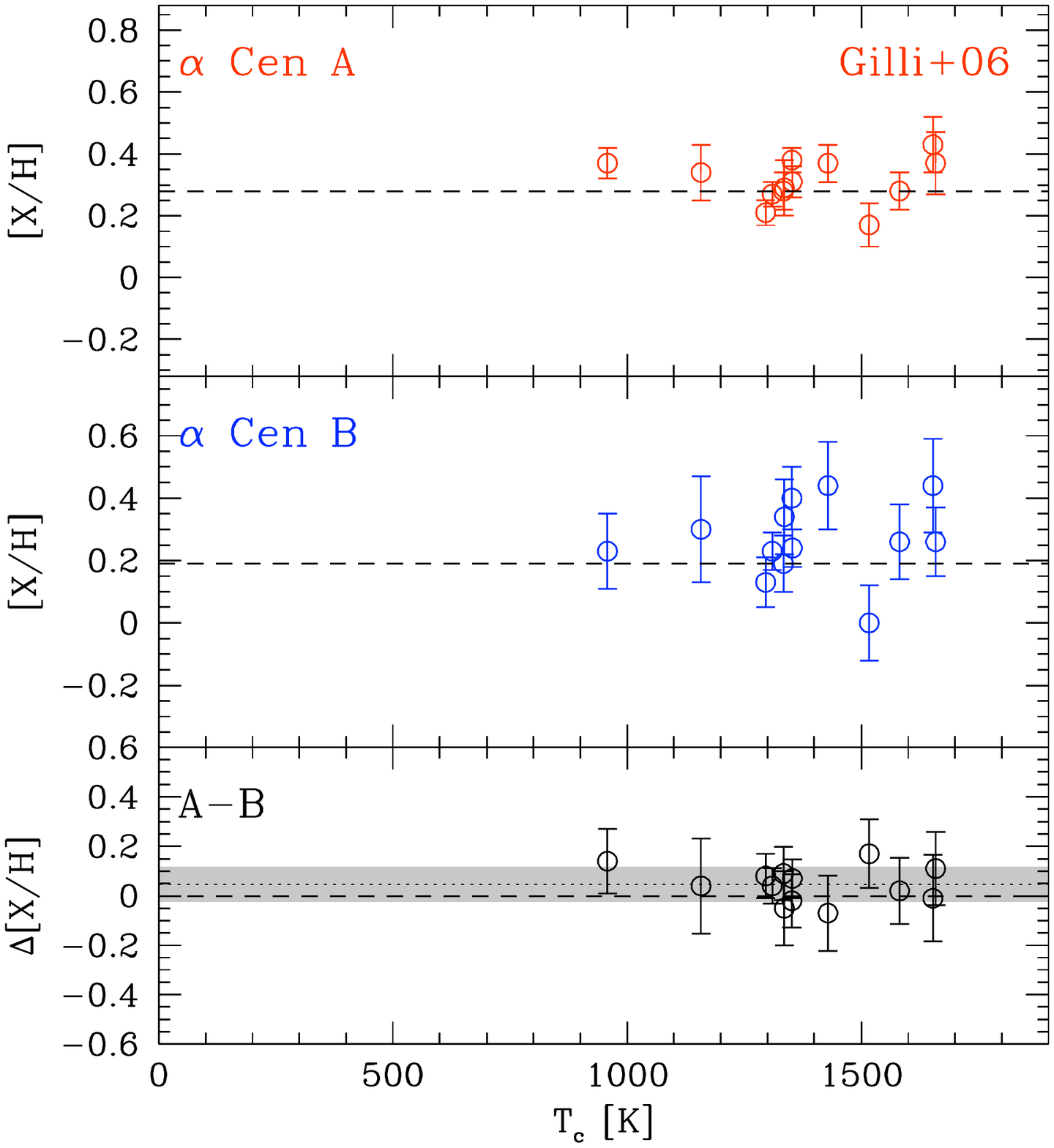}
\end{minipage}
\begin{minipage}[t]{0.33\textwidth}
\centering
\includegraphics[trim=52 150 65 70,clip,width=0.95\textwidth]{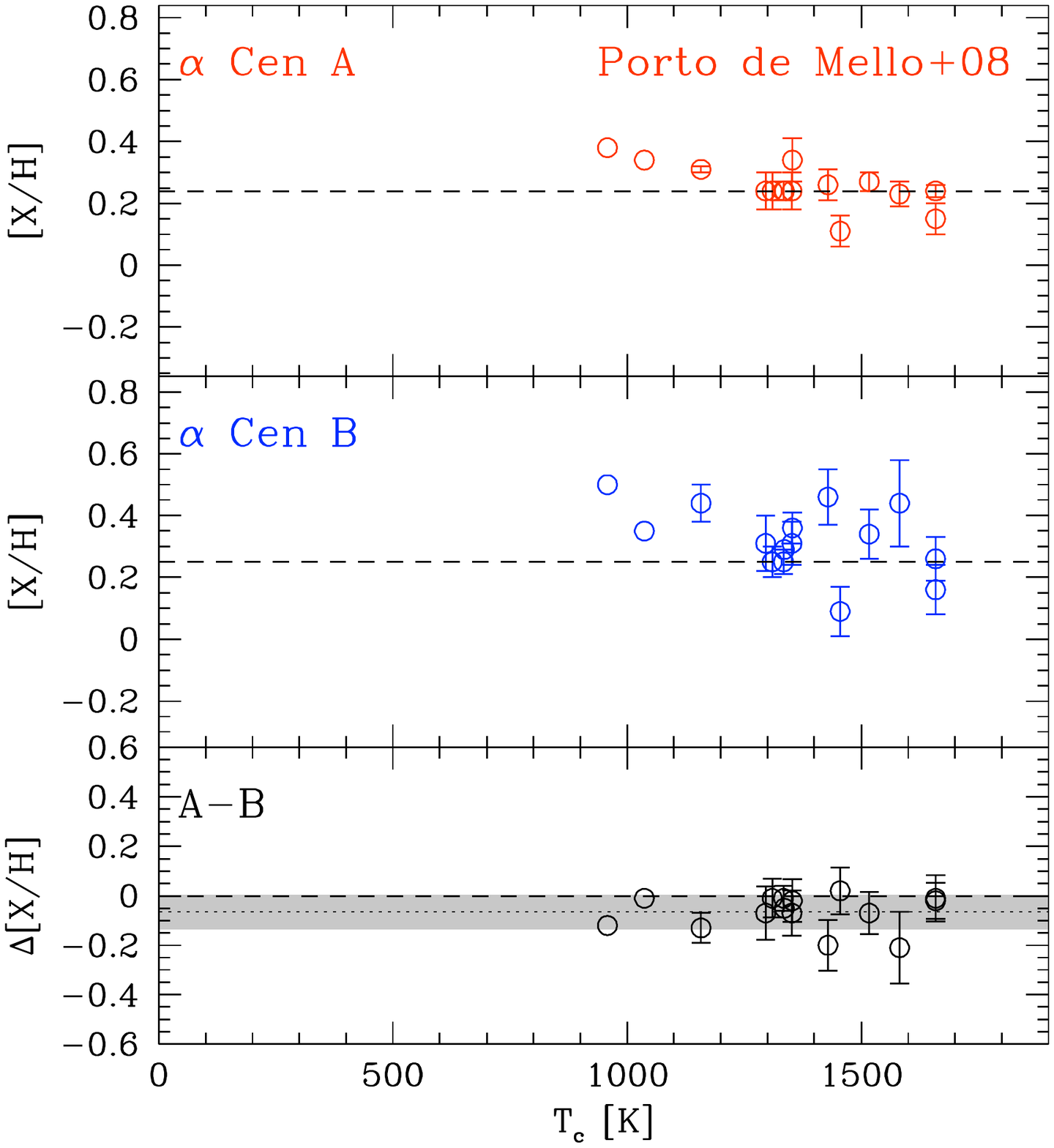}
\end{minipage}
\begin{minipage}[t]{0.33\textwidth}
\centering
\includegraphics[trim=52 150 65 70,clip,width=0.95\textwidth]{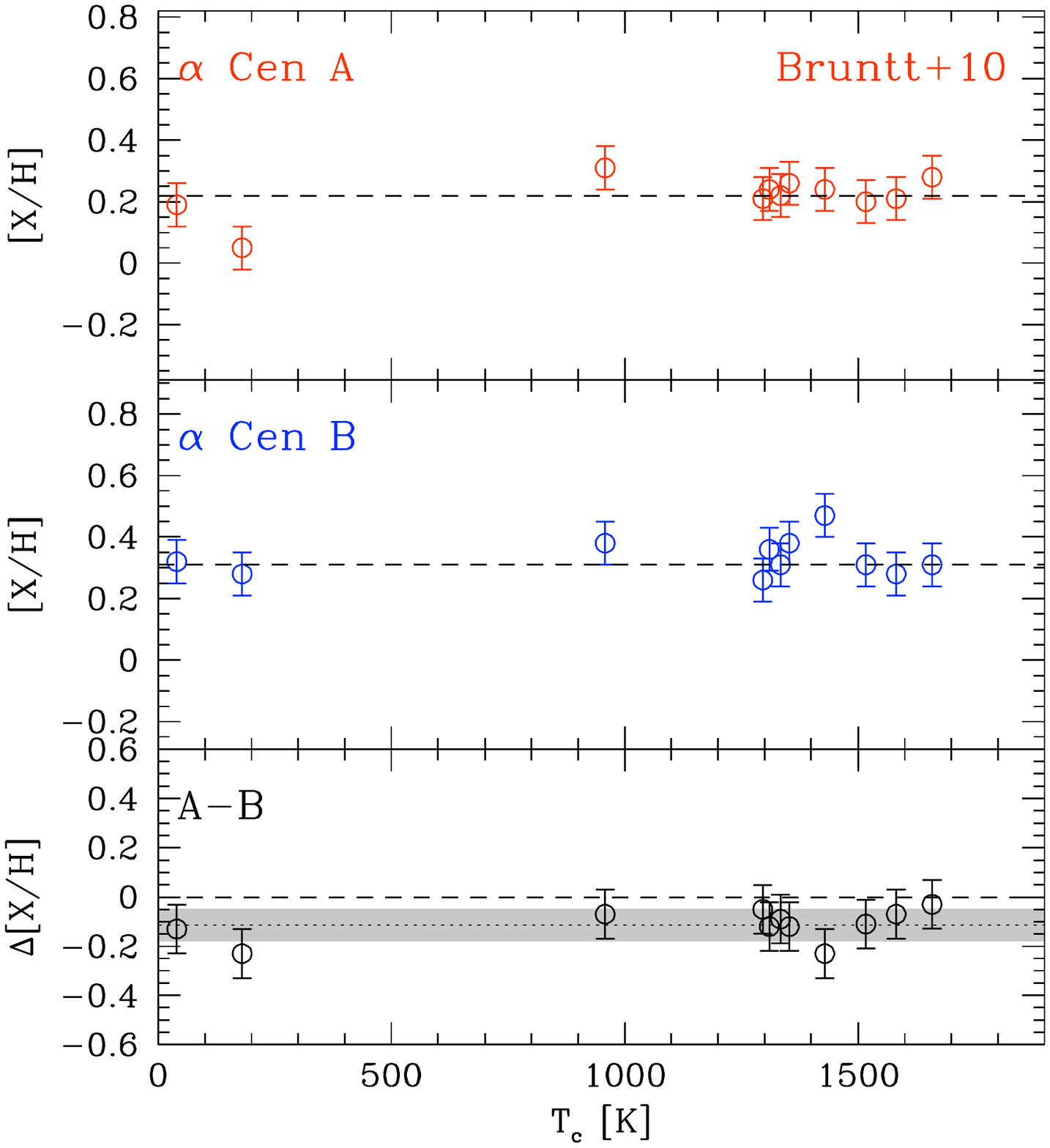}
\end{minipage}
\begin{minipage}[t]{0.33\textwidth}
\centering
\includegraphics[trim=52 150 65 70,clip,width=0.95\textwidth]{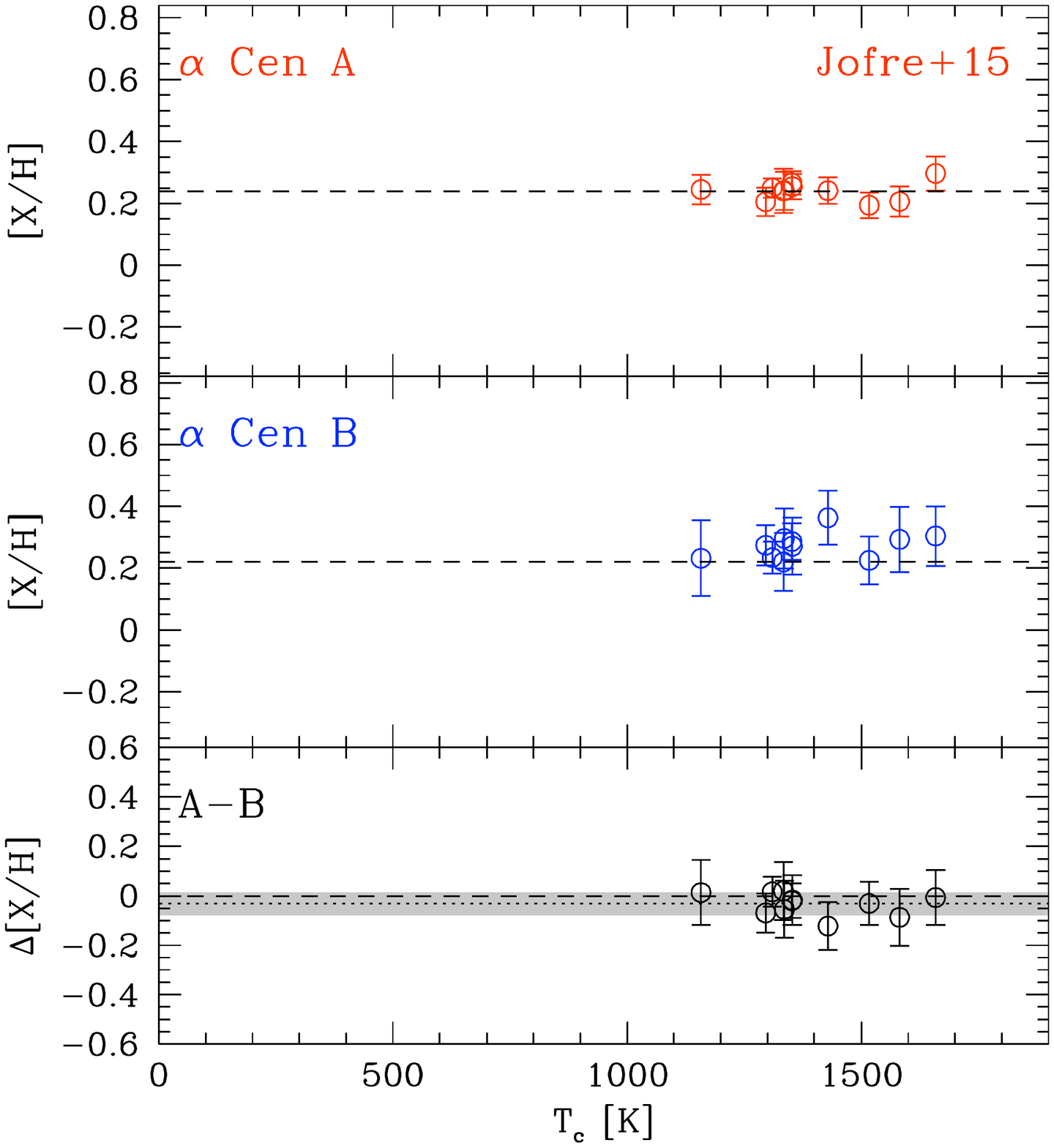}
\end{minipage}
\begin{minipage}[t]{0.33\textwidth}
\centering
\includegraphics[trim=52 150 65 70,clip,width=0.95\textwidth]{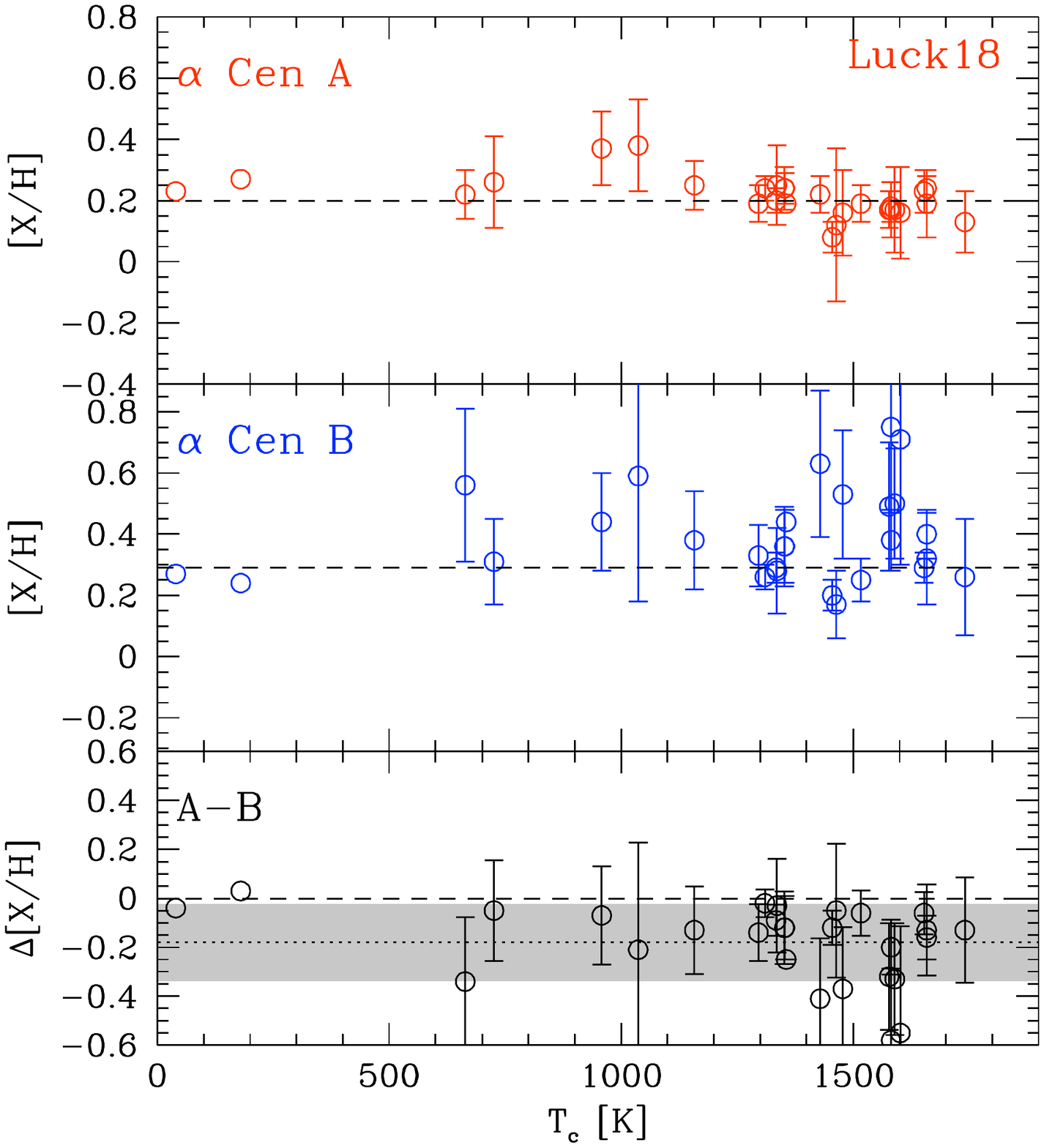}
\end{minipage}
\begin{minipage}[t]{0.33\textwidth}
\centering
\includegraphics[trim=52 150 65 70,clip,width=0.95\textwidth]{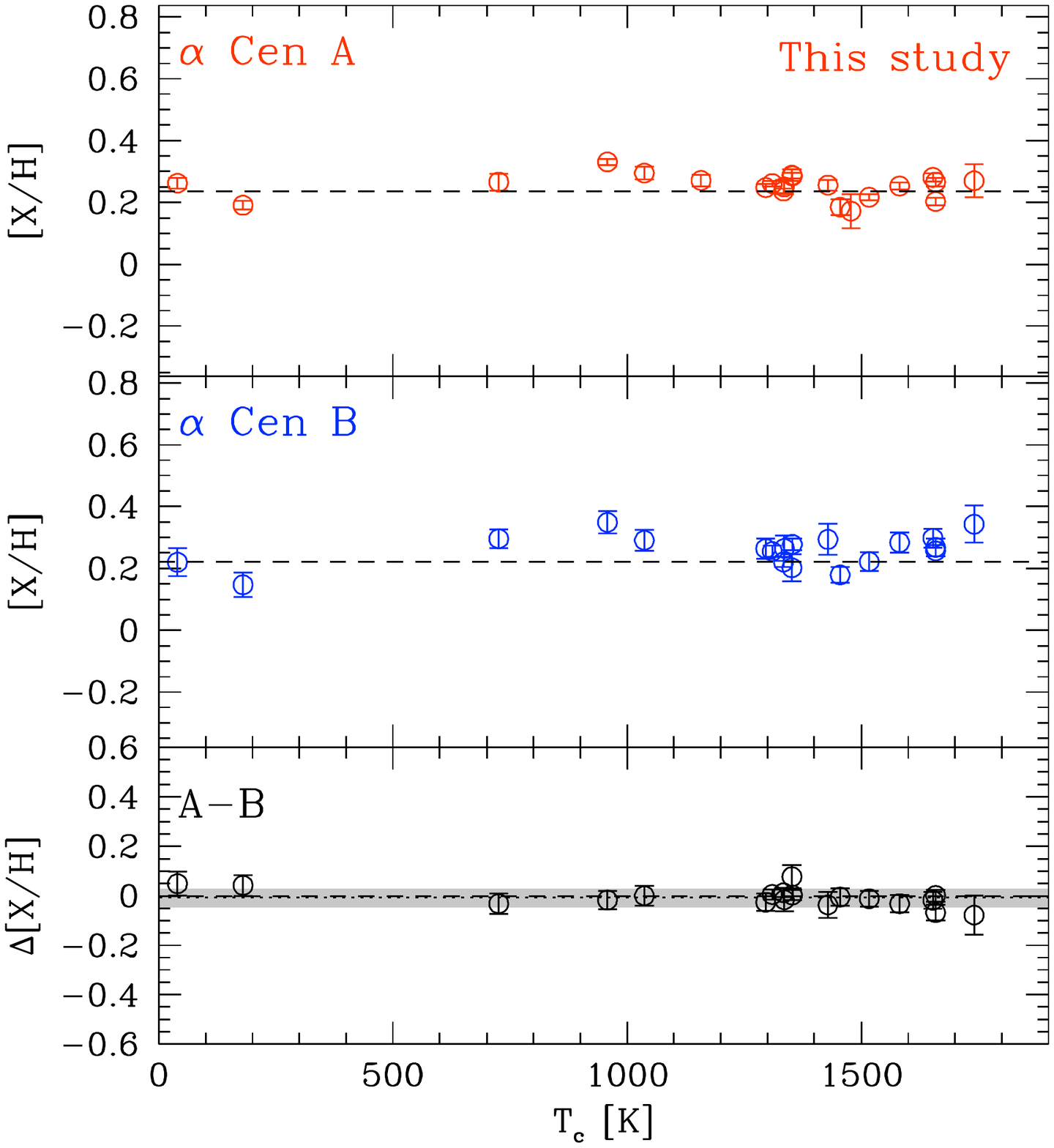}
\end{minipage}
\caption{Comparison between our abundance patterns and previous ones in the literature \citep{edvardsson88,neuforge97,luck18,allende_prieto04,gilli06,porto_de_mello08,bruntt10,jofre15}. For elements with abundances corresponding to two ionisation stages, we followed the procedure outlined in Sect.~\ref{sect_results_chemical_abundances}. The results are shown as a function of $T_\mathrm{c}$ \citep{lodders03}. For each study, the metal abundances of $\alpha$ Cen A and B relative to hydrogen, [X/H], are shown in the top and middle panel, respectively. A dashed line is drawn at [Fe/H]. The bottom panel shows the abundance differences, $\Delta$[X/H], between $\alpha$ Cen A and B. To guide the eye, a dashed line is drawn at $\Delta$[X/H] = 0. The dotted line indicates the mean difference, while the grey strip shows the corresponding 1-$\sigma$ uncertainties. For \citet{edvardsson88}, \citet{porto_de_mello08} and our study, the uncertainties for elements other than iron refer to [X/Fe], not [X/H]. However, they are expected to be representative.}
\label{fig_abundances_literature}
\end{figure*}

Let us now compare our results to previous ones in the literature. We restrict ourselves to spectroscopic studies that derived the chemical composition of {\it both} components \citep{edvardsson88,neuforge97,luck18,allende_prieto04,gilli06,porto_de_mello08,bruntt10,jofre15}. We ignore the early study of \citet{england80}, as it is plagued by large uncertainties and does not yield useful constraints. On the other hand, other works \citep[e.g.][]{valenti05} only derived the abundances of a few elements. Finally, we do not discuss the study of $\alpha$ Cen by \citet{feltzing01} because it duplicates to a large extent that of \citet{neuforge97}, using the same EWs, atomic data, and atmospheric parameters. The stellar parameters adopted by the eight selected studies are summarised in Table \ref{tab_parameters_literature}. It should be noted that several analyses are not self-consistent, as the atmospheric parameters could be taken from other works \citep[e.g.][]{edvardsson88} or not derived from spectroscopy \citep[e.g.][]{allende_prieto04}. The abundance patterns (including that from our study) are shown for both stars in Fig.~\ref{fig_abundances_literature} as a function of the 50\% condensation temperature for a solar-system-composition gas \citep[$T_\mathrm{c}$;][]{lodders03}. Adopting $T_\mathrm{c}$ values appropriate to more metal-rich material is unlikely to affect our conclusions, as they are systematically and only slightly offset (\citealt{lodders03}; but see \citealt{bond10}). Also shown are the abundance differences between the two components as a function of $T_\mathrm{c}$. 

\begin{table*}[h!]
\scriptsize
\caption{Stellar parameters adopted by previous abundance studies in the literature.}\label{tab_parameters_literature} 
\centering
\begin{tabular}{l|cccc|cccc}
\hline\hline
      & \multicolumn{4}{c}{$\alpha$ Cen A} & \multicolumn{4}{c}{$\alpha$ Cen B}\\
Study & $T_\mathrm{eff}$ & $\log g$ & $\xi$         & [Fe/H] & $T_\mathrm{eff}$ & $\log g$ & $\xi$         & [Fe/H]\\
      & [K]             & [cgs]    & [km s$^{-1}$] &         & [K]             & [cgs]    & [km s$^{-1}$] &       \\\hline
Reference values         & 5795$\pm$19 & 4.32$\pm$0.02 & ... & ... & 5231$\pm$21 & 4.53$\pm$0.02 & ... & ...\\
\hline
\citet{edvardsson88}\tablefootmark{a}     & 5750         & 4.42$\pm$0.11 & 1.5           & 0.20          & 5250         & 4.65$\pm$0.11 & 1.3           & 0.26\\
\citet{neuforge97}                        & 5830$\pm$30  & 4.34$\pm$0.05 & 1.09$\pm$0.11 & 0.25$\pm$0.02 & 5255$\pm$50  & 4.51$\pm$0.08 & 1.00$\pm$0.08 & 0.24$\pm$0.03\\
\citet{allende_prieto04}\tablefootmark{b} & 5519$\pm$123 & 4.26$\pm$0.10 & 1.04          & 0.12$\pm$0.05 & 4970$\pm$180 & 4.59$\pm$0.04 & 0.81          & 0.27$\pm$0.07\\
\citet{gilli06}\tablefootmark{c}          & 5844$\pm$42  & 4.30$\pm$0.19 & 1.18$\pm$0.05 & 0.28$\pm$0.06 & 5199$\pm$80  & 4.37$\pm$0.27 & 1.05$\pm$0.10 & 0.19$\pm$0.09\\
\citet{porto_de_mello08}                  & 5847$\pm$27  & 4.34$\pm$0.12 & 1.46$\pm$0.03 & 0.24$\pm$0.03 & 5316$\pm$28  & 4.44$\pm$0.15 & 1.28$\pm$0.12 & 0.25$\pm$0.04\\
\citet{bruntt10}                          & 5745$\pm$80  & 4.31$\pm$0.06 & 1.00$\pm$0.07 & 0.22$\pm$0.07 & 5145$\pm$80  & 4.52$\pm$0.04 & 0.83$\pm$0.07 & 0.30$\pm$0.07\\
\citet{jofre15}\tablefootmark{d}          & 5792$\pm$16  & 4.30$\pm$0.01 & 1.20$\pm$0.07 & 0.24$\pm$0.08 & 5231$\pm$20  & 4.53$\pm$0.03 & 0.99$\pm$0.31 & 0.22$\pm$0.10\\
\citet{luck18}\tablefootmark{e}           & 5753         & 4.26          & 1.07          & 0.20$\pm$0.04 & 5242         & 4.57          & 0.25          & 0.29$\pm$0.05\\
This study \tablefootmark{f}              & 5829$\pm$6   & 4.35$\pm$0.02 & 1.26$\pm$0.02 & 0.24$\pm$0.01 & 5189$\pm$18  & 4.30$\pm$0.05 & 0.95$\pm$0.04 & 0.22$\pm$0.02\\
\hline
\end{tabular}
\tablefoot{The reference $T_\mathrm{eff}$ and $\log g$ values are obtained from interferometric and asteroseismic measurements, and are discussed in Sect.~\ref{sect_parameter_determination}.\\
\tablefoottext{a}{$T_\mathrm{eff}$ and $\xi$ values taken from \citet{england80} and \citet{smith86}, respectively.}
\tablefoottext{b}{$T_\mathrm{eff}$ and $\log g$ derived from photometric indices and isochrone fitting, respectively. $\xi$ determined from calibrations.}
\tablefoottext{c}{All values taken from \citet{santos05}.}
\tablefoottext{d}{$T_\mathrm{eff}$ and $\log g$ values taken from \citet{heiter15}. $\xi$ and LTE [Fe/H] values taken from \citet{jofre14}.}
\tablefoottext{e}{$T_\mathrm{eff}$ and $\log g$ derived from photometric indices and isochrone fitting, respectively.}
\tablefoottext{f}{Mean values from differential, unconstrained analysis (see Table \ref{tab_appendix_differential_parameters}).}
}
\end{table*}

One of our most important results is that the abundance pattern of the two components is identical within the errors. The largest discrepancy is observed for [Y/Fe] at a significance level of $\sim$2.5$\sigma$ (Table \ref{tab_final_results}). The mean abundance differences (A -- B) are: $<$$\Delta$[X/H]$>$ = --0.008$\pm$0.037 dex and $<$$\Delta$[X/Fe]$>$ = --0.021$\pm$0.038 dex, where X represents a given species. The false alarm probability for the [X/Fe] data of the two stars to be uncorrelated is $\sim$0.4\% based on the computation of the generalised Kendall's $\tau$ correlation coefficient. The size of the sample is exceedingly small, but we also find that the distribution of the $\Delta$[X/Fe] values is statistically indistinguishable from a normal distribution centred at zero and with a standard deviation corresponding to our median uncertainty ($\sigma$ $\sim$ 0.037 dex). 

The effects of microscopic diffusion are larger in $\alpha$ Cen A because of its thinner outer convective envelope. As gravitational settling largely dominates over radiative levitation, $\alpha$ Cen B should be less depleted in metals at the surface. Namely, we expect the mean abundance difference, $<$$\Delta$[X/H]$>$, defined above to be negative. This is only hinted at by our data, but any changes arising from diffusion can probably be accommodated by our uncertainties. The models of \citet{deal15} for C, Mg, and Fe indicate abundance differences of only $\sim$0.002 dex for the binary components of 16 Cyg with a mass difference of 0.04 M$_{\sun}$ and an age (6.4 Gyrs) comparable to that of $\alpha$ Cen (see Sect.~\ref{sect_ages}). Deviations of $\sim$0.02 dex for Mg, Ti, Fe, and Ni are suggested by the computations of \citet{michaud04} for solar-metallicity, FG dwarfs with the same age as above and a mass difference more appropriate to our case \citep[$\sim$0.17 M$_\sun$;][]{kervella16}. Finally, the uncalibrated models of $\alpha$ Cen AB by \citet{turcotte98} also lead to differences of this magnitude for the numerous metals they investigated.

The similarity between the chemical pattern of the two components was also pointed out by \citet{neuforge97} based on data of similar precision (see Fig.~\ref{fig_abundances_literature}), but our larger number of elements puts this conclusion on a firmer footing. We do not see any clear reasons to disregard their results, as was done by \citet{hinkel13} who combined a selection of abundance results in the literature. We note that \citet{hinkel13} inferred iron abundances ([Fe/H]=+0.28 and +0.31 dex for $\alpha$ Cen A and B, respectively) that are significantly larger than ours or, more generally, those in the literature (see Table \ref{tab_parameters_literature}). The studies of \citet{allende_prieto04} and \citet{bruntt10} suggest that $\alpha$ Cen B is overabundant in metals at the $\sim$0.12 dex level (see Fig.~\ref{fig_abundances_literature}). However, the $T_\mathrm{eff}$ values adopted by \citet{allende_prieto04} and, to a much lesser extent, \citet{bruntt10} are lower than our or previous spectroscopic estimates (Table \ref{tab_parameters_literature}). Probably more telling is the fact that they are also lower than the interferometric measurements (Sect.~\ref{sect_parameter_determination}) by $\sim$270 and $\sim$70 K, respectively. It is therefore tempting to associate these putative metallicity differences to a $T_\mathrm{eff}$ scale that is too cool; see \citet{ramirez10} for further evidence indicating that the $T_\mathrm{eff}$ values derived from colour indices by \citet{allende_prieto04} are underestimated. The even larger metallicity offset ($\sim$0.18 dex) found by \citet{luck18} requires another explanation. It may arise from the very low microturbulence adopted for $\alpha$ Cen B ($\xi$ = 0.25 km s$^{-1}$).

\begin{figure*}[h!]
\centering
\includegraphics[trim= -87 165 -108 80 ,clip,width=\hsize]{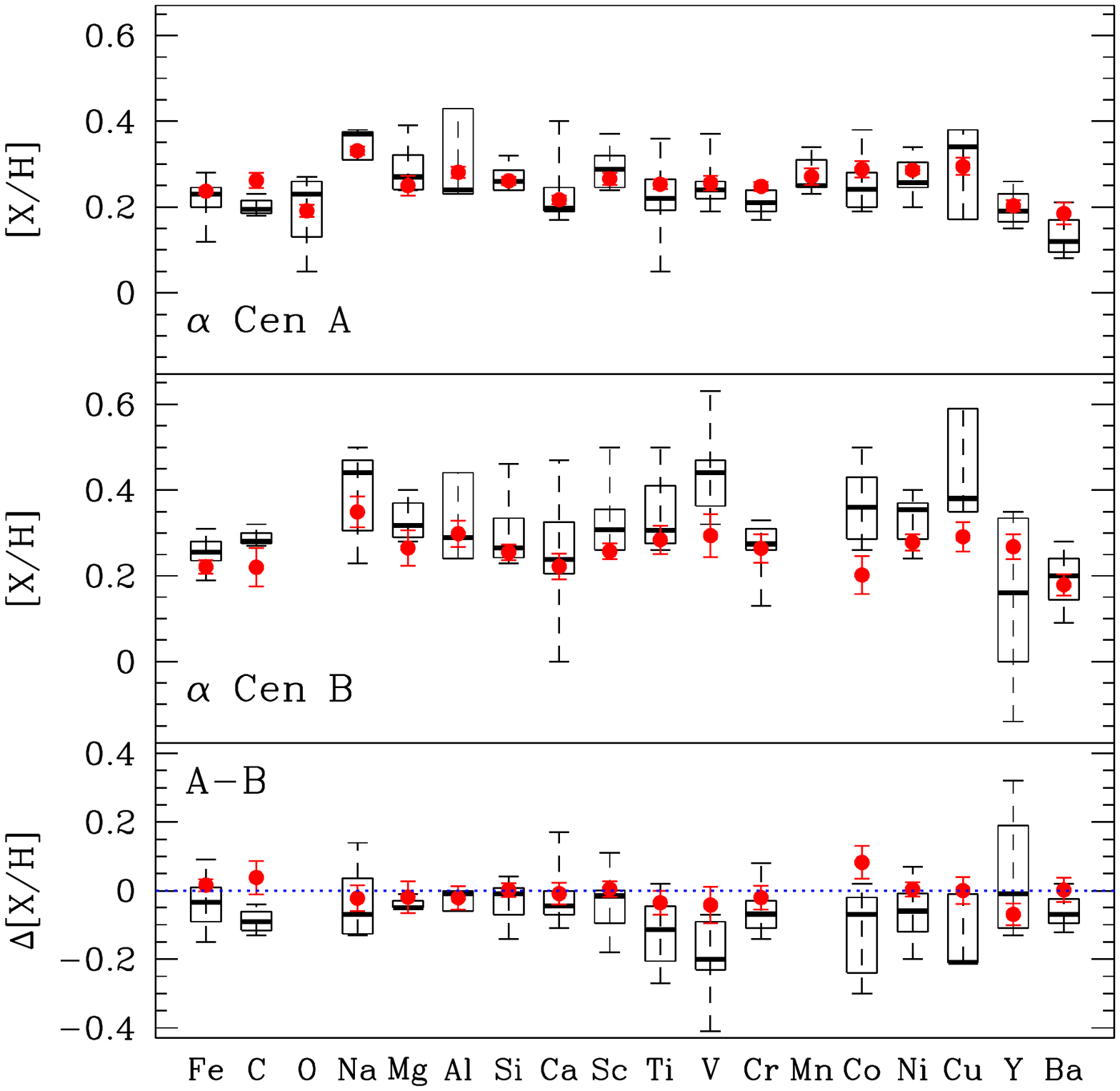}
\caption{Comparison between our abundances and those from previous studies (listed in Table \ref{tab_parameters_literature}) for elements with at least three measurements available. The vertical dashed line connects the extreme values found. The box covers the first to third quartile of the literature data, while the thick horizontal line inside the box shows the median. Our results are overplotted as red, filled circles. To guide the eye, a dotted line is drawn in the bottom panel at $\Delta$[X/H] = 0. Our uncertainties for elements other than iron refer to [X/Fe], not [X/H]. However, they are expected to be representative.}
\label{fig_statistics}
\end{figure*}

Another way to look at differences between our results and previous ones is provided by Fig.~\ref{fig_statistics}. We only consider here elements with at least three measurements available from the studies listed in Table \ref{tab_parameters_literature}. Noticeable discrepancies are apparent for C, V, Co, and Cu. However, for carbon we find a better agreement between the abundance of the two components. The last three elements are affected by HFS effects, which sometimes seem to have been ignored \citep[e.g.][]{gilli06}. This may explain the V, Co, and Cu overabundances reported in $\alpha$ Cen B where HFS corrections are particularly large.

\subsection{Age of the system from abundance indicators}\label{sect_ages}
Following the work of \citet{da_silva12}, recent high-precision studies of solar twins/analogues at near-solar metallicities (--0.15 $\lesssim$ [Fe/H] $\lesssim$ +0.15) have unveiled remarkable correlations between some abundance ratios and isochrone ages \citep[e.g.][]{nissen15,nissen16}. Their robustness is supported by the fact that independent studies making use of different abundance data and sets of isochrones provide nearly identical relations. Ages derived from isochrone fitting are notoriously known to be model-dependent and prone to large uncertainties. However, quantitatively similar correlations are also found for stars in the {\it Kepler} LEGACY sample \citep{lund17,silva_aguirre17} with asteroseismic ages uncertain to within 10-20\% \citep{nissen17}.\footnote{The new relations of \citet{nissen17} are not considered further because they are based on a mixture of ages derived from different techniques (isochrone fitting and asteroseismology). However, they do not noticeably differ from those assumed in our paper. Indeed, they lead to similar ages within the errors.} The distinct behaviour shown by elements produced through different nucleosynthesis channels strongly suggests that the abundance-age trends are intimately linked to the chemical evolution of the Galaxy \citep[e.g.][]{spina16a}. 

Of particular usefulness as age indicators are [Y/Mg] and [Y/Al], which show a tight and steep decline as a function of look-back time \citep{nissen15,nissen16,spina16a,spina18,tucci_maia16}. The correlation  is interpreted as arising from a varying enrichment of the interstellar medium (ISM) in these two elements as the Galaxy evolves. For instance, the $\alpha$-element magnesium is produced through core-collapse supernova events in the early Galaxy, while the $s$-process element yttrium is released in the ISM mainly through the winds of low-mass AGB stars \citep[e.g.][]{bisterzo14} over much longer time scales. The correlation extends over $\sim$10 Gyrs and the age scatter is at most 1 Gyr for a given abundance ratio. Furthermore, the relations seem to be obeyed by distinct stellar populations \citep[e.g. thin- and thick-disc stars;][]{nissen15}, although this needs to be investigated further. 

We considered various isochrone age-abundance calibrations to estimate the age of $\alpha$ Cen AB based on the [Y/Mg] and [Y/Al] ratios \citep{nissen16,tucci_maia16,spina18}. The quadratic relations of \citet{spina18} were used, as the authors claim that they provide a significantly better fit to their data. The results are provided in Table \ref{tab_ages}. We regard the ages derived for $\alpha$ Cen A ($\sim$6.2 Gyrs) as more reliable, not only because of the smaller error bars, but also because they are likely less affected by systematic effects (e.g. non-LTE corrections) arising from departures from the solar parameters.

\begin{table*}[h!]
%\scriptsize
\caption{Stellar ages derived from [Y/Mg] and [Y/Al] abundance ratios.}\label{tab_ages} 
\centering
\begin{tabular}{lcrccc}
\hline\hline
Star           & \multicolumn{1}{c}{Abundance ratio} & \multicolumn{1}{c}{Value} & \multicolumn{3}{c}{Age [Gyr]}               \\
               &            &                    & N16           & TM16          & S18           \\
\hline                                                                          
$\alpha$ Cen A & $[$Y/Mg$]$ & --0.048$\pm$0.035  & 5.88$\pm$1.00 & 5.66$\pm$0.85 & 5.89$\pm$0.69 \\
               & $[$Y/Al$]$ & --0.086$\pm$0.021  & 6.60$\pm$0.58 & ...           & 6.51$\pm$0.38 \\
$\alpha$ Cen B & $[$Y/Mg$]$ &   0.003$\pm$0.061  & 4.50$\pm$1.67 & 4.43$\pm$1.48 & 4.82$\pm$1.40 \\
               & $[$Y/Al$]$ & --0.025$\pm$0.053  & 5.18$\pm$1.27 & ...           & 5.44$\pm$1.01 \\
\hline
\end{tabular}
\tablefoot{Keywords for age-abundance calibrations --- N16: \citet{nissen16}; TM16: \citet{tucci_maia16}; S18: \citet{spina18}. The uncertainties in the abundance ratios and calibrations are propagated into the age estimates.}
\end{table*}

Although the dependencies are more complex and not as clear, ages determined from abundance ratios relative to iron can valuably complement the [Y/Mg]- and [Y/Al]-based values. For ten metals with a reasonably well-defined behaviour and covering a sufficient abundance range (C, Mg, Al, Si, Sc, Ti, Cu, Zn, Y, and Ba), we determine an age for $\alpha$ Cen A of 5.7$\pm$0.8 Gyrs from the linear relations of \citet{nissen16}. A different interpretation of the data was proposed by \citet{spina16a} who favoured hyperbolic fits with a turnover at intermediate ages to the abundance-age relations of many elements. Using the appropriate fitting functions (either linear or hyperbolic), we obtain an age of 6.6$\pm$1.2 Gyrs for seven elements. No hyperbolic solutions could be obtained for three elements in common with \citet{nissen16}: Si, Cu, and Zn.\footnote{In a few cases, the ages we derive slightly exceed (by less than 0.8 Gyr) the domains of validity of the calibrations, which are 6 and 8 Gyrs for \citet{nissen16} and \citet{spina16a}, respectively. Because of the high Na and Ni abundances in $\alpha$ Cen AB compared to solar analogues (see Sect.~\ref{sect_abundance_patterns}), the [Na/Fe] and [Ni/Fe] ratios cannot be used. In any case, these ratios are poorly correlated with age \citep{spina16a}.} Linear relations based on a larger sample of solar analogues were recently proposed by \citet{bedell18}. For five of the ten species above (Mg, Al, Ti, Y, and Ba), we obtain an age of 6.3$\pm$1.3 Gyrs, while for the others the values exceed the validity range of the calibrations (8 Gyrs) by 1.3 Gyr on average. This further strengthens the case for a system older than solar.

In summary, we infer an age of $\sim$6 Gyrs for $\alpha$ Cen. This value is fully compatible with that determined by several theoretical studies that performed an asteroseismic modelling of both components. These works are based on $p$-mode frequencies derived from CORALIE \citep{thoul03,eggenberger04,miglio05} or UVES/UCLES \citep{yildiz11} RV time series. However, an approximately solar age was found from CORALIE \citep{thevenin02} and HARPS \citep{bazot16} data. Our results are not in sharp disagreement with this conclusion considering the relatively large error bars, but they tend to support instead a system slightly more evolved than the Sun. Discrepant ages for the two components are found by \citet{lundkvist14}. It should be noted that the asteroseismic age of $\alpha$ Cen A is quite uncertain and depends on whether the optimal model presents a small convective core or not \citep[see, e.g.][]{bazot16}. A seismic study cannot be performed for Proxima Cen since M stars are not known to present any observables that arise from pulsations \citep[e.g.][]{rodriguez16}. A wide range of ages is inferred for $\alpha$ Cen AB from gyrochronology \citep[4.0-9.2 Gyrs;][]{mamajek08,delorme11,barnes07,angus15,epstein14}. The age we derive is broadly consistent with the rotation period of Proxima Cen \citep[$\sim$83 d;][]{benedict98} according to the calibrations for late M dwarfs by \citet{engle18}.

We caution that the accuracy of ages relying on spectroscopic indicators is still not well established. In particular, as discussed by \citet{feltzing17}, there is a significant spread in [Y/Mg] for a given age in nearby dwarfs, with more metal-poor stars displaying lower ratios. We therefore anticipate that, in virtue of their metal-rich nature, the [Y/Mg]-based age we derive for $\alpha$ Cen AB is affected by this effect and is likely to be revised upwards. However, with the caveat that the metallicity dependence of the calibrations remains to be fully understood and quantified, we speculate that the corrections are slight (tentatively of the order of $\sim$0.5 Gyr). No discernable variation in the [Y/Mg]-age relation is indeed found when splitting samples of solar analogues in two metallicity bins separated by $\Delta$[Fe/H] $\sim$ 0.15 dex \citep{tucci_maia16}. Another concern is that $\alpha$ Cen B is clearly not a solar analogue. The data of \citet{nissen17} show that stars significantly warmer (up to 6350 K) and more evolved ($\log g$ down to 3.95 dex) than the Sun follow the relations defined by solar analogues, albeit with a larger scatter \citep[see also][]{adibekyan16}. This suggests that the calibrations may reasonably be used for stars falling formally outside the solar analogue category provided the mismatch in terms of parameters is not too large (but see \citealt{slumstrup17} who argue that the relations might also be valid for solar-metallicity, core-helium burning giants). The \ion{Mg}{i} and \ion{Al}{i} abundances are largely insensitive to non-LTE effects: the differential corrections do not exceed 0.01 dex. The departures from LTE for \ion{Y}{ii} are unknown, but are probably small if one considers that this ionisation stage is by far the most populated.

\subsection{Abundance patterns in the context of planetary formation}\label{sect_discussion_planetary_formation}

\subsubsection{Planets in $\alpha$ Cen}\label{sect_discussion_planets}

Before discussing the implications of our results in the context of planetary formation in $\alpha$ Cen, let us review the attempts made to detect substellar-mass companions in the system.

Being our closest neighbour, the system is a prime target for the detection of planets that could possibly host life. Issues related to the accretion of planetesimals and stability of planets in close binaries (the semi-major axis is only about 23.4 AU) cannot be ignored \cite[e.g.][]{kraus16}. However, circumstellar planets in tight binaries do exist \citep[e.g.][and references therein]{ortiz16}. It has also been claimed that the presence of Proxima Cen orbiting the inner pair is unlikely to prevent planet formation \citep[][]{worth16}.

If one accepts that the conditions are indeed favourable to planet formation and stability in $\alpha$ Cen AB, then it might be expected that the components harbour Jupiter-like companions in view of the higher occurrence of close-in giant planets around metal-rich stars \citep[e.g.][]{gonzalez97}. However, such planets have remained elusive in spite of considerable observational effort \cite[e.g.][]{endl01,kervella06,zhao18}. The failure of high-resolution imaging and sensitive RV monitoring, which probe different regions around the stars, to detect giant planets down to a few Jupiter masses casts serious doubts on their existence.

Evidence for lower-mass planets in $\alpha$ Cen AB is also inconclusive. The discovery of a planet with a minimum mass of 1.3 $M_{\oplus}$ orbiting Proxima Cen was recently announced \citep{anglada_escude16}. It was followed by the detection of cold dust belts and possibly a warm dust reservoir that could be leftovers from a past planetary formation episode \citep{anglada17,ribas17}. Furthermore, the presence of a Keplerian signal was confirmed by \citet{damasso17} from a re-analysis of the RV data of \citet[][]{anglada_escude16}. Hints of transit-like events were also found \citep[][and references therein]{blank18}. However, there is currently little observational evidence for planets of this kind in $\alpha$ Cen AB. There have been claims of a close-in, Earth-mass planet in $\alpha$ Cen B ($\alpha$ Cen Bb) through the detection  of very low-amplitude RV variations in HARPS data \citep[$K$ $\sim$ 0.5 m s$^{-1}$;][]{dumusque12}. Unfortunately, independent re-analyses of these data demonstrated that the weak planetary signal, which is close to the detection capability of the instrument and buried  in ``jitter'' noise arising from magnetic activity, is very likely spurious \citep{hatzes13,rajpaul15,rajpaul16}. The possible existence of a small-size transiting planet, $\alpha$ Cen Bc, was recently reported \citep{demory15}, but needs confirmation. Infrared emission from dusty debris discs has not been detected in $\alpha$ Cen AB \citep{wiegert14}. From the theoretical side, several studies have investigated the formation and dynamics of planets in the binary system, but quite often reached different conclusions owing to the complexity of the problem \citep[e.g.][and references therein]{thebault08}. The recognition that the detection of terrestrial planets might be within reach through an extremely intensive and precise RV monitoring \citep[e.g.][]{guedes08,eggl13} has triggered a number of ambitious campaigns \cite[e.g.][]{endl15}. 

To summarise, even though the existence of a low-mass planet in Proxima Cen seems quite secure, there are only observational hints of a transiting planet in $\alpha$ Cen B and no indications that $\alpha$ Cen A hosts a planet at all.

\subsubsection{Trends with condensation temperature}\label{sect_discussion_Tc_trends}

\citet{melendez09} and \citet{ramirez09} convincingly demonstrated that the Sun is depleted at the 20\% level in species that can easily condensate in dust grains (refractory elements) relative to volatiles when compared to most solar analogues. They showed that only $\sim$15\% of all the stars in their samples have an abundance pattern closely resembling that of the Sun or are poorer in refractories. Moreover, the level of depletion is an increasing function of $T_\mathrm{c}$. They proposed that a similar behaviour in other stars might provide indirect evidence for the existence of rocky material trapped in terrestrial planets  \citep[see also,
e.g.][]{ramirez10}. \citet{chambers10} and \citet{melendez12} went a step further by claiming that an extremely precise abundance analysis can help to constrain the total mass and relative amount of Earth-like and meteoritic material around the star. How gas giant planets fit into this scenario is not completely clear, but it has been postulated that their formation could lead to a global metal deficiency in the parent star and not necessarily a discernable trend in the [X/Fe]-$T_\mathrm{c}$ relation \citep{ramirez14}.

 Over the last decade, a great number of studies have investigated whether a correlation in solar-like dwarfs between metal abundances and condensation temperature could indeed be a relic of a past planetary formation episode. Although appealing, this claim has not received wide support \citep[e.g.][]{gonzalez_hernandez13}. The biggest blow against this interpretation is arguably the lack of any clear refractory depletion in a small sample of {\it Kepler} targets that are known to host Earth-size planets \citep{schuler15}. Nonetheless, the authors warned that their conclusions are still subject to a number of uncertainties, such as those related to the architecture of the planetary systems or the chemical evolution of the Galaxy (the age spread in their sample is at least 3 Gyrs). Conversely, \citet{liu16b} claimed that such a depletion of refractory elements is, as expected, present in the terrestrial planet host {\it Kepler}-10. 

One of the major obstacles that prevents one from clearly establishing a causal link between the depletion of refractory elements and the formation of terrestrial planets is that it is very difficult to define clear-cut and well-defined samples of bona fide single stars and planet hosts. Even in confirmed hosts, the planet census is not complete and there is a very strong bias against finding Earth-like planets. To complicate the matter further, the predictions are sensitive to a number of uncertain assumptions; for example, the size of the convective envelope when the circumstellar material was accreted \citep[e.g.][]{chambers10} or the chemical composition of super-Earth and Neptune-like planets \citep[e.g.][]{rogers15}. Furthermore, the opposite effect (i.e. an excess of refractory elements) can result from the infall of planets onto the star \citep[e.g.][]{spina15}. All these difficulties conspire to make a clear interpretation of the abundance pattern in exoplanet host stars difficult and very often ambiguous.
Although the debate is still far from being settled, some general conclusions are, however, emerging. First and foremost, trends shown by solar twins/analogues between [X/Fe] and $T_\mathrm{c}$ appear to primarily be an age effect imprinted by the chemical evolution of the Galaxy \citep[e.g.][]{adibekyan14,spina16b,nissen15}. Superimposed on these [X/Fe]-$T_\mathrm{c}$ trends are second-order effects unrelated to age that introduce additional scatter. They arise either from the sequestration/ingestion of planetary material \citep[e.g.][]{melendez09,ramirez14}, radiative cleansing of dust in the primordial gas cloud \citep[e.g.][]{onehag14,liu16a}, and/or dust-gas segregation in protoplanetary discs \citep[e.g.][]{gaidos15}. 

\begin{figure*}[h!]
\begin{minipage}[t]{0.5\textwidth}
\centering
\includegraphics[trim=52 165 65 85,clip,width=0.95\textwidth]{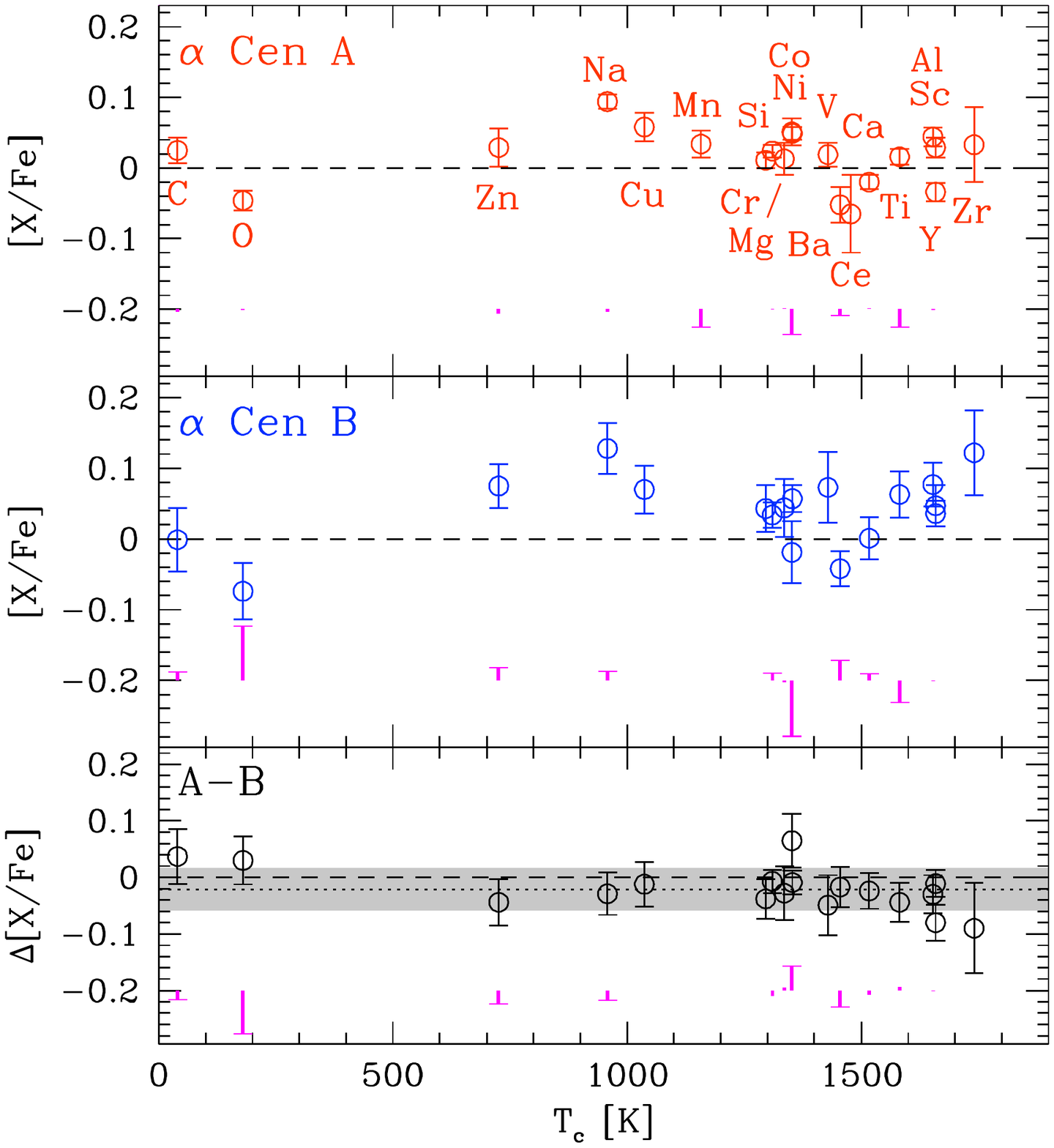}
\end{minipage}
\begin{minipage}[t]{0.5\textwidth}
\centering
\includegraphics[trim=52 165 65 85,clip,width=0.95\textwidth]{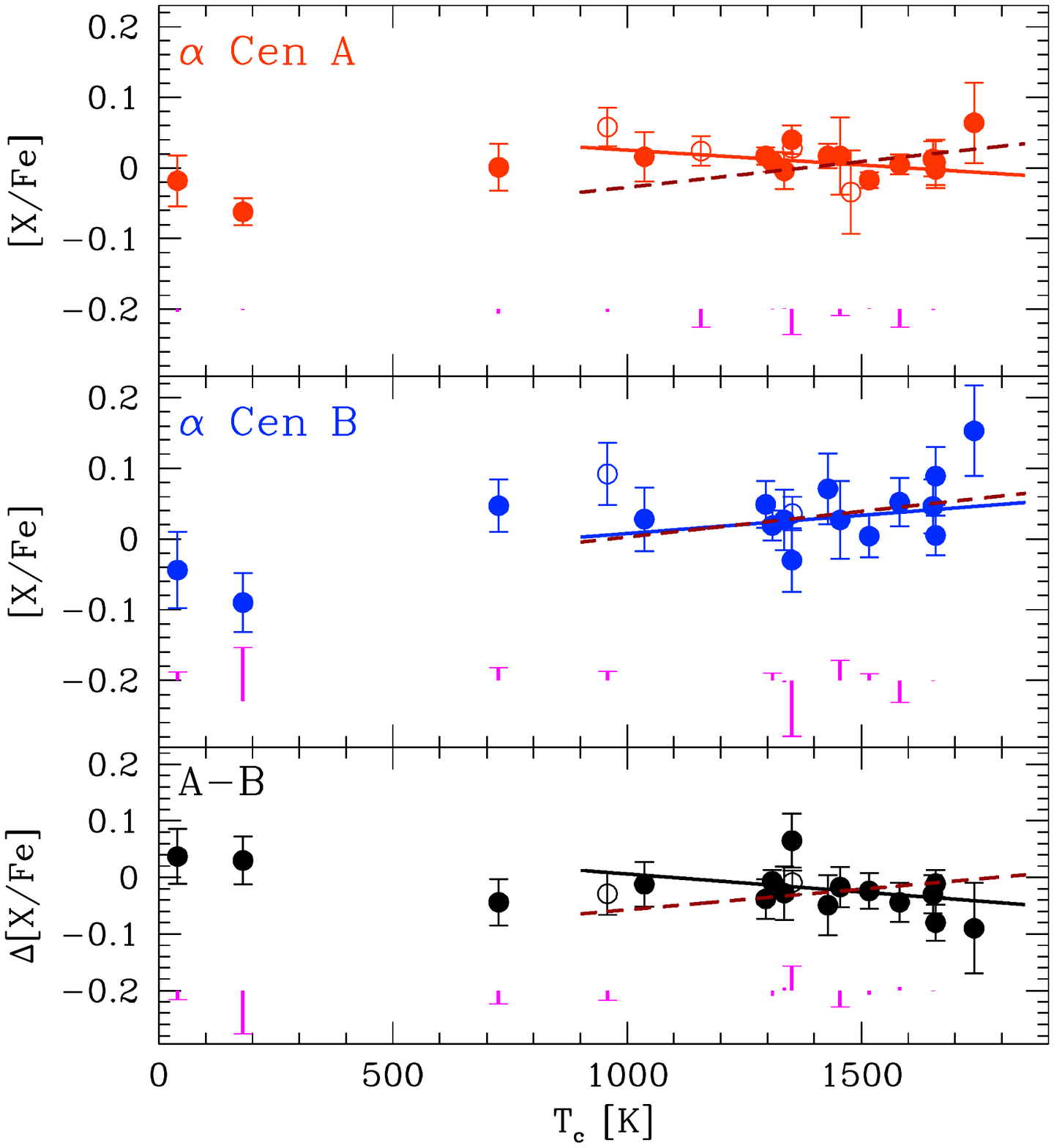}
\end{minipage}
\begin{minipage}[t]{0.5\textwidth}
\end{minipage}
\caption{{\it Left panels:} Abundance patterns as a function of $T_\mathrm{c}$ \citep{lodders03}. The metal abundances of $\alpha$ Cen A and B relative to iron, [X/Fe], are shown in the top and middle panel, respectively. The bottom panel shows the abundance differences, $\Delta$[X/Fe], between $\alpha$ Cen A and B. To guide the eye, a dashed line is drawn at $\Delta$[X/Fe] = 0. The dotted line in the bottom panel indicates the mean difference, while the grey strip shows the corresponding 1-$\sigma$ uncertainties. The vertical lines in the bottom of each panel show the non-LTE corrections (non-LTE minus LTE). A horizontal tick at the bottom or top of the line indicates a negative or positive correction, respectively. Values lower than $\sim$0.01 dex are shown with a dot. {\it Right panels:} Same as left panels, but after correction for GCE. The solid lines are the best linear fits to the refractory elements ($T_\mathrm{c}$ $>$ 900 K; Table \ref{tab_Tc_slopes}). The elements not included in the fit (Na, Mn, Ni, and Ce) are shown as open circles. The dashed line shows the typical GCE-corrected relationship found for solar analogues \citep{bedell18} after adjustment of the intercept to minimise the residuals. }
\label{fig_abundances}
\end{figure*}

To remove the effect of Galactic chemical evolution (GCE), we corrected the abundances using the age-[X/Fe] relations of \citet{nissen16}. For V, Co, Zr, and Ce, which were not included in this study, we adopted the relations of \citet{bedell18}. We assumed an age of 6$\pm$1 Gyrs (Sect.~\ref{sect_ages}). The corrected abundances are given in Table \ref{tab_final_results}. Although the corrections are small given the similar age of $\alpha$ Cen and the Sun, they are noticeable at this level of precision and lead to a reduced scatter (e.g. from 0.040 to 0.029 dex in $\alpha$ Cen A), as can be seen in Fig.~\ref{fig_abundances}. 

To examine the relationship between [X/Fe] corrected for GCE effects and $T_\mathrm{c}$, we only consider refractory elements with $T_\mathrm{c}$ $>$ 900 K, as these species are better indicators of phenomena related to planetary formation \citep[e.g.][]{bedell18}. In addition, the abundances of volatiles (C, O, and Zn) are based on a few lines that are particularly sensitive to some physical effects not considered here (e.g. departures from LTE for the \ion{O}{i} triplet). These elements with uncertain abundances and much lower $T_\mathrm{c}$ values would strongly bias the slopes obtained. We also ignore Na and Ni whose exceptionally high abundance in $\alpha$ Cen (Fig.~\ref{fig_Na_Ni}) is unlikely to be related to planetary formation. For consistency, we finally discard Mn and Ce in order to have exactly the same set of elements for $\alpha$ Cen A and B. We finally end up with 13 elements. The best-fit parameters obtained from a linear regression taking the errors in the abundance ratios into account are given in Table \ref{tab_Tc_slopes}. A linear fit is expected to constitute a good representation of the data, as some differential studies of stellar twins in binaries (where the highest precision is reached) have shown that the element-to-element scatter relative to the regression line is comparable to the measurement errors \citep[e.g.][]{ramirez15}. As seen in Fig.~\ref{fig_abundances}, the non-LTE effects discussed in Sect.~\ref{sect_results_chemical_abundances} are generally small and unlikely to significantly bias the [X/Fe]-$T_\mathrm{c}$ trends.

\begin{table}[h!]
%\scriptsize
\caption{Results of the weighted, linear [X/Fe]-$T_\mathrm{c}$ fits after correction for GCE effects.}\label{tab_Tc_slopes} 
\centering
\begin{tabular}{l|ccc}
\hline\hline
\multicolumn{1}{c}{} & \multicolumn{1}{c}{Slope}                            & \multicolumn{1}{c}{Intercept} & $\chi^2_\mathrm{r}$\\
                     & \multicolumn{1}{c}{[10$^{-5}$ dex K$^{-1}$]}  & \multicolumn{1}{c}{[10$^{-2}$ dex]} & \\
\hline
$\alpha$ Cen A       & --4.201$\pm$3.607 &  +6.735$\pm$5.213   & 0.79\\
$\alpha$ Cen B       &  +5.162$\pm$5.649 & --4.392$\pm$8.257   & 0.92\\
A -- B               & --6.382$\pm$5.253 &  +6.995$\pm$7.755   & 0.68\\
\hline
\end{tabular}
\end{table}

The data for $\alpha$ Cen A hint at an even larger depletion of refractory elements compared to the Sun. However, the [X/Fe]-$T_\mathrm{c}$ slope is very close (at the 1.2$\sigma$ level) to solar and a linear fit with a zero slope is nearly as good based on $\chi^2$ statistics. In contrast, the slope differs from that typical of solar analogues over the same $T_\mathrm{c}$ range and corrected for GCE effects \citep[$\sim$+7.3 $\times$ 10$^{-5}$ dex K$^{-1}$;][]{bedell18} at a much higher confidence level ($\sim$3.2$\sigma$).\footnote{A deviation slightly more pronounced is found if Mn and Ce are taken into account.} Indeed, the slope for $\alpha$ Cen A is comparable to the lowest found among the 68 solar analogues studied by \citet{bedell18}. However, although a fit to our data using the relation for solar analogues after adjusting the intercept to minimise the residuals is significantly worse ($\chi^2$ ratio of $\sim$2.2), it cannot be ruled out on statistical grounds. For $\alpha$ Cen B, no conclusions can be drawn in view of the larger error bars: trends similar to that in the Sun or in solar analogues are both possible. The situation for the abundance differences is similar to that encountered for $\alpha$ Cen A. Although a similar behaviour for the two components is preferred, a fit using the typical relation for solar analogues is also statistically acceptable: slope deviating by 2.6$\sigma$ from the best-fit one and increase in $\chi^2$ by a factor of about 1.9. It is worth recalling that the data discussed here are sensitive to the treatment of GCE effects. However, similar conclusions are obtained when using  the GCE corrections of \citet{bedell18} for all elements.

\begin{figure*}[h!]
\centering
\includegraphics[trim= -5 253 -17 80,clip,width=\hsize]{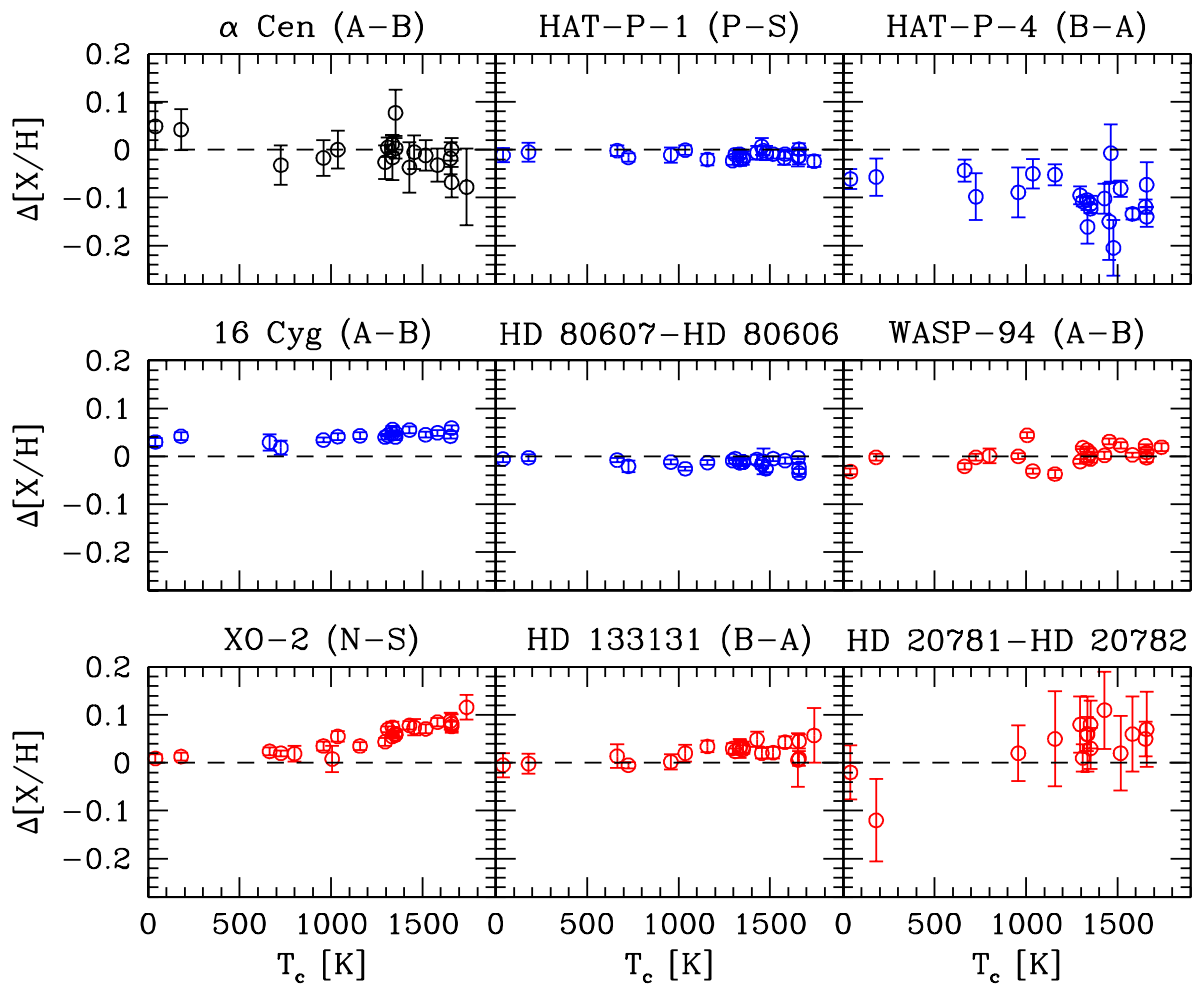}
\caption{Comparison between our abundance differences and those reported in the literature for planet-host binaries also analysed differentially. Systems with only one or the two components known to host planets are shown with blue and red symbols, respectively. For systems with only one planet host, the convention is star without planet minus star with planet. Source of the data: HAT-P-1 \citep{liu14}, HAT-P-4 \citep{saffe17}, 16 Cyg \citep{tucci_maia14}, HD 80606/HD 80607 \citep{liu18}, WASP-94 \citep{teske16a}, XO-2 \citep{ramirez15}, HD 133131 \citep{teske16b}, and HD 20781/HD 20782 \citep{mack14}. For elements with abundances corresponding to two ionisation stages, we followed the procedure outlined in Sect.~\ref{sect_results_chemical_abundances}. For WASP-94, we considered the dataset with strong lines excluded \citep[see][]{teske16a}.}
\label{fig_binaries_literature}
\end{figure*}

Significant progress in our understanding of the [X/Fe]-$T_\mathrm{c}$ trends may come from comparing the abundance pattern of binary components because any differences found are free from environmental or age effects (the latter assuming coevality). To put our results in perspective, we compare in Fig.~\ref{fig_binaries_literature} the abundance differences we find in $\alpha$ Cen to those reported in the literature for planet-host binaries also analysed differentially. Some of them only have one star that is known to host a planet (\object{HAT-P-1}, \object{HAT-P-4}, \object{16 Cyg}, and \object{HD 80606}/\object{HD 80607}), while others have planets orbiting both the primary and secondary (\object{WASP-94}, \object{XO-2}, \object{HD 133131}, and \object{HD 20781}/\object{HD 20782}). These are Jupiter-mass giants in all cases. The only exception is HD 20781, which harbours two super-Earth and two Neptune-like companions \citep{udry17}. A noticeable element-to-element scatter is observed for systems where the $T_\mathrm{eff}$ mismatch between the stars is the largest: $\alpha$ Cen ($\Delta T_\mathrm{eff}$ = 640 K) and HD 20781/HD 20782 ($\Delta T_\mathrm{eff}$ = 465 K). The deviations are much lower for most other systems that all have $\Delta T_\mathrm{eff}$ $\lesssim$ 200 K. This suggests that much of the scatter in $\alpha$ Cen arises from the mismatch in spectral type between the binary components. If present, a clear metallicity offset and/or a well-defined [X/H]-$T_\mathrm{c}$ behaviour, as seen in XO-2, must be buried in the noise. Narrowing down the uncertainties to the appropriate level would require a full treatment of 3D/non-LTE effects and diffusion for all elements; a formidable task well beyond our current capabilities. However, a large offset in bulk metallicity of $\sim$0.1 dex, as claimed in HAT-P-4 and attributed to planet engulfment \citep{saffe17} can be ruled out. The large scatter observed for this system is puzzling considering that the components are extremely close in terms of $T_\mathrm{eff}$ and $\log g$. Although departures in the chemical patterns of binary components can likely be ascribed to the existence of planets, as shown in Fig.~\ref{fig_binaries_literature}, the reverse is {\it not} necessarily true. This is nicely illustrated by HAT-P-1 and HD 80606/HD 80607; despite a close-in giant planet orbiting one of the two stars, the abundance patterns are strikingly similar \citep[to within 0.01 dex:][]{liu14,liu18}.

We now explore the possibility that a putative planet that initially formed in either $\alpha$ Cen A or B was swallowed during the star's evolution. This can occur because of planet-planet scattering \citep[e.g.][]{mustill15} or secular perturbations in multiple stellar systems \citep[e.g.][]{petrovich15}. The latter process operates on timescales that are dramatically longer than the former and is therefore more likely to leave an imprint on the surface abundances because of the much thinner stellar convective zone (CZ) at old ages (see below).

\begin{figure}[h!]
\centering
\includegraphics[trim= 90 160 88 100 ,clip,width=\hsize]{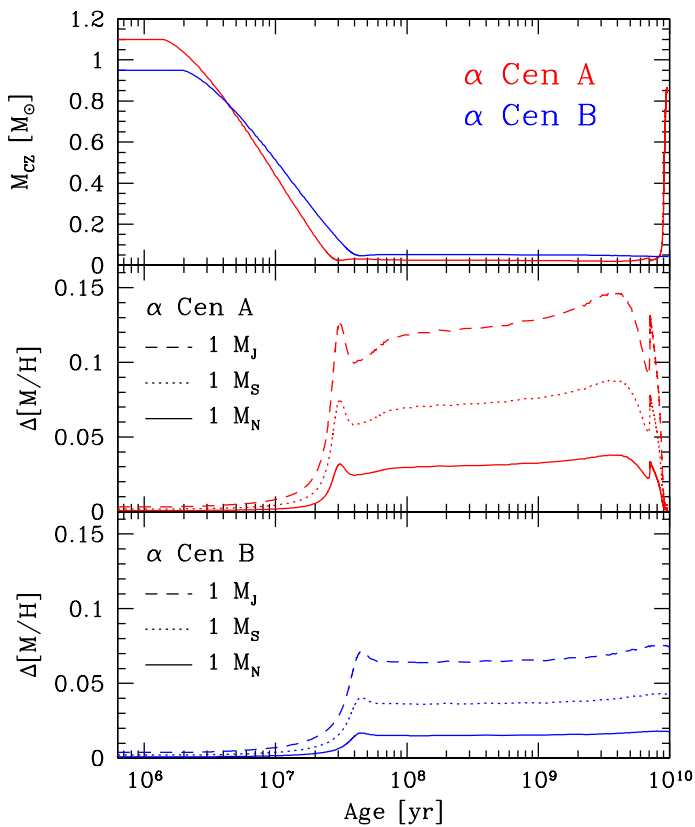}
\caption{{\it Upper panel:} variation of the mass of the CZ as a function of stellar age. {\it Middle and bottom panel:} variation in bulk metallicity corresponding to the ingestion of a planet as a function of stellar age. Three illustrative cases are shown:  Jupiter-, Saturn-, and Neptune-mass planets.}
\label{fig_Mcz_and_Delta_M_H_vs_age}
\end{figure}

The change in metallicity, $\Delta$[M/H], resulting from planet ingestion can be written as \citep[e.g.][]{teske16b}
\begin{equation}
\Delta \mathrm{[M/H]} = \log \left[ \frac{M_\mathrm{CZ}+ M_\mathrm{p} \times [(Z/X)_\mathrm{p}/(Z/X)_\mathrm{CZ}]}{M_\mathrm{CZ}+M_\mathrm{p}} \right],
\label{equat_Mcz_and_Delta_M_H_vs_age}
\end{equation}
where $M_\mathrm{p}$ and $M_\mathrm{CZ}$ are the masses of the planet and of the CZ at the time of accretion, respectively. $(Z/X)_\mathrm{p}$ and $(Z/X)_\mathrm{CZ}$ are the corresponding metal content relative to hydrogen. The resulting change in metallicity is a crude estimate (likely an upper limit) assuming no readjustment of the whole internal structure and that the metal-rich accreted material is instantaneously diluted within the convective envelope and does not sink because of, for example, thermohaline convection \citep[e.g.][]{garaud11,theado12}. To estimate the variation of the CZ mass as the stars evolve, we use Yale models \citep{spada13} computed for a mixing-length parameter $\alpha_\mathrm{MLT}$ = 1.875, [Fe/H] = +0.30, and masses of 0.95 and 1.10  M$_{\sun}$ \citep[the measured values are 1.1055 and 0.9373 M$_{\sun}$ for $\alpha$ Cen A and B, respectively;][]{kervella16}. In addition, we follow \citet{thorngren16} and assume that the planet and stellar metallicities are related via $(Z/X)_\mathrm{p}$/$(Z/X)_\mathrm{CZ}$ $\sim$ 9.7 $\times$ $M_\mathrm{p}^{-0.45}$. Three illustrative cases are examined with planet masses corresponding to that of Neptune, Saturn, and Jupiter. The results are shown in Fig.~\ref{fig_Mcz_and_Delta_M_H_vs_age}. 

A modification of the surface bulk metallicity would be easier to detect in $\alpha$ Cen A because of its larger mass and therefore thinner CZ. Standard evolutionary models indicate that the late accretion of at least one Saturn mass would lead to detectable changes. One would therefore be tempted to conclude that the ingestion of such an amount of planetary material since the time when the stars reached their present-day CZs (30-40 Myrs after star formation) can be ruled out. This conclusion can be extended to earlier times if alternative models where the CZ shrinks faster are considered \citep{baraffe10}. An anomalous abundance for the fragile elements Li and Be may also be taken as evidence for a swallowing event, although the quantitative effect is still very much controversial \citep[e.g.][]{deal15}. No attempts were made here to derive the lithium abundance from \ion{Li}{i} $\lambda$670.8 because of its extreme weakness, but other studies indicate that the Li content in the two stars differs by $\sim$0.7 dex, with $\alpha$ Cen B being more depleted \citep[][]{luck18}. For Be, it amounts to $\sim$0.6 dex \citep{primas97}. These discrepancies may be accounted for by the different internal structure, but detailed calculations are warranted. Accretion of circumstellar material is the best candidate to account for the clear departures between the abundance patterns of the components in the old 16 Cyg and XO-2 systems (Fig.~\ref{fig_binaries_literature}). Although this suggests that extra-mixing mechanisms are perhaps less efficient at erasing the imprint of such processes than currently thought, they are still likely to play an important role and to affect the conclusions above \citep[e.g.][]{garaud11,theado12}. The uncertain impact of mixing on the predictions should be kept in mind.

\section{Conclusions}\label{sect_conclusions}

We derive a robust estimate for the metallicity of the system: [Fe/H] $\sim$ +0.23 dex. The 3D hydrodynamical simulations of $\alpha$ Cen A by \citet{bigot08} intriguingly suggest a lower metallicity ([Fe/H] $\sim$ +0.16 dex) than found by most 1D spectroscopic studies, but the results are regarded by the authors as still preliminary. Further calculations are urgently needed. An inspection of the iron abundances reported in the literature for Proxima Cen reveals an uncomfortably wide spread: +0.16$\pm$0.20 \citep{neves14}, --0.03$\pm$0.09 \citep{maldonado15}, --0.07$\pm$0.14 \citep{passegger16}, and +0.08$\pm$0.12 \citep{zhao18}. Abundances of a few elements (including Na, Ti, and Fe) were also reported by \citet{pavlenko17} to be roughly consistent with solar. The relatively large study-to-study scatter illustrates the difficulties in properly modelling the spectra and atmospheres of very cool stars, even though it is one of the rare M stars benefiting from a $T_\mathrm{eff}$ constraint coming from  interferometric measurements \citep{demory09}. Our study suggests with a high degree of confidence that the iron abundance of the main pair lies in the range +0.20 $<$ [Fe/H] $<$ +0.25. At first glance, this appears to be at odds with the values determined for Proxima Cen, although the lack of consensus for the M star prevents one from reaching definitive conclusions. Further investigations are needed to clarify whether the iron abundance of Proxima Cen fulfils the constraint imposed by $\alpha$ Cen AB. Otherwise, the identification of $\alpha$ Cen as a triple system formed out of the same natal cloud may need to be questioned.

We find evidence in $\alpha$ Cen A for a [X/Fe]-$T_\mathrm{c}$ trend much more similar to what is observed in the Sun than in the majority of solar analogues (no conclusion can be drawn for $\alpha$ Cen B owing to the larger error bars). However, we refrain from associating this finding to the sequestration of rocky material in small bodies orbiting the star. First, a behaviour similar to that of solar analogues cannot be firmly ruled out on statistical grounds. More importantly, a causal link between planetary formation and [X/Fe]-$T_\mathrm{c}$ trends is still hotly debated \citep[e.g.][]{schuler15}. Another difficulty is related to the removal of GCE effects. The system has thin-disc kinematics \citep[e.g.][]{ramirez07} and a Galactic orbit that lies within the solar annulus: mean Galactocentric distance of $\sim$8.3 kpc, eccentricity of $\sim$0.07, and maximum distance above the Galactic plane of $\sim$310 pc \citep{casagrande11}. However, its high metallicity is only shared by a few per cent of all stars in the solar neighbourhood \citep[e.g.][]{casagrande11} and cannot be attributed to a young age. It is conceivable that $\alpha$ Cen migrated from the inner Galaxy \citep[e.g.][]{kordopatis15} where the nucleosynthesis history could have been different from that experienced by the bulk of the nearby solar analogues used to define the GCE trends. It might also imply that the Sun and $\alpha$ Cen formed in different environments and suffered various amounts of dust radiative cleansing \citep[see, e.g.][]{onehag14,liu16a}.

We find that the abundance patterns of $\alpha$ Cen A and B are remarkably similar considering that both line-formation (e.g. 3D/non-LTE) or other physical (e.g. diffusion) effects likely introduce significant random and systematic differences. The similarity between the abundance patterns of the two stars suggests that the late (i.e. after the stars have reached their present-day CZs) accretion of large amounts (above about one Saturn mass) of metal-rich material by either star did not occur, but this conclusion is very sensitive to the details of how this material was subsequently transported downwards.

Assuming coeval formation, we also suggest an age of about $\sim$6 Gyrs for the cool tertiary Proxima Cen and its planet. Obviously, this conclusion fully relies on the assumption that the three stars formed together and Proxima Cen was not captured, an issue which is still not completely settled \citep{feng18,reipurth12}.

\begin{acknowledgements}
The author acknowledges financial support from Belspo for contract PRODEX PLATO. He wishes to thank Megan Bedell for sending her abundance data prior to publication and Chris Sneden for his help with the {\tt blends} driver in MOOG. The useful comments from the anonymous referee are also greatly appreciated. This research made use of NASA's Astrophysics Data System Bibliographic Services and the SIMBAD database operated at CDS, Strasbourg (France).
\end{acknowledgements}

\bibliographystyle{aa} % style aa.bst
\bibliography{33125_corr} % your references Yourfile.bib

\begin{appendix}

\section{Lines selected and EW measurements}\label{sect_appendixA}

Table \ref{tab_appendix_EWs} provides the spectral features selected for each line list and the EW measurements. 

\newpage

\begin{table*}[h!]
%\scriptsize
\scriptsize
\caption{Lines selected and EW measurements.}\label{tab_appendix_EWs} 
\centering
\begin{tabular}{lccc|r|rl|rl}
\hline\hline
    &           &          &      & \multicolumn{1}{c}{Sun}    & \multicolumn{2}{c}{$\alpha$ Cen A} & \multicolumn{2}{c}{$\alpha$ Cen B} \\
Ion & $\lambda$ & Flag HFS & LEP  & \multicolumn{1}{c|}{EW}     & \multicolumn{1}{c}{EW}     & Line lists                & \multicolumn{1}{c}{EW}     & Line lists                \\
    & [nm]      &          & [eV] & \multicolumn{1}{|c|}{[m\AA]} & \multicolumn{1}{c}{[m\AA]} &                           & \multicolumn{1}{c}{[m\AA]} &                           \\
\hline
 \ion{C}{i}     & 505.217   &   N  &  7.685  &   36.0   &    53.1  &  Me14, R03                                        &     36.2  &  Me14, R03                                   \\                    
 \ion{C}{i}     & 538.034   &   N  &  7.685  &   21.6   &    35.7  &  Me14, R03                                        &     16.9  &  Me14, R03                                   \\         
 \ion{O}{i}     & 777.194   &   N  &  9.140  &   73.8   &    91.3  &  Be14, C00, FG01, Me14, Mo14, R03                 &     41.5  &  Be14, C00, FG01, Me14, Mo14, R03                \\         
 \ion{O}{i}     & 777.416   &   N  &  9.140  &   62.8   &    79.6  &  Be14, C00, FG01, Me14, Mo14, R03                 &     38.4  &  Be14, C00, FG01, Me14, Mo14, R03                \\         
 \ion{O}{i}     & 777.539   &   N  &  9.140  &   50.6   &    63.8  &  Be14, C00, FG01, Me14, Mo14, R03                 &     26.7  &  Be14, C00, FG01, Me14, Mo14, R03                \\         
 \ion{Na}{i}    & 615.423   &   N  &  2.100  &   38.6   &    58.6  &  Be14, Bi12, C00, FG01, Me14, Mo14, R03           &     84.1  &  Be14, Bi12, C00, FG01, Me14, Mo14, R03           \\         
 \ion{Na}{i}    & 616.075   &   N  &  2.100  &   59.5   &    81.1  &  Be14, Bi12, C00, Me14, Mo14, R03                 &     ...   &  ...                                      \\    
 \ion{Mg}{i}    & 571.109   &   N  &  4.340  &  106.0   &   121.2  &  Be14, C00, Me14, Mo14                            &    152.1  &  Be14, C00, Me14, Mo14                         \\         
 \ion{Mg}{i}    & 631.871   &   N  &  5.108  &   48.2   &    64.8  &  J15, R03                                         &     83.2  &  J15, R03                                    \\         
 \ion{Al}{i}    & 669.602   &   N  &  3.140  &   39.2   &    56.4  &  Be14, Bi12, Me14, Mo14                           &     80.0  &  Be14, Bi12, Me14, Mo14                        \\         
 \ion{Al}{i}    & 669.867   &   N  &  3.140  &   23.0   &    35.1  &  Be14, Bi12, C00, FG01, Me14, Mo14, R03           &     55.0  &  Be14, Bi12, C00, FG01, Me14, Mo14, R03           \\         
 \ion{Si}{i}    & 551.754   &   N  &  5.080  &   14.3   &    21.7  &  Me14                                             &     21.2  &  Me14                                       \\         
 \ion{Si}{i}    & 564.561   &   N  &  4.930  &   36.2   &    49.0  &  Be14, J15, Me14                                  &     48.7  &  Be14, J15, Me14                              \\         
 \ion{Si}{i}    & 566.556   &   N  &  4.920  &   41.4   &    55.8  &  Be14, C00, FG01, J15, Me14                       &     55.2  &  Be14, C00, FG01, J15, Me14                     \\         
 \ion{Si}{i}    & 568.448   &   N  &  4.950  &   63.2   &    78.6  &  Be14, J15, Me14                                  &     72.7  &  Be14, J15, Me14                              \\         
 \ion{Si}{i}    & 569.042   &   N  &  4.930  &   51.1   &    64.8  &  Be14, C00, J15, Me14                             &     60.7  &  Be14, C00, J15, Me14                          \\         
 \ion{Si}{i}    & 570.841   &   N  &  4.950  &   77.5   &    ...   &  ...                                                &     87.9  &  C00, J15                                    \\ 
 \ion{Si}{i}    & 577.215   &   N  &  5.080  &   53.8   &    70.2  &  Be14, C00, J15                                   &     65.9  &  Be14, C00, J15                               \\         
 \ion{Si}{i}    & 579.307   &   N  &  4.930  &   44.2   &    60.9  &  Be14, C00, FG01, J15, Me14, Mo14, R03            &     57.5  &  Be14, C00, FG01, J15, Me14, Mo14, R03            \\         
 \ion{Si}{i}    & 579.786   &   N  &  4.950  &   44.3   &    59.1  &  Be14, C00                                        &     ...   &  ...                                      \\                    
 \ion{Si}{i}    & 612.502   &   N  &  5.610  &   33.2   &    48.6  &  Be14, Bi12, C00, FG01, Me14, R03                 &     43.3  &  Be14, Bi12, C00, FG01, Me14, R03                \\         
 \ion{Si}{i}    & 614.248   &   N  &  5.610  &   36.3   &    50.4  &  Be14, Bi12, C00, FG01, R03                       &     45.0  &  Be14, Bi12, C00, FG01, R03                     \\         
 \ion{Si}{i}    & 614.502   &   N  &  5.610  &   40.1   &    55.2  &  Be14, Bi12, C00, Me14, R03                       &     47.9  &  Be14, Bi12, C00, Me14, R03                     \\         
 \ion{Si}{i}    & 615.569   &   N  &  5.619  &    9.3   &    15.8  &  FG01                                             &     14.3  &  FG01                                       \\         
 \ion{Si}{i}    & 623.732   &   N  &  5.610  &   62.0   &    82.8  &  Be14, FG01                                       &     ...   &  ...                                      \\                    
 \ion{Si}{i}    & 624.381   &   N  &  5.610  &   48.8   &    64.3  &  Be14, Me14                                       &     57.3  &  Be14, Me14                                  \\         
 \ion{Si}{i}    & 624.447   &   N  &  5.610  &   47.6   &    63.0  &  Be14, Me14, R03                                  &     57.8  &  Be14, Me14, R03                              \\         
 \ion{Si}{i}    & 641.498   &   N  &  5.871  &   48.7   &    68.3  &  Bi12                                             &     57.2  &  Bi12                                       \\         
 \ion{Si}{ii}   & 637.137   &   N  &  8.121  &   30.3   &    43.0  &  J15, R03                                         &     ...   &  ...                                     \\                    
 \ion{Ca}{i}    & 526.039   &   N  &  2.520  &   33.9   &    43.8  &  Be14, J15, Me14                                  &     64.2  &  Be14, J15, Me14                              \\         
 \ion{Ca}{i}    & 586.756   &   N  &  2.930  &   26.0   &    33.3  &  Be14, Bi12, C00, FG01, J15, Me14, R03            &     54.7  &  Be14, Bi12, C00, FG01, J15, Me14, R03            \\         
 \ion{Ca}{i}    & 616.130   &   N  &  2.520  &   67.2   &    78.9  &  Be14, Bi12, C00                                  &     ...   &  ...                                      \\                    
 \ion{Ca}{i}    & 616.644   &   N  &  2.520  &   71.9   &    82.8  &  Be14, Bi12, C00, FG01, J15, Me14, Mo14, R03      &     ...   &  ...                                      \\            
 \ion{Ca}{i}    & 645.560   &   N  &  2.520  &   58.7   &    69.8  &  Be14, Bi12, C00, FG01, J15, Me14, Mo14, R03      &     92.4  &  Be14, Bi12, C00, FG01, J15, Me14, Mo14, R03       \\         
 \ion{Ca}{i}    & 649.965   &   N  &  2.520  &   88.0   &    98.9  &  Be14, Bi12, C00, J15, Me14, Mo14                 &     ...   &  ...                                      \\                    
 \ion{Ca}{ii}   & 645.687   &   N  &  8.438  &   17.6   &    28.1  &  J15                                              &     15.4  &  J15                                        \\         
 \ion{Sc}{i}    & 508.157   &   Y  &  1.448  &    9.1   &    14.7  &  Me14                                             &     ...   &  ...                                       \\         
 \ion{Sc}{i}    & 548.463   &   Y  &  1.851  &    4.3   &    ...   &  ...                                              &     16.8  &  FG01                                       \\  \ion{Sc}{i}    & 552.050   &   Y  &  1.865  &    6.8   &    10.2  &  Me14                                             &     21.4  &  Me14                                       \\         
 \ion{Sc}{i}    & 567.182   &   Y  &  1.448  &   15.5   &    23.5  &  Me14                                             &     60.3  &  Me14                                       \\         
 \ion{Sc}{ii}   & 552.679   &   Y  &  1.768  &   77.8   &    ...   &  ...                                              &     80.5  &  FG01, J15, Me14                              \\ 
 \ion{Sc}{ii}   & 565.789   &   Y  &  1.507  &   68.8   &    84.9  &  J15, Me14                                        &     71.6  &  J15, Me14                                   \\         
 \ion{Sc}{ii}   & 624.563   &   Y  &  1.507  &   37.0   &    49.7  &  Me14, R03                                        &     44.8  &  Me14, R03                                   \\         
 \ion{Sc}{ii}   & 630.070   &   Y  &  1.507  &    8.7   &    12.8  &  FG01                                             &     12.9  &  FG01                                       \\         
 \ion{Sc}{ii}   & 632.084   &   Y  &  1.500  &    8.9   &    ...   &  ...                                              &     11.3  &  Mo14                                  \\ 
 \ion{Sc}{ii}   & 660.460   &   Y  &  1.357  &   37.9   &    51.0  &  FG01, J15, Me14, R03                             &     45.9  &  FG01, J15, Me14, R03                          \\         
 \ion{Ti}{i}    & 480.542   &   N  &  2.345  &   33.8   &    ...   &  ...                                              &     71.1  &  Bi12                                       \\  \ion{Ti}{i}    & 482.041   &   N  &  1.500  &   44.3   &    55.1  &  Be14, Bi12, Me14                                 &     ...   &  ...                                         \\                    
 \ion{Ti}{i}    & 491.361   &   N  &  1.870  &   53.0   &    63.7  &  Be14, Bi12, J15, Me14                            &     ...   &  ...                                      \\                    
 \ion{Ti}{i}    & 492.614   &   N  &  0.818  &    6.3   &    10.0  &  J15                                              &     37.0  &  J15                                        \\         
 \ion{Ti}{i}    & 496.471   &   N  &  1.969  &    9.0   &    12.8  &  J15                                              &     37.7  &  J15                                        \\         
 \ion{Ti}{i}    & 499.709   &   N  &  0.000  &   33.3   &    44.2  &  J15                                              &     ...   &  ...                                    \\                    
 \ion{Ti}{i}    & 511.344   &   N  &  1.440  &   28.7   &    39.5  &  Be14, J15, Me14, R03                             &     74.1  &  Be14, J15, Me14, R03                          \\         
 \ion{Ti}{i}    & 514.546   &   N  &  1.460  &   37.9   &    49.7  &  J15                                              &     ...   &  ...                                    \\                    
 \ion{Ti}{i}    & 514.747   &   N  &  0.000  &   39.8   &    50.6  &  J15, Me14                                        &     ...   &  ...                                     \\                    
 \ion{Ti}{i}    & 515.218   &   N  &  0.021  &   36.9   &    47.6  &  J15                                              &     ...   &  ...                                    \\                    
 \ion{Ti}{i}    & 521.970   &   N  &  0.020  &   28.6   &    38.2  &  Be14, Bi12, J15, Me14, R03                       &     80.8  &  Be14, Bi12, J15, Me14, R03                     \\         
 \ion{Ti}{i}    & 522.430   &   N  &  2.134  &   44.3   &    54.7  &  J15                                              &     ...   &  ...                                    \\                    
 \ion{Ti}{i}    & 529.577   &   N  &  1.067  &   13.4   &    19.2  &  J15, Me14                                        &     51.3  &  J15, Me14                                   \\         
 \ion{Ti}{i}    & 530.001   &   N  &  1.050  &   18.6   &    26.1  &  Be14                                             &     49.7  &  Be14                                       \\         
 \ion{Ti}{i}    & 542.625   &   N  &  0.020  &    6.4   &    10.9  &  Be14, J15                                        &     43.5  &  Be14, J15                                   \\         
 \ion{Ti}{i}    & 547.119   &   N  &  1.440  &    7.5   &     9.6  &  Be14                                             &     38.9  &  Be14                                       \\         
 \ion{Ti}{i}    & 547.422   &   N  &  1.460  &   11.4   &    14.9  &  Be14                                             &     49.3  &  Be14                                       \\         
 \ion{Ti}{i}    & 549.015   &   N  &  1.460  &   23.4   &    31.5  &  Be14, FG01, J15, Me14                            &     62.7  &  Be14, FG01, J15, Me14                         \\         
 \ion{Ti}{i}    & 566.215   &   N  &  2.318  &   24.2   &    32.7  &  J15                                              &     61.5  &  J15                                        \\         
 \ion{Ti}{i}    & 568.946   &   N  &  2.297  &   14.7   &    21.6  &  J15                                              &     44.8  &  J15                                        \\         
 \ion{Ti}{i}    & 570.265   &   N  &  2.292  &    8.4   &    12.7  &  J15                                              &     33.8  &  J15                                        \\         
 \ion{Ti}{i}    & 571.644   &   N  &  2.297  &    6.8   &    ...   &  ...                                              &     28.5  &  J15                                        \\ 
 \ion{Ti}{i}    & 573.947   &   N  &  2.249  &    8.6   &    15.0  &  FG01, Me14                                       &     33.3  &  FG01, Me14                                  \\         
 \ion{Ti}{i}    & 573.998   &   N  &  2.236  &    8.5   &    13.2  &  FG01                                             &     25.4  &  FG01                                       \\         
 \ion{Ti}{i}    & 576.633   &   N  &  3.294  &    9.3   &    15.3  &  Mo14                                             &     31.1  &  Mo14                                       \\         
 . &  . & . & . & . & . & . & . & .\\
 . &  . & . & . & . & . & . & . & .\\
\hline         
\hline
\end{tabular}
\tablefoot{The third column indicates for a given element whether HFS effects were taken into account (``Y'') or not (``N''). Keywords for line lists --- Be14: \citet{bensby14}; Bi12: \citet{biazzo12}; C00: \citet{chen00}; FG01: \citet{feltzing01}; J14: \citet{jofre14}; J15: \citet{jofre15}; Me14: \citet{melendez14}; Mo14: \citet{morel14}; R03: \citet{reddy03}; S08: \citet{sousa08}. A line was excluded from the analysis if, in the star or in the Sun, it was too strong, affected by telluric features, or not adequately fit by a Gaussian profile (Sect.~\ref{sect_line_selection}). Lines yielding discrepant abundances were also not considered further. The table is available in its entirety through the CDS. A portion is shown here for guidance regarding its form and content.\\
}
\end{table*}

\newpage

\section{Stellar parameters and chemical abundances obtained for each line list}\label{sect_appendixB}

Tables \ref{tab_appendix_classical_parameters} and \ref{tab_appendix_differential_parameters} provide for each line list the stellar parameters obtained from the classical and differential analysis, respectively. The chemical abundances of $\alpha$ Cen A and B are given in Tables \ref{tab_appendix_abundances_alfCenA} and \ref{tab_appendix_abundances_alfCenB}, respectively. 

\begin{sidewaystable*}[h!]
\scriptsize
\caption{Stellar parameters obtained from the classical analysis.}\label{tab_appendix_classical_parameters} 
\centering
\begin{tabular}{l|rr|cccc|ccccc|ccc}
\hline\hline
{\bf Star} & \multicolumn{2}{c|}{Number}      & \multicolumn{4}{c}{Unconstrained} & \multicolumn{5}{c}{Constrained --- excitation balance} & \multicolumn{3}{c}{Constrained --- ionisation balance}  \\
Line list  & \multicolumn{2}{c|}{of lines}    & $T_\mathrm{eff}$ & $\log g$ & $\xi$        & [Fe/H] & $T_\mathrm{eff}$ & $\xi$        & [\ion{Fe}{i}/H] & [\ion{Fe}{ii}/H] & [Fe/H] & $T_\mathrm{eff}$ & $\xi$        & [Fe/H]\\
           & \multicolumn{1}{c}{\ion{Fe}{i}} & \multicolumn{1}{c|}{\ion{Fe}{ii}} & [K]             & [cgs]    & [km s$^{-1}$] & & [K]             & [km s$^{-1}$] & & & & [K]             & [km s$^{-1}$] &        \\\hline{\bf $\alpha$ Cen A} &&&&&&&&&&&&&&\\
\citet{bensby14}   & 119 &  13 & 5885$\pm$63 & 4.36$\pm$0.14 & 1.130$\pm$0.097 & 0.251$\pm$0.073 & 5905$\pm$61 & 1.165$\pm$0.084 & 0.274$\pm$0.104 & 0.162$\pm$0.091 & 0.211$\pm$0.068 & 5865$\pm$26 & 1.148$\pm$0.080 & 0.213$\pm$0.065\\
\citet{biazzo12}   &  46 &   9 & 5860$\pm$59 & 4.47$\pm$0.14 & 1.261$\pm$0.100 & 0.263$\pm$0.063 & 5905$\pm$58 & 1.348$\pm$0.080 & 0.275$\pm$0.077 & 0.159$\pm$0.088 & 0.225$\pm$0.058 & 5785$\pm$26 & 1.301$\pm$0.085 & 0.214$\pm$0.053\\
\citet{chen00}     &  55 &   5 & 5795$\pm$49 & 4.34$\pm$0.12 & 1.195$\pm$0.099 & 0.232$\pm$0.050 & 5805$\pm$55 & 1.217$\pm$0.086 & 0.244$\pm$0.070 & 0.150$\pm$0.065 & 0.194$\pm$0.048 & 5785$\pm$16 & 1.207$\pm$0.085 & 0.202$\pm$0.040\\
\citet{feltzing01} &  44 &   7 & 5880$\pm$49 & 4.40$\pm$0.12 & 1.314$\pm$0.071 & 0.248$\pm$0.042 & 5900$\pm$48 & 1.359$\pm$0.069 & 0.254$\pm$0.055 & 0.213$\pm$0.059 & 0.235$\pm$0.040 & 5835$\pm$21 & 1.327$\pm$0.064 & 0.234$\pm$0.035\\
\citet{jofre14}    &  60 &   4 & 5800$\pm$83 & 4.28$\pm$0.17 & 1.094$\pm$0.113 & 0.216$\pm$0.064 & 5785$\pm$84 & 1.068$\pm$0.098 & 0.209$\pm$0.096 & 0.255$\pm$0.079 & 0.236$\pm$0.061 & 5810$\pm$26 & 1.081$\pm$0.090 & 0.230$\pm$0.047\\
\citet{melendez14} &  58 &   7 & 5775$\pm$44 & 4.20$\pm$0.12 & 1.194$\pm$0.075 & 0.247$\pm$0.049 & 5745$\pm$45 & 1.135$\pm$0.076 & 0.250$\pm$0.071 & 0.249$\pm$0.062 & 0.249$\pm$0.047 & 5820$\pm$22 & 1.172$\pm$0.075 & 0.246$\pm$0.044\\
\citet{morel14}    &  30 &   7 & 5785$\pm$42 & 4.28$\pm$0.10 & 1.372$\pm$0.094 & 0.206$\pm$0.042 & 5775$\pm$48 & 1.350$\pm$0.090 & 0.204$\pm$0.055 & 0.239$\pm$0.064 & 0.219$\pm$0.042 & 5800$\pm$21 & 1.364$\pm$0.088 & 0.221$\pm$0.036\\
\citet{reddy03}    &  41 &   5 & 5825$\pm$45 & 4.49$\pm$0.14 & 1.321$\pm$0.063 & 0.248$\pm$0.042 & 5850$\pm$45 & 1.380$\pm$0.065 & 0.255$\pm$0.066 & 0.191$\pm$0.050 & 0.214$\pm$0.040 & 5760$\pm$16 & 1.312$\pm$0.062 & 0.224$\pm$0.031\\
\citet{sousa08}    & 144 &  16 & 5810$\pm$29 & 4.35$\pm$0.07 & 1.223$\pm$0.048 & 0.247$\pm$0.034 & 5825$\pm$23 & 1.250$\pm$0.039 & 0.260$\pm$0.042 & 0.194$\pm$0.046 & 0.230$\pm$0.031 & 5795$\pm$14 & 1.240$\pm$0.038 & 0.224$\pm$0.030\\
%\hline
$<>$               &     &     & 5817$\pm$15 & 4.35$\pm$0.04 & 1.246$\pm$0.025 & 0.239$\pm$0.016 & 5828$\pm$14 & 1.263$\pm$0.023 & 0.247$\pm$0.021 & 0.203$\pm$0.021 & 0.225$\pm$0.015 & 5799$\pm$6  & 1.247$\pm$0.022 & 0.224$\pm$0.013\\\hline
{\bf $\alpha$ Cen B} &&&&&&&&&&&&&&\\
\citet{bensby14}   &  85 &  10 & 5135$\pm$106 & 4.26$\pm$0.22 & 0.588$\pm$0.253 & 0.223$\pm$0.088 & ...         & ...             & ...             & ...             & ...             & ...         & ...             & ...            \\
\citet{biazzo12}   &  30 &   5 & 5310$\pm$86  & 4.61$\pm$0.18 & 1.082$\pm$0.144 & 0.255$\pm$0.061 & 5370$\pm$83 & 1.184$\pm$0.136 & 0.255$\pm$0.080 & 0.156$\pm$0.092 & 0.212$\pm$0.060 & 5275$\pm$23 & 1.101$\pm$0.129 & 0.238$\pm$0.045\\
\citet{chen00}     &  38 &   5 & 5185$\pm$141 & 4.43$\pm$0.19 & 0.992$\pm$0.250 & 0.181$\pm$0.077 & ...         & ...             & ...             & ...             & ...             & 5225$\pm$26 & 0.977$\pm$0.175 & 0.176$\pm$0.059\\
\citet{feltzing01} &  44 &   5 & 5265$\pm$76  & 4.50$\pm$0.17 & 0.931$\pm$0.128 & 0.235$\pm$0.054 & 5245$\pm$65 & 0.888$\pm$0.092 & 0.235$\pm$0.050 & 0.295$\pm$0.111 & 0.245$\pm$0.046 & 5275$\pm$30 & 0.912$\pm$0.080 & 0.254$\pm$0.042\\
\citet{jofre14}    &  46 &   4 & 5060$\pm$202 & 4.14$\pm$0.30 & 0.532$\pm$0.454 & 0.232$\pm$0.114 & ...         & ...             & ...             & ...             & ...             & ...         & ...             & ...            \\
\citet{melendez14} &  49 &   7 & 5140$\pm$67  & 4.41$\pm$0.17 & 0.835$\pm$0.165 & 0.243$\pm$0.059 & 5000$\pm$98 & 0.524$\pm$0.239 & 0.271$\pm$0.083 & 0.402$\pm$0.160 & 0.299$\pm$0.074 & 5190$\pm$22 & 0.737$\pm$0.148 & 0.242$\pm$0.052\\
\citet{morel14}    &  29 &   6 & 5165$\pm$61  & 4.27$\pm$0.15 & 1.032$\pm$0.131 & 0.190$\pm$0.053 & ...         & ...             & ...             & ...             & ...             & 5270$\pm$26 & 0.830$\pm$0.175 & 0.271$\pm$0.057\\
\citet{reddy03}    &  33 &   6 & 5210$\pm$64  & 4.57$\pm$0.19 & 1.024$\pm$0.097 & 0.261$\pm$0.051 & 5235$\pm$51 & 1.064$\pm$0.089 & 0.260$\pm$0.063 & 0.258$\pm$0.091 & 0.259$\pm$0.052 & 5200$\pm$19 & 1.030$\pm$0.082 & 0.271$\pm$0.043\\
\citet{sousa08}    & 124 &  13 & 5160$\pm$49  & 4.29$\pm$0.09 & 0.830$\pm$0.104 & 0.230$\pm$0.046 & ...         & ...             & ...             & ...             & ...             & 5260$\pm$25 & 0.583$\pm$0.112 & 0.294$\pm$0.054\\
%\hline
$<>$               &     &     & 5187$\pm$25  & 4.37$\pm$0.06 & 0.940$\pm$0.048 & 0.230$\pm$0.020 & 5232$\pm$34 & 0.989$\pm$0.056 & 0.250$\pm$0.032 & 0.248$\pm$0.053 & 0.250$\pm$0.028 & 5236$\pm$9  & 0.901$\pm$0.042 & 0.252$\pm$0.019\\\hline
\hline
\end{tabular}
\tablefoot{The last row of each subtable gives the average value weighted by the inverse variance. The blanks indicate that the analysis did not converge (see Sect.~\ref{sect_results_parameters}). The quoted [Fe/H] values are the differences with respect to those obtained from a similar analysis of the solar spectrum.\\
%\tablefoottext{a}{}
%\tablefoottext{b}{}
%\tablefoottext{c}{}
}
\end{sidewaystable*}
%\end{table*}

%\addtocounter{table}{-1}
%\addtocounter{subtable}{1}
\begin{sidewaystable*}[h!]
%\begin{table*}
\scriptsize
\caption{Stellar parameters obtained from the differential analysis.}\label{tab_appendix_differential_parameters} 
\centering
\begin{tabular}{l|rr|cccc|ccccc|ccc}
\hline\hline
{\bf Star} & \multicolumn{2}{c|}{Number}      & \multicolumn{4}{c}{Unconstrained} & \multicolumn{5}{c}{Constrained --- excitation balance} & \multicolumn{3}{c}{Constrained --- ionisation balance}  \\
Line list  & \multicolumn{2}{c|}{of lines}    & $T_\mathrm{eff}$ & $\log g$ & $\xi$        & [Fe/H] & $T_\mathrm{eff}$ & $\xi$        & [\ion{Fe}{i}/H] & [\ion{Fe}{ii}/H] & [Fe/H] & $T_\mathrm{eff}$ & $\xi$        & [Fe/H]\\
           & \multicolumn{1}{c}{\ion{Fe}{i}} & \multicolumn{1}{c|}{\ion{Fe}{ii}} & [K]             & [cgs]    & [km s$^{-1}$] & & [K]             & [km s$^{-1}$] & & & & [K]             & [km s$^{-1}$] &        \\\hline{\bf $\alpha$ Cen A} &&&&&&&&&&&&&&\\
\citet{bensby14}   & 117 &  14 & 5840$\pm$16 & 4.36$\pm$0.04 & 1.145$\pm$0.033 & 0.243$\pm$0.021 & 5855$\pm$15 & 1.177$\pm$0.028 & 0.249$\pm$0.027 & 0.210$\pm$0.028 & 0.230$\pm$0.019 & 5815$\pm$12 & 1.159$\pm$0.027 & 0.228$\pm$0.018\\
\citet{biazzo12}   &  49 &   9 & 5835$\pm$23 & 4.37$\pm$0.06 & 1.350$\pm$0.040 & 0.241$\pm$0.024 & 5845$\pm$23 & 1.374$\pm$0.035 & 0.244$\pm$0.028 & 0.206$\pm$0.034 & 0.229$\pm$0.022 & 5805$\pm$12 & 1.360$\pm$0.034 & 0.224$\pm$0.019\\
\citet{chen00}     &  56 &   6 & 5835$\pm$28 & 4.36$\pm$0.05 & 1.218$\pm$0.048 & 0.242$\pm$0.023 & 5855$\pm$19 & 1.253$\pm$0.038 & 0.247$\pm$0.028 & 0.204$\pm$0.027 & 0.225$\pm$0.019 & 5815$\pm$12 & 1.230$\pm$0.038 & 0.227$\pm$0.018\\
\citet{feltzing01} &  47 &   7 & 5835$\pm$25 & 4.36$\pm$0.08 & 1.341$\pm$0.040 & 0.232$\pm$0.023 & 5845$\pm$31 & 1.362$\pm$0.075 & 0.236$\pm$0.029 & 0.202$\pm$0.033 & 0.221$\pm$0.022 & 5815$\pm$13 & 1.346$\pm$0.035 & 0.220$\pm$0.016\\
\citet{jofre14}    &  61 &   5 & 5845$\pm$23 & 4.36$\pm$0.06 & 1.138$\pm$0.039 & 0.244$\pm$0.022 & 5865$\pm$21 & 1.176$\pm$0.031 & 0.251$\pm$0.026 & 0.206$\pm$0.030 & 0.232$\pm$0.020 & 5820$\pm$17 & 1.152$\pm$0.032 & 0.229$\pm$0.019\\
\citet{melendez14} &  58 &   9 & 5810$\pm$13 & 4.32$\pm$0.05 & 1.283$\pm$0.032 & 0.220$\pm$0.020 & 5815$\pm$13 & 1.287$\pm$0.029 & 0.223$\pm$0.023 & 0.213$\pm$0.030 & 0.219$\pm$0.018 & 5805$\pm$12 & 1.285$\pm$0.028 & 0.217$\pm$0.018\\
\citet{morel14}    &  33 &   7 & 5835$\pm$20 & 4.36$\pm$0.05 & 1.272$\pm$0.043 & 0.245$\pm$0.019 & 5845$\pm$19 & 1.304$\pm$0.039 & 0.248$\pm$0.025 & 0.213$\pm$0.025 & 0.230$\pm$0.018 & 5810$\pm$12 & 1.278$\pm$0.037 & 0.232$\pm$0.016\\
\citet{reddy03}    &  38 &   5 & 5825$\pm$19 & 4.35$\pm$0.06 & 1.356$\pm$0.030 & 0.234$\pm$0.018 & 5835$\pm$16 & 1.373$\pm$0.031 & 0.237$\pm$0.025 & 0.210$\pm$0.018 & 0.219$\pm$0.015 & 5810$\pm$12 & 1.356$\pm$0.030 & 0.224$\pm$0.017\\
\citet{sousa08}    & 140 &  17 & 5830$\pm$18 & 4.33$\pm$0.05 & 1.254$\pm$0.034 & 0.233$\pm$0.022 & 5835$\pm$14 & 1.267$\pm$0.026 & 0.235$\pm$0.027 & 0.220$\pm$0.029 & 0.228$\pm$0.020 & 5820$\pm$12 & 1.262$\pm$0.026 & 0.226$\pm$0.019\\
%\hline
$<>$               &     &     & 5829$\pm$6  & 4.35$\pm$0.02 & 1.265$\pm$0.012 & 0.237$\pm$0.007 & 5840$\pm$6  & 1.272$\pm$0.011 & 0.241$\pm$0.009 & 0.210$\pm$0.009 & 0.225$\pm$0.006 & 5812$\pm$4  & 1.265$\pm$0.010 & 0.225$\pm$0.006\\\hline
{\bf $\alpha$ Cen B} &&&&&&&&&&&&&&\\
\citet{bensby14}   &  80 &   8 & 5100$\pm$54  & 4.18$\pm$0.11 & 0.646$\pm$0.040 & 0.226$\pm$0.045 & ...         & ...             & ...             & ...             & ...             & ...         & ...             & ...            \\
\citet{biazzo12}   &  29 &   6 & 5235$\pm$71  & 4.49$\pm$0.16 & 1.096$\pm$0.142 & 0.231$\pm$0.053 & 5200$\pm$65 & 1.037$\pm$0.099 & 0.234$\pm$0.049 & 0.291$\pm$0.092 & 0.247$\pm$0.043 & 5250$\pm$16 & 1.073$\pm$0.086 & 0.242$\pm$0.031\\
\citet{chen00}     &  35 &   5 & 5165$\pm$126 & 4.33$\pm$0.17 & 0.920$\pm$0.277 & 0.197$\pm$0.078 & ...         & ...             & ...             & ...             & ...             & ...         & ...             & ...            \\
\citet{feltzing01} &  44 &   5 & 5220$\pm$59  & 4.44$\pm$0.15 & 0.943$\pm$0.114 & 0.238$\pm$0.049 & 5150$\pm$62 & 0.800$\pm$0.100 & 0.243$\pm$0.047 & 0.366$\pm$0.099 & 0.266$\pm$0.042 & 5255$\pm$16 & 0.883$\pm$0.077 & 0.262$\pm$0.030\\
\citet{jofre14}    &  47 &   4 & 5105$\pm$93  & 4.16$\pm$0.15 & 0.706$\pm$0.197 & 0.223$\pm$0.055 & ...         & ...             & ...             & ...             & ...             & ...         & ...             & ...            \\
\citet{melendez14} &  43 &   7 & 5210$\pm$40  & 4.30$\pm$0.09 & 0.885$\pm$0.106 & 0.225$\pm$0.039 & ...         & ...             & ...             & ...             & ...             & 5305$\pm$28 & 0.423$\pm$0.296 & 0.338$\pm$0.068\\
\citet{morel14}    &  30 &   7 & 5195$\pm$73  & 4.28$\pm$0.13 & 1.002$\pm$0.129 & 0.206$\pm$0.041 & ...         & ...             & ...             & ...             & ...             & 5310$\pm$25 & 0.910$\pm$0.118 & 0.275$\pm$0.047\\
\citet{reddy03}    &  26 &   5 & 5225$\pm$36  & 4.44$\pm$0.12 & 1.060$\pm$0.067 & 0.232$\pm$0.038 & 5195$\pm$37 & 0.999$\pm$0.064 & 0.241$\pm$0.040 & 0.310$\pm$0.062 & 0.261$\pm$0.034 & 5250$\pm$14 & 1.035$\pm$0.059 & 0.257$\pm$0.026\\
\citet{sousa08}    & 112 &  12 & 5145$\pm$50  & 4.23$\pm$0.11 & 0.851$\pm$0.123 & 0.195$\pm$0.051 & ...         & ...             & ...             & ...             & ...             & 5285$\pm$19 & 0.445$\pm$0.181 & 0.312$\pm$0.051\\
%\hline
$<>$               &     &     & 5189$\pm$18  & 4.30$\pm$0.05 & 0.950$\pm$0.039 & 0.221$\pm$0.016 & 5186$\pm$29 & 0.963$\pm$0.047 & 0.240$\pm$0.026 & 0.317$\pm$0.046 & 0.259$\pm$0.023 & 5265$\pm$7  & 0.958$\pm$0.038 & 0.265$\pm$0.015\\\hline
\hline
%\tablefootmark{}
\end{tabular}
\tablefoot{The last row of each subtable gives the average value weighted by the inverse variance. The blanks indicate that the analysis did not converge (see Sect.~\ref{sect_results_parameters}).\\
%\tablefoottext{a}{}
%\tablefoottext{b}{}
%\tablefoottext{c}{}
}
\end{sidewaystable*}
%\end{table*}

%\newpage

%\addtocounter{table}{-1}
%\addtocounter{subtable}{1}
\begin{sidewaystable*}[h!]
%\begin{table*}
\scriptsize
\caption{Chemical abundances of $\alpha$ Cen A obtained from the differential, unconstrained analysis.}\label{tab_appendix_abundances_alfCenA} 
%\centering
\begin{tabular}{l|cccccccccc}
\hline\hline
Line list           &  $[$\ion{Fe}{i}/H$]$   & $[$\ion{Fe}{ii}/H$]$ &  $[$\ion{C}{i}/Fe$]$ & $[$\ion{O}{i}/Fe$]$   & $[$\ion{Na}{i}/Fe$]$ & $[$\ion{Mg}{i}/Fe$]$ & $[$\ion{Al}{i}/Fe$]$ & $[$\ion{Si}{i}/Fe$]$ & $[$\ion{Si}{ii}/Fe$]$ & $[$\ion{Ca}{i}/Fe$]$   \\\hline 
\citet{bensby14}    &  0.243$\pm$0.028 (117) & 0.243$\pm$0.032 (14) &  ...                 & --0.064$\pm$0.034 (3) & 0.089$\pm$0.023 (2)  & 0.002$\pm$0.057 (1)  & 0.042$\pm$0.026 (2)  & 0.016$\pm$0.023 (11) &  ...                  & --0.022$\pm$0.024 (5)  \\        
\citet{biazzo12}    &  0.241$\pm$0.030 (49)  & 0.242$\pm$0.039 (9)  &  ...                 &   ...                 & 0.094$\pm$0.027 (2)  &  ...                 & 0.046$\pm$0.031 (2)  & 0.024$\pm$0.028 (4)  &  ...                  & --0.032$\pm$0.031 (5)  \\ 
\citet{chen00}      &  0.241$\pm$0.029 (56)  & 0.243$\pm$0.038 (6)  &  ...                 & --0.052$\pm$0.040 (3) & 0.098$\pm$0.029 (2)  & 0.013$\pm$0.060 (1)  & 0.036$\pm$0.058 (1)  & 0.023$\pm$0.024 (7)  &  ...                  & --0.015$\pm$0.030 (4)  \\ 
\citet{feltzing01}  &  0.232$\pm$0.031 (47)  & 0.232$\pm$0.035 (7)  &  ...                 & --0.045$\pm$0.042 (3) & 0.102$\pm$0.055 (1)  &  ...                 & 0.042$\pm$0.055 (1)  & 0.040$\pm$0.028 (6)  &  ...                  & --0.026$\pm$0.031 (3)  \\            
\citet{jofre14,jofre15}     &  0.244$\pm$0.027 (61)  & 0.244$\pm$0.039 (5)  &  ...                 &   ...                 &  ...                 & 0.016$\pm$0.057 (1)  &  ...                 & 0.018$\pm$0.029 (6)  & --0.014$\pm$0.062 (1) & --0.020$\pm$0.027 (4)  \\ 
\citet{melendez14}  &  0.220$\pm$0.024 (58)  & 0.219$\pm$0.036 (9)  & 0.033$\pm$0.027 (2)  & --0.027$\pm$0.033 (3) & 0.102$\pm$0.023 (2)  & 0.018$\pm$0.057 (1)  & 0.056$\pm$0.028 (2)  & 0.016$\pm$0.022 (8)  &  ...                  & --0.015$\pm$0.023 (4)  \\ 
\citet{morel14}     &  0.245$\pm$0.023 (33)  & 0.246$\pm$0.034 (7)  &  ...                 & --0.056$\pm$0.032 (3) & 0.085$\pm$0.025 (2)  & 0.006$\pm$0.057 (1)  & 0.037$\pm$0.028 (2)  & 0.052$\pm$0.053 (1)  &  ...                  & --0.013$\pm$0.025 (3)  \\  
\citet{reddy03}     &  0.234$\pm$0.025 (38)  & 0.234$\pm$0.026 (5)  & 0.019$\pm$0.023 (2)  & --0.037$\pm$0.032 (3) & 0.095$\pm$0.020 (2)  & 0.022$\pm$0.054 (1)  & 0.037$\pm$0.053 (1)  & 0.030$\pm$0.025 (5)  &   0.021$\pm$0.056 (1) & --0.027$\pm$0.029 (3)  \\ 
\citet{sousa08}     &  0.233$\pm$0.027 (140) & 0.233$\pm$0.037 (17) &  ...                 &   ...                 &  ...                 &  ...                 &  ...                 &  ...                 &  ...                  &   ...                  \\ 
$<>$                &  0.237$\pm$0.009       & 0.237$\pm$0.011      & 0.025$\pm$0.018      & --0.046$\pm$0.014     & 0.094$\pm$0.010      & 0.013$\pm$0.023      & 0.044$\pm$0.013      & 0.024$\pm$0.009      &   0.005$\pm$0.042     & --0.020$\pm$0.010      \\
\hline       
\multicolumn{1}{c}{} &  & &  &  & & & & & \\
\hline
Line list           &   $[$\ion{Ca}{ii}/Fe$]$ &   $[$\ion{Sc}{i}/Fe$]$ & $[$\ion{Sc}{ii}/Fe$]$ & $[$\ion{Ti}{i}/Fe$]$ & $[$\ion{Ti}{ii}/Fe$]$ & $[$\ion{V}{i}/Fe$]$  & $[$\ion{Cr}{i}/Fe$]$   & $[$\ion{Cr}{ii}/Fe$]$ & $[$\ion{Mn}{i}/Fe$]$  & $[$\ion{Co}{i}/Fe$]$\\\hline 
\citet{bensby14}    &   ...                   &   ...                  &  ...                  & 0.014$\pm$0.027 (10) & --0.009$\pm$0.020 (5) &  ...                 &   0.010$\pm$0.029 (6)  & 0.045$\pm$0.055 (1)   &  ...                  &  ...                \\        
\citet{biazzo12}    &   ...                   &   ...                  &  ...                  & 0.007$\pm$0.032 (7)  &   0.001$\pm$0.055 (1) &  ...                 &   0.012$\pm$0.029 (2)  &  ...                  &  ...                  &  ...                \\ 
\citet{chen00}      &   ...                   &   ...                  &  ...                  & 0.018$\pm$0.035 (3)  &   ...                 & 0.025$\pm$0.042 (3)  &   0.010$\pm$0.040 (2)  &  ...                  &  ...                  &  ...                \\ 
\citet{feltzing01}  &   ...                   &   ...                  & 0.014$\pm$0.036 (2)   & 0.018$\pm$0.031 (7)  &   ...                 & 0.024$\pm$0.038 (9)  &   0.020$\pm$0.030 (5)  &  ...                  &  ...                  & 0.064$\pm$0.028 (5) \\            
\citet{jofre14,jofre15}     & --0.036$\pm$0.059 (1)   &   ...                  & 0.054$\pm$0.041 (2)   & 0.019$\pm$0.036 (23) &   0.022$\pm$0.049 (7) & 0.030$\pm$0.041 (12) &   0.009$\pm$0.030 (7)  & 0.035$\pm$0.055 (1)   & 0.053$\pm$0.042 (2)   & 0.051$\pm$0.039 (8) \\ 
\citet{melendez14}  &   ...                   &  0.021$\pm$0.028 (3)   & 0.027$\pm$0.035 (3)   & 0.008$\pm$0.030 (12) &   0.031$\pm$0.048 (5) & 0.018$\pm$0.039 (8)  &   0.013$\pm$0.027 (10) & 0.044$\pm$0.023 (2)   & 0.029$\pm$0.022 (2)   & 0.030$\pm$0.039 (6) \\ 
\citet{morel14}     &   ...                   &   ...                  &  ...                  & 0.042$\pm$0.056 (1)  &   ...                 &  ...                 & --0.014$\pm$0.055 (1)  &  ...                  &  ...                  &  ...                \\  
\citet{reddy03}     &   ...                   &   ...                  & 0.029$\pm$0.019 (2)   & 0.021$\pm$0.025 (5)  &   ...                 & 0.005$\pm$0.033 (6)  &   0.011$\pm$0.028 (3)  &  ...                  &  ...                  & 0.036$\pm$0.076 (2) \\ 
\citet{sousa08}     &   ...                   &   ...                  &  ...                  &  ...                 &   ...                 &  ...                 &   ...                  &  ...                  &  ...                  &  ...                \\ 
$<>$                & --0.036$\pm$0.059       &  0.021$\pm$0.028       & 0.029$\pm$0.014       & 0.016$\pm$0.011      &   0.000$\pm$0.016     & 0.019$\pm$0.017      &   0.011$\pm$0.011      & 0.043$\pm$0.020       & 0.034$\pm$0.019       & 0.051$\pm$0.019     \\
\hline       
\multicolumn{1}{c}{} &  & &  &  & & & & & \\
\hline
Line list           & $[$\ion{Ni}{i}/Fe$]$  &   $[$\ion{Cu}{i}/Fe$]$ & $[$\ion{Zn}{i}/Fe$]$ & $[$\ion{Y}{ii}/Fe$]$  & $[$\ion{Zr}{ii}/Fe$]$ & $[$\ion{Ba}{ii}/Fe$]$ & $[$\ion{Ce}{ii}/Fe$]$ & $[$\ion{Y}{ii}/\ion{Mg}{i}$]$ &  $[$\ion{Y}{ii}/\ion{Al}{i}$]$ & \\\hline 
\citet{bensby14}    & 0.047$\pm$0.025 (30)  &    ...                 & 0.016$\pm$0.054 (1)  & --0.039$\pm$0.023 (6) &  ...                  & --0.080$\pm$0.056 (1) &   ...                 & --0.041$\pm$0.059             & --0.081$\pm$0.030              & \\        
\citet{biazzo12}    & 0.046$\pm$0.028 (19)  &    ...                 & 0.026$\pm$0.055 (1)  &   ...                 &  ...                  &   ...                 &   ...                 &   ...                         &   ...                          & \\ 
\citet{chen00}      & 0.049$\pm$0.032 (16)  &    ...                 &  ...                 &   ...                 &  ...                  & --0.048$\pm$0.056 (1) &   ...                 &   ...                         &   ...                          & \\ 
\citet{feltzing01}  & 0.053$\pm$0.025 (19)  &    ...                 &  ...                 &   ...                 &  ...                  &   ...                 &   ...                 &   ...                         &   ...                          & \\            
\citet{jofre14,jofre15}     & 0.061$\pm$0.027 (13)  &    ...                 &  ...                 &   ...                 &  ...                  &   ...                 &   ...                 &   ...                         &   ...                  & \\ 
\citet{melendez14}  & 0.046$\pm$0.026 (14)  &   0.057$\pm$0.029 (2)  & 0.027$\pm$0.054 (1)  & --0.044$\pm$0.023 (4) &  ...                  & --0.073$\pm$0.055 (1) & --0.065$\pm$0.055 (1) & --0.062$\pm$0.061             & --0.100$\pm$0.033              & \\ 
\citet{morel14}     & 0.051$\pm$0.025 (7)   &    ...                 &  ...                 &   ...                 &  ...                  & --0.027$\pm$0.055 (1) &   ...                 &   ...                         &   ...                          & \\  
\citet{reddy03}     & 0.044$\pm$0.023 (12)  &   0.058$\pm$0.028 (2)  & 0.048$\pm$0.054 (1)  & --0.021$\pm$0.022 (4) & 0.033$\pm$0.053 (1)   & --0.035$\pm$0.053 (1) &   ...                 & --0.043$\pm$0.060             & --0.058$\pm$0.059              & \\ 
\citet{sousa08}     &  ...                  &    ...                 &  ...                 &   ...                 &  ...                  &   ...                 &   ...                 &   ...                         &   ...                          & \\ 
$<>$                & 0.049$\pm$0.009       &   0.058$\pm$0.020      & 0.029$\pm$0.027      & --0.034$\pm$0.013     & 0.033$\pm$0.053       & --0.052$\pm$0.025     & --0.065$\pm$0.055     & --0.048$\pm$0.035             &   --0.086$\pm$0.021            & \\
\hline\hline       
%\tablefootmark{}
\end{tabular}
\tablefoot{The number in parentheses gives the number of lines used for a given ion. The last row of each subtable gives the average value weighted by the inverse variance.\\
%\tablefoottext{a}{}
%\tablefoottext{b}{}
%\tablefoottext{c}{}
}
\end{sidewaystable*}
%\end{table*}

%\addtocounter{table}{-1}
%\addtocounter{subtable}{1}
\begin{sidewaystable*}[h!]
%\begin{table*}
\scriptsize
\caption{Chemical abundances of $\alpha$ Cen B obtained from the differential, unconstrained analysis.}\label{tab_appendix_abundances_alfCenB} 
%\centering
\begin{tabular}{l|cccccccccc}
\hline\hline
Line list           &  $[$\ion{Fe}{i}/H$]$   & $[$\ion{Fe}{ii}/H$]$ &  $[$\ion{C}{i}/Fe$]$  & $[$\ion{O}{i}/Fe$]$   & $[$\ion{Na}{i}/Fe$]$ & $[$\ion{Mg}{i}/Fe$]$  & $[$\ion{Al}{i}/Fe$]$ & $[$\ion{Si}{i}/Fe$]$ & $[$\ion{Si}{ii}/Fe$]$ & $[$\ion{Ca}{i}/Fe$]$   \\\hline 
\citet{bensby14}    &  0.225$\pm$0.053 (80)  & 0.227$\pm$0.085 (8)  &  ...                  &   0.003$\pm$0.095 (3) & 0.091$\pm$0.090 (1)  & --0.001$\pm$0.095 (1) & 0.035$\pm$0.074 (2)  & 0.046$\pm$0.044 (11) &  ...                  &   0.003$\pm$0.078 (3)  \\        
\citet{biazzo12}    &  0.231$\pm$0.063 (29)  & 0.230$\pm$0.099 (6)  &  ...                  &   ...                 & 0.136$\pm$0.098 (1)  &  ...                  & 0.099$\pm$0.081 (2)  & 0.017$\pm$0.052 (4)  &  ...                  & --0.021$\pm$0.085 (2)  \\ 
\citet{chen00}      &  0.197$\pm$0.088 (35)  & 0.197$\pm$0.166 (5)  &  ...                  & --0.030$\pm$0.138 (3) & 0.143$\pm$0.157 (1)  &   0.045$\pm$0.155 (1) & 0.085$\pm$0.152 (1)  & 0.052$\pm$0.069 (8)  &  ...                  & --0.009$\pm$0.139 (2)  \\ 
\citet{feltzing01}  &  0.238$\pm$0.058 (44)  & 0.237$\pm$0.090 (5)  &  ...                  & --0.102$\pm$0.108 (3) & 0.125$\pm$0.093 (1)  &  ...                  & 0.069$\pm$0.087 (1)  & 0.035$\pm$0.046 (5)  &  ...                  & --0.028$\pm$0.083 (2)  \\            
\citet{jofre14,jofre15}     &  0.223$\pm$0.059 (47)  & 0.225$\pm$0.147 (4)  &  ...                  &   ...                 &  ...                 &   0.055$\pm$0.114 (1) &  ...                 & 0.025$\pm$0.062 (7)  &  ...                  & --0.035$\pm$0.122 (3)  \\ 
\citet{melendez14}  &  0.225$\pm$0.044 (43)  & 0.223$\pm$0.089 (7)  & --0.014$\pm$0.063 (1) & --0.107$\pm$0.087 (3) & 0.152$\pm$0.079 (1)  &   0.080$\pm$0.093 (1) & 0.110$\pm$0.070 (1)  & 0.016$\pm$0.048 (10) &  ...                  &   0.044$\pm$0.072 (3)  \\ 
\citet{morel14}     &  0.206$\pm$0.043 (30)  & 0.207$\pm$0.122 (7)  &  ...                  & --0.073$\pm$0.105 (3) & 0.132$\pm$0.112 (1)  &   0.033$\pm$0.122 (1) & 0.076$\pm$0.096 (2)  & 0.064$\pm$0.065 (1)  &  ...                  &   0.082$\pm$0.104 (1)  \\  
\citet{reddy03}     &  0.233$\pm$0.046 (26)  & 0.230$\pm$0.067 (5)  &   0.012$\pm$0.063 (1) & --0.102$\pm$0.084 (3) & 0.123$\pm$0.077 (1)  &   0.048$\pm$0.075 (1) & 0.071$\pm$0.072 (1)  & 0.031$\pm$0.043 (5)  & ...                   & --0.020$\pm$0.060 (2)  \\ 
\citet{sousa08}     &  0.195$\pm$0.061 (112) & 0.196$\pm$0.094 (12) &  ...                  &   ...                 &  ...                 &  ...                  &  ...                 &  ...                 &  ...                  &   ...                  \\ 
$<>$                &  0.221$\pm$0.018       & 0.223$\pm$0.032      & --0.001$\pm$0.045     & --0.074$\pm$0.040     & 0.128$\pm$0.036      &   0.044$\pm$0.041     & 0.077$\pm$0.031      & 0.034$\pm$0.018      & ...                   &   0.001$\pm$0.030     \\
\hline       
\multicolumn{1}{c}{} &  & &  &  & & & & & \\
\hline
Line list           &  $[$\ion{Ca}{ii}/Fe$]$ &  $[$\ion{Sc}{i}/Fe$]$ & $[$\ion{Sc}{ii}/Fe$]$ & $[$\ion{Ti}{i}/Fe$]$ & $[$\ion{Ti}{ii}/Fe$]$ & $[$\ion{V}{i}/Fe$]$   & $[$\ion{Cr}{i}/Fe$]$  & $[$\ion{Cr}{ii}/Fe$]$ & $[$\ion{Mn}{i}/Fe$]$ & $[$\ion{Co}{i}/Fe$]$  \\\hline 
\citet{bensby14}    &  ...                   &  ...                  &  ...                  & 0.050$\pm$0.095 (9)  & --0.048$\pm$0.080 (5) &  ...                  &   0.051$\pm$0.085 (6) & 0.181$\pm$0.047 (2)   &  ...                 &  ...                  \\        
\citet{biazzo12}    &  ...                   &  ...                  &  ...                  & 0.084$\pm$0.110 (6)  &  ...                  &  ...                  & ...                   &  ...                  &  ...                 &  ...                  \\ 
\citet{chen00}      &  ...                   &  ...                  &  ...                  & 0.063$\pm$0.155 (3)  &   ...                 &   0.085$\pm$0.162 (1) &   0.041$\pm$0.136 (1) &  ...                  &  ...                 &  ...                  \\ 
\citet{feltzing01}  &  ...                   & --0.055$\pm$0.105 (1) &   0.068$\pm$0.039 (3) & 0.047$\pm$0.096 (7)  &   ...                 &   0.060$\pm$0.112 (5) &   0.053$\pm$0.089 (5) &  ...                  &  ...                 &   0.016$\pm$0.074 (5) \\            
\citet{jofre14,jofre15}     &  0.210$\pm$0.097 (1)   &  ...                  & --0.025$\pm$0.054 (3) & 0.018$\pm$0.146 (14) & --0.053$\pm$0.087 (5) &   0.004$\pm$0.161 (9) & --0.018$\pm$0.115 (3) &  ...                  &  ...                 & --0.076$\pm$0.084 (7) \\ 
\citet{melendez14}  &  ...                   & --0.060$\pm$0.119 (2) &   0.017$\pm$0.038 (4) & 0.089$\pm$0.075 (6)  & --0.008$\pm$0.080 (4) &   0.086$\pm$0.083 (5) &   0.066$\pm$0.061 (4) & 0.182$\pm$0.063 (1)   & ...                  & --0.021$\pm$0.079 (4) \\ 
\citet{morel14}     &  ...                   &  ...                  & --0.050$\pm$0.070 (1) & 0.089$\pm$0.111 (1)  &   ...                 &  ...                  &   0.038$\pm$0.110 (1) &  ...                  &  ...                 &  ...                  \\  
\citet{reddy03}     &  ...                   &  ...                  &   0.062$\pm$0.029 (2) & 0.052$\pm$0.063 (3)  &   ...                 &   0.086$\pm$0.100 (3) &   0.022$\pm$0.079 (1) &  ...                  &  ...                 &   0.052$\pm$0.196 (2) \\ 
\citet{sousa08}     &  ...                   &  ...                  &  ...                  &  ...                 &   ...                 &  ...                  & ...                   &  ...                  &  ...                 &  ...                  \\ 
$<>$                &  0.210$\pm$0.097       & --0.057$\pm$0.079     &   0.036$\pm$0.018     & 0.063$\pm$0.033      & --0.035$\pm$0.047     &   0.073$\pm$0.050     &   0.043$\pm$0.033     & 0.181$\pm$0.038       & ...                  & --0.019$\pm$0.044     \\
\hline       
\multicolumn{1}{c}{} &  & &  &  & & & & & \\
\hline
Line list           & $[$\ion{Ni}{i}/Fe$]$ &   $[$\ion{Cu}{i}/Fe$]$ & $[$\ion{Zn}{i}/Fe$]$ & $[$\ion{Y}{ii}/Fe$]$  & $[$\ion{Zr}{ii}/Fe$]$ & $[$\ion{Ba}{ii}/Fe$]$ & $[$\ion{Ce}{ii}/Fe$]$ & $[$\ion{Y}{ii}/\ion{Mg}{i}$]$ &  $[$\ion{Y}{ii}/\ion{Al}{i}$]$ & \\\hline 
\citet{bensby14}    & 0.064$\pm$0.064 (29) &    ...                 & 0.081$\pm$0.059 (1)  & 0.034$\pm$0.071 (4)   &  ...                  & --0.061$\pm$0.065 (2) &   ...                 & 0.035$\pm$0.128               & --0.001$\pm$0.107              & \\        
\citet{biazzo12}    & 0.061$\pm$0.052 (14) &    ...                 & 0.048$\pm$0.066 (1)  &   ...                 &  ...                  &   ...                 &   ...                 &   ...                         &   ...                          & \\ 
\citet{chen00}      & 0.064$\pm$0.067 (13) &    ...                 &  ...                 &   ...                 &  ...                  & --0.088$\pm$0.085 (2) &   ...                 &   ...                         &   ...                          & \\ 
\citet{feltzing01}  & 0.047$\pm$0.046 (17) &    ...                 &  ...                 &   ...                 &  ...                  &   ...                 &   ...                 &   ...                         &   ...                          & \\            
\citet{jofre14,jofre15}     & 0.072$\pm$0.076 (14) &    ...                 &  ...                 &   ...                 &  ...                  &   ...                 &   ...                 &   ...                         &   ...                          & \\ 
\citet{melendez14}  & 0.057$\pm$0.050 (11) &   0.081$\pm$0.063 (2)  & 0.089$\pm$0.061 (1)  & 0.070$\pm$0.067 (3)   &  ...                  & --0.040$\pm$0.049 (2) & ...                   & --0.010$\pm$0.119             & --0.040$\pm$0.098              & \\ 
\citet{morel14}     & 0.058$\pm$0.056 (8)  &    ...                 &  ...                 &   ...                 &  ...                  & --0.022$\pm$0.074 (1) &   ...                 &   ...                         &   ...                          & \\  
\citet{reddy03}     & 0.052$\pm$0.047 (9)  &   0.065$\pm$0.041 (2)  & 0.077$\pm$0.062 (1)  & 0.043$\pm$0.036 (3)   & 0.122$\pm$0.060 (1)   & --0.031$\pm$0.040 (2) &   ...                 & --0.005$\pm$0.085             & --0.028$\pm$0.079              & \\ 
\citet{sousa08}     &  ...                 &    ...                 &  ...                 &   ...                 &  ...                  &   ...                 &   ...                 &   ...                         &   ...                          & \\ 
$<>$                & 0.057$\pm$0.019      &   0.070$\pm$0.034      & 0.075$\pm$0.031      & 0.047$\pm$0.029       & 0.122$\pm$0.060       & --0.042$\pm$0.025     &              ...      &   0.003$\pm$0.061             & --0.025$\pm$0.053              & \\
\hline\hline       
%\tablefootmark{}
\end{tabular}
\tablefoot{The number in parentheses gives the number of lines used for a given ion. The last row of each subtable gives the average value weighted by the inverse variance.\\
%\tablefoottext{a}{}
%\tablefoottext{b}{}
%\tablefoottext{c}{}
}
\end{sidewaystable*}
%\end{table*}

\end{appendix}

\end{document}